\documentclass[reprint,aps,prb,english,superscriptaddress]{revtex4-2}

\usepackage{graphicx}
\usepackage{bm,amssymb,amsmath,mathrsfs,amstext,latexsym,physics}
\usepackage[T1]{fontenc}

\usepackage[usenames,dvipsnames]{xcolor}
\usepackage[colorlinks=true,citecolor=blue,urlcolor=blue,linkcolor=blue]{hyperref}
\usepackage[normalem]{ulem}

\usepackage{soul}
\setstcolor{red}

\begin{document}

\title{Dynamical properties of particulate composites derived from ultradense stealthy hyperuniform sphere packings}

\author{Carlo Vanoni}
\affiliation{Department of Physics, Princeton University, Princeton, New Jersey, 08544, USA}

\author{Jaeuk Kim}
\affiliation{Department of Physics, Princeton University, Princeton, New Jersey, 08544, USA}
\affiliation{Department of Chemistry, Princeton University, Princeton, New Jersey, 08544, USA}
\affiliation{Princeton Institute for the Science and Technology of Materials,
Princeton University, Princeton, New Jersey 08544, USA}

\author{Paul J. Steinhardt}
\affiliation{Department of Physics, Princeton University, Princeton, New Jersey, 08544, USA}

\author{Salvatore Torquato}
\email{torquato@princeton.edu}
\affiliation{Department of Physics, Princeton University, Princeton, New Jersey, 08544, USA}
\affiliation{Department of Chemistry, Princeton University, Princeton, New Jersey, 08544, USA}
\affiliation{Princeton Institute for the Science and Technology of Materials,
Princeton University, Princeton, New Jersey 08544, USA}
\affiliation{Program in Applied and Computational Mathematics,
Princeton University, Princeton, New Jersey 08544, USA}

\date{\today}


\begin{abstract}
Stealthy hyperuniform (SHU) many-particle systems are distinguished by a structure factor that vanishes not only at zero wave number (as in ``standard'' hyperuniform systems) but also across an extended range of wave numbers near the origin.
We generate disordered SHU packings of identical and `nonoverlapping' spheres in $d$-dimensional Euclidean space using a modified collective-coordinate optimization algorithm that incorporates a soft-core repulsive potential between particles in addition to the standard stealthy pair potential. 
Compared to SHU packings without soft-core repulsions, these SHU packings are ultradense with packing fractions ranging from 0.67-0.86 for $d=2$ and 0.47-0.63 for $d=3$, spanning a broad spectrum of structures depending on the stealthiness parameter $\chi$.
We consider two-phase media composed of hard particles derived from ultradense SHU packings (phase 2) embedded in a matrix phase (phase 1), with varying stealthiness parameter $\chi$ and packing fractions $\phi$.
Our main objective is the estimation of the dynamical physical properties of such two-phase media, namely, the effective dynamic dielectric constant and the time-dependent diffusion spreadability, which is directly related to nuclear magnetic relaxation in fluid-saturated porous media. We show through spreadability that two-phase media derived from ultradense SHU packings exhibit faster interphase diffusion due to the higher packing fractions achievable compared to media obtained without soft-core repulsion. 
The imaginary part of the effective dynamic dielectric constant of SHU packings vanishes at a small wave number, implying perfect transparency for the corresponding wavevectors. While a larger packing fraction yields a smaller transparency interval, we show that it also displays a reduced height of the attenuation peak.
We also obtain cross-property relations between transparency characteristics and long-time behavior of the spreadability for such two-phase media, showing that one leads to information about the other and vice versa. 
Our results demonstrate that disordered two-phase media derived from ultradense SHU packings exhibit advantageous transport and optical behaviors of both theoretical and experimental significance.
\end{abstract}

\maketitle

\section{Introduction} 
\label{sec:intro}

Since the paper by Torquato and Stillinger~\cite{To03a}, hyperuniformity has found a wide variety of contexts and applications and has been at the center of intense studies in the last two decades~\cite{Ga02,To18a,Og17}. 
Roughly defined as the anomalous suppression of large-scale density fluctuations in a many-body system, the notion of hyperuniformity provides a framework to distinguish and characterize a wide variety of systems, including crystals, quasicrystals, and special disordered point configurations. 
A hyperuniform many-particle system in $d$-dimensional Euclidean space $\mathbb{R}^d$ is defined so that the number variance $\sigma_N^2(R) = \langle N(R)^2 \rangle - \langle N(R) \rangle^2$ of particles within a spherical window of radius $R$ grows slower than the volume of the window $R^d$, for large $R$. 

For instance, typical nonhyperuniform systems have number variance with an asymptotic volume scaling $\sigma_N^2(R) \sim R^d$~\cite{To03a}, while crystals and many perfect quasicrystals~\cite{Le84,Og17} have an asymptotic surface-area scaling $\sigma_N^2(R) \sim R^{d-1}$, and are therefore hyperuniform. Nowadays there are many known examples of disordered hyperuniform systems, including perfect glasses~\cite{Zh16a}, one-component plasmas~\cite{Le00}, critical absorbing states of random organization models~\cite{He15,Wiese2024Hyperuniformity,Ma19}, maximally random jammed states~\cite{To00b, Ma23}, random matrices and number theory~\cite{To08b,Mon73}, cold atom processes~\cite{Le14}, active particle systems \cite{Le19b,backofen_nonequilibrium_2024}, biological systems~\cite{Ji14,Ma15,Hu21,ge_hidden_2023} and multihyperuniform systems~\cite{Lo18a,christogeorgos2024computational} (see Ref.~\cite{To18a} for a review and references therein).

For all crystals and disordered hyperuniform systems, hyperuniformity can be equivalently stated as the vanishing of the structure factor $S(\mathbf{k})$ as the wave number $k=|\mathbf{k}|$ goes to zero, i.e., $\lim_{|\mathbf{k}| \to 0} S(\mathbf{k}) = 0$.
A special class of hyperuniform systems is the \emph{stealthy} hyperuniform (SHU) variety~\cite{Uc04b,batten_classical_2008,torquato_ensemble_2015,Zh15a}, defined by a structure factor that vanishes for a range of wave numbers around the origin, i.e., 
\begin{equation}
\label{eq:Sk_SHU}
    S(\mathbf{k}) = 0 \quad \mathrm{for} \quad 0 < k \leq K.
\end{equation} 
SHU systems can be ordered (e.g., perfect crystals) or disordered (i.e., statistically isotropic with no Bragg peaks)~\cite{Uc04b,torquato_ensemble_2015,Zh15a}. The latter are of central interest in the present work for their exotic properties. In particular, we will consider disordered SHU particulate composites, that is, monodisperse sphere packings embedded in a matrix, a special class of two-phase media.
Stealthy configurations are characterized by the stealthiness parameter $\chi$ (see Eq.~\eqref{eq:chi1}), corresponding to the fraction of independently constrained wave vectors relative to the total degrees of freedom \cite{batten_classical_2008}.

Disordered SHU materials are isotropic and have no Bragg peaks. Yet they have characteristics of crystals, including no single scattering  from intermediate to infinite
wavelengths [see Eq.~\eqref{eq:Sk_SHU}] and disallow arbitrarily large holes in the thermodynamic limit~\cite{zhang_can_2017,ghosh_generalized_2017,To18a}.
Consequently, the hybrid crystal/liquid nature of disordered SHU materials endows them with unprecedented and superior transport, elasticity, and wave propagation properties among isotropic amorphous states of matter~\cite{To18a}. 
Examples include wave transparency~\cite{Le16,Fr17,To21a,Fr23,Kl22,kim_effective_2023,alhaitz_experimental_2023,Kim_2024_extraordinary},
enhanced wave absorption~\cite{Bi19},
improved solar cell efficiency~\cite{merkel_stealthy_2023},
tunable localization and diffusive regimes~\cite{Fr17,Sg22,Sc22},
phononic characteristics~\cite{Gk17,Ro19,Roh20},
Luneberg lenses with minimized back-scattering~\cite{Zh19},
extraordinary phased arrays~\cite{Ch21,tang2023hyperuniform},
optimal sampling arrays for 3D ultrasound imaging~\cite{tamraoui_hyperuniform_2023},
high-quality optical cavities~\cite{granchi_nearfield_2023},
and network materials achieving nearly maximal effective electrical conductivities and elastic moduli~\cite{To18c}, and particulate media with nearly maximal effective diffusion coefficients~\cite{zhang_transport_2016}.

\begin{figure}
    \centering
    \includegraphics[width=0.49\linewidth]{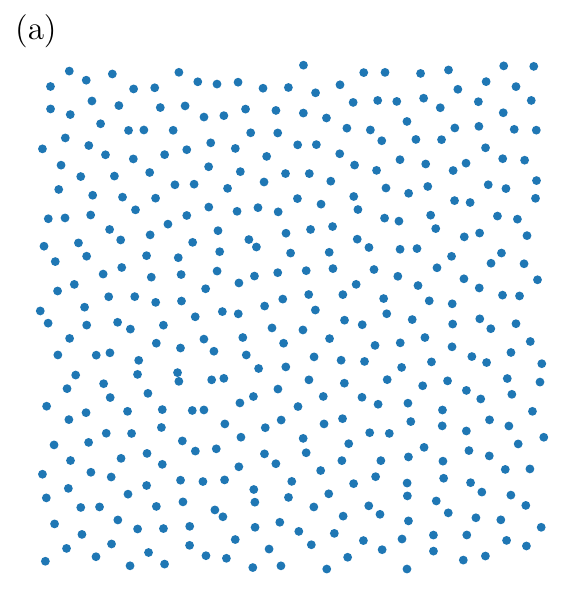}
    \includegraphics[width=0.49\linewidth]{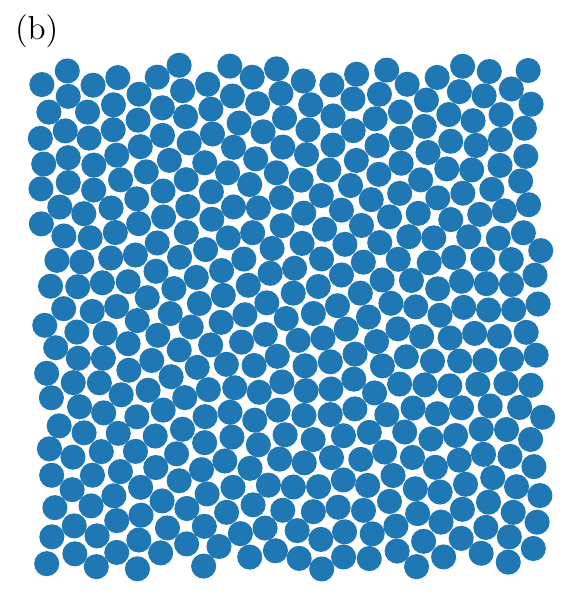}
    \caption{Stealthy disordered hyperuniform packing of $N=400$ identical disks, with stealthiness parameter $\chi=0.35$, as adapted from Ref.~\cite{Kim_2025_DenseSphere}. (a) The packing configuration is generated using the collective coordinate optimization procedure without soft-core repulsion, leading to a packing fraction $\phi \simeq 0.095$. (b) For the same value of $\chi$, adding the soft-core repulsion allows achieving ultradense packings, in the present case with $\phi = 0.76$.}
    \label{fig:fig_1}
\end{figure}

As anticipated, our primary interest in this paper is the study of the dynamical physical properties of specific two-phase media consisting of highly dense disordered SHU sphere packings within a continuous matrix. 
In particular, we examine the diffusive transport properties as quantified by the time-dependent diffusion spreadability~\cite{To21d}, and optical properties as quantified by the effective dynamic dielectric constant~\cite{To21a}. In addition, we address the cross-property relations between these two properties in ultradense disordered SHU particulate composites.
We follow the recent investigation by Kim and Torquato~\cite{Kim_2025_DenseSphere} to generate ultradense SHU monodisperse sphere packings by using a modified collective-coordinate optimization procedure to study the aforementioned properties. This recent optimization scheme incorporates a two-body short-range soft-core repulsive potential between particles in addition to the standard
stealthy potential, as detailed in Sec.~\ref{sec:backgr:collcoor}. This modification leads to packings whose densities far exceed those achieved without the soft-core repulsion; see Fig.~\ref{fig:fig_1}
for a vivid comparative illustration. We quantify the density of the packing using the packing fraction $\phi = \rho \, v_1(R)$, defined as the fraction of space covered by the spheres, where $\rho$ is the number density and $v_1(R)$ the volume of a sphere of radius $R$.

\begin{figure}
    \centering
    \includegraphics[width=\linewidth]{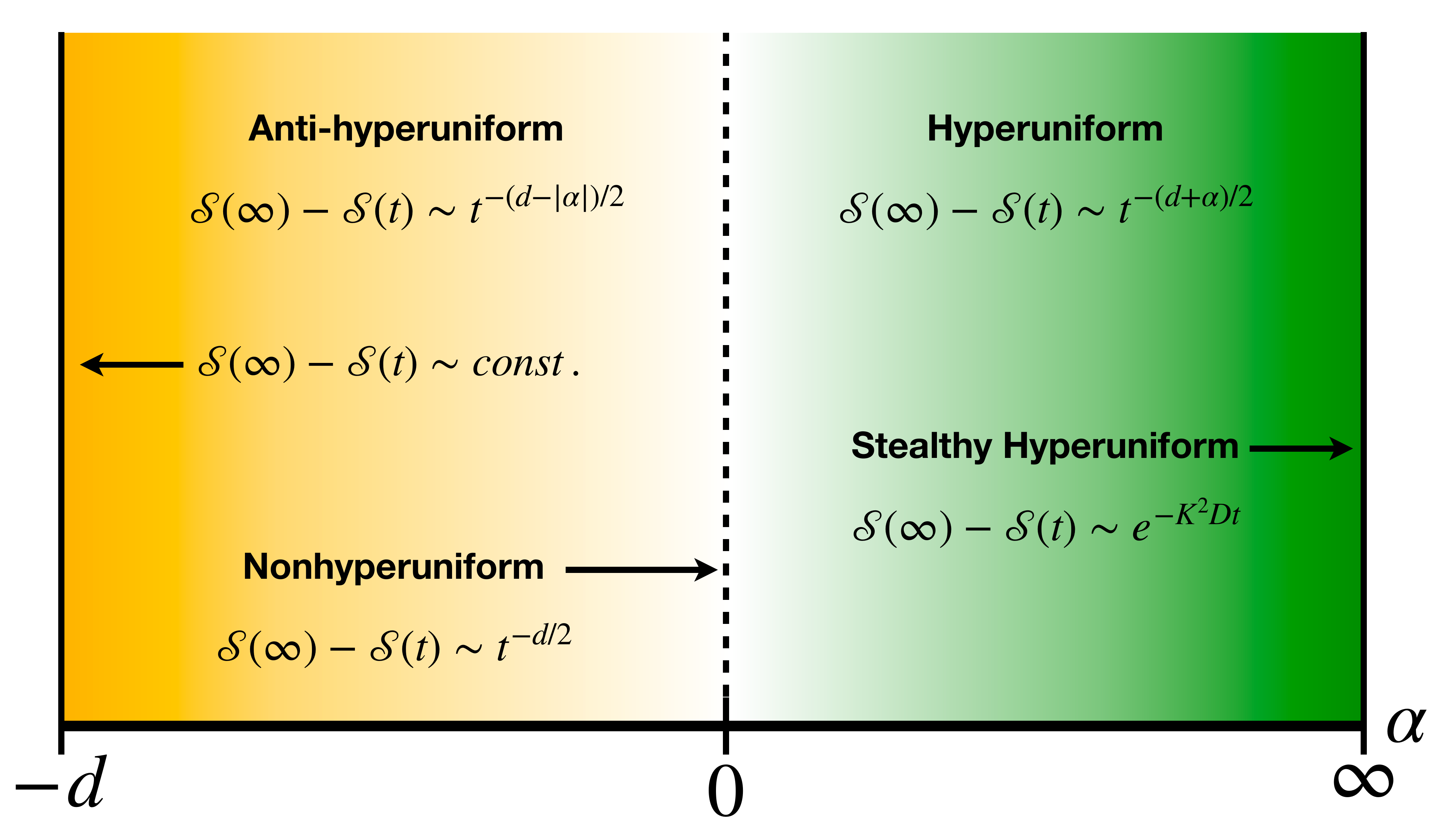}
    \caption{Schematic phase diagram of the excess spreadability $\mathcal{S}(\infty)- \mathcal{S}(t)$ spectrum~\cite{To21d}, as a function of $\alpha$, the hyperuniformity exponent. For $\alpha = 0$ the system is neither hyperuniform nor anti-hyperuniform, and the usual behavior is recovered. The excess spreadability decays as a power law for all finite values $\alpha > -d$, approaching a constant in the extreme anti-hyperuniform limit $\alpha = -d$. In the other limit $\alpha \to \infty$, the system is hyperuniform and the decay of the excess spreadability is exponential.}
    \label{fig:phase_diag_HU}
\end{figure}

The time-dependent diffusion spreadability $\mathcal{S}(t)$ quantifies the efficiency of a two-phase medium in diffusing a substance from one phase to the other. 
The larger the value of $\mathcal{S}(t)$, the faster the diffusion of the substance in the medium~\cite{To21d}. 
It has been shown that, in the long-time limit, hyperuniform media have considerably larger spreadability compared to non-hyperuniform systems, and such a difference becomes exponentially large in disordered SHU systems~\cite{To21d,Kim_2024_extraordinary}, as summarized in Fig.~\ref{fig:phase_diag_HU}. 
Compared with previous works that were limited in the packing fraction that could be achieved, our results show that the addition of the soft-core repulsion allows us to tune the rate of such exponential decay by adjusting the radius $\sigma$ of the repulsive potential. Specifically, the larger the packing fraction, the faster the diffusion. 
As a consequence, the addition of soft-core repulsion provides a way to design SHU microstructures with extremely efficient spreadability. Moreover, we show that for fixed values of the stealthiness parameter $\chi$ and packing fraction $\phi$, the microstructure of configurations obtained with larger $\sigma$ leads to faster interphase diffusion.

The effective dynamic dielectric constant tensor $\mathbf{\epsilon}_e(\mathbf{k},\omega)$ relates the externally applied electric field to the field inside the medium. 
Recently Torquato and Kim~\cite{To21a} derived exact general nonlocal strong-contrast expansions for a linear fractional transformation of the effective dielectric constant tensor $\mathbf{\epsilon}_e(\mathbf{k},\omega)$. 
This expansion offers the significant benefit of leading to the rapid convergence of the series and hence their lower-order truncations can accurately approximate higher-order functionals of the exact series across all orders using lower-order diagrams.
In the present work, we will consider specific fully 3D isotropic media consisting of 3D SHU sphere packings and transversely isotropic media, whose cross sections are obtained from the 2D disk packings as detailed in Sec.~\ref{sec:back:EDDC}. 
We then apply the truncated strong-contrast formulas to estimate $\mathbf{\epsilon}_e(\mathbf{k},\omega)$ and compare disordered SHU media generated with and without soft-core repulsion. 
We find that, in agreement with previous results~\cite{To21a, kim_effective_2023, kim_accurate_2024}, the smaller the packing fraction, the larger the size of the transparency interval at a small wave number. 
On the other hand, the height of the attenuation peak decreases for increasing packing fraction, as the medium becomes progressively more ordered. 

We find that disordered ultradense SHU two-phase media present advantageous physical properties compared to SHU obtainable without soft-core repulsion, such as increased diffusion spreadability and reduced attenuation of electromagnetic waves. Furthermore, SHU packings generated with soft-core repulsion with larger values of $\sigma$, for fixed stealthiness parameter and packing fraction, exhibit the same favorable physical properties compared to smaller values of $\sigma$, emphasizing the importance of the microstructural arrangement. 

Importantly, the diffusion spreadability and the effective dynamic dielectric constant are determined by the microstructural information contained in the spectral density only, as explained in Sec.~\ref{sec:back:spread} and~\ref{sec:back:EDDC}. The spectral density can be obtained from scattering data of two-phase media and allows us to determine the exact result for the spreadability~\cite{To21d} and highly accurate approximations for the effective dynamic dielectric constant~\cite{To21a}. It is noteworthy that our property predictions enable the tuning and design of material properties by selecting the desired spectral densities.

A previous study showed that the imaginary parts of the effective dynamic dielectric constant at small wave numbers and the spreadability at long times are positively correlated as the structures span from nonhyperuniform, nonstealthy hyperuniform, and stealthy hyperuniform media~\cite{Kim_2024_extraordinary}.
In this work, we explore how the unprecedented possibility of achieving ultradense disordered SHU packings enables new cross-property relations between analogous quantities and reveals remarkable effects of high packing fractions on physical properties not reported in Ref.~\cite{Kim_2024_extraordinary}. 
In particular, we will show correlation between the decay rate of the excess diffusion spreadability and the size of the optical transparency interval. 
The existence of such cross-property relations will be linked to the dependence of the physical properties we will investigate on the spectral density only, as we will detail.

The rest of the paper is organized as follows. In Sec.~\ref{sec:backgr:main} we provide an introduction to the fundamental concepts that will be necessary to understand our results, and in particular, we will give a background on SHU systems in Sec.~\ref{sec:backgr:hyperun}, we will describe the collective-coordinate optimization procedure in Sec.~\ref{sec:backgr:collcoor}, and in Sec.~\ref{sec:backgr:twophase} we will give a brief discussion of two-phase media, focusing on diffusion spreadability in Sec.~\ref{sec:back:spread} and effective dynamic dielectric constant in Sec.~\ref{sec:back:EDDC}. We will discuss the results for the spectral density in Sec.~\ref{sec:res:spectr_dens}, time-dependent diffusion spreadability in Sec.~\ref{sec:diff_spread}, and the effective dynamic dielectric constant in Sec.~\ref{sec:dieleconst}. We will then draw our conclusions in Sec.~\ref{sec:conclusions}. In the Supplemental Material (SM), we discuss the effect of different microstructures, obtained using different soft-core repulsion potentials.


\section{Background and methods}
\label{sec:backgr:main}

In this Section, we provide some background materials that are useful to understand the results presented in subsequent sections of the paper.



\subsection{Hyperuniformity}
\label{sec:backgr:hyperun}

As anticipated in the Introduction Sec.~\ref{sec:intro}, a hyperuniform system is characterized by a suppression of the large-scale density fluctuations compared to those of typical disordered systems. To be more precise, let us consider a many-particle system in $d$-dimensional Euclidean space $\mathbb{R}^d$ and consider a hyperspherical observation window $\Omega$ of radius $R$. 
The large-$R$ behavior of the number variance $\sigma^2_N(R) \equiv \langle N(R)^2 \rangle - \langle N(R) \rangle^2 $ of particles within $\Omega$ is used to distinguish hyperuniform systems from typical disordered systems.

The structure factor  $S(\mathbf{k})$ is directly related to the number variance through the integral equation~\cite{To03a,To18a}
\begin{equation}
\label{eq:sigma_F_space}
    \sigma^2_N(R) = \langle N(R) \rangle \left[\frac{1}{(2\pi)^d} \int_{\mathbb{R}^d} S(\mathbf{k}) \tilde{\alpha}(k;R) d {\bf k}\right]
\end{equation}
where $\tilde{\alpha}(k;R) = 2^d \pi^{d/2} \Gamma(1+d/2)\ [J_{d/2}(kR)]^2/k^d$.

According to Eq.~\eqref{eq:sigma_F_space}, the large-$R$ behavior of the number variance $\sigma_N^2(R)$ is governed by the small-$k$ scaling of structure factor $S(\mathbf{k})$. 
Specifically, if we consider hyperuniform systems for which $S(\mathbf{k}) \sim |\mathbf{k}|^{\alpha}$ at small wave numbers, the corresponding large-scale behavior of $\sigma^2_N(R)$ \cite{Za09,To18a} is given by 
\begin{align*}
\sigma^2_N(R) = 
\begin{cases}
    R^{d-1},  & \text{if } \alpha > 1 \quad \text{(Class I)},\\
    R^{d-1} \ln R,  & \text{if } \alpha = 1 \quad \text{(Class II)},\\
    R^{d-\alpha}, & \text{if } 0 < \alpha < 1 \quad \text{(Class III)}.
\end{cases}
\end{align*}
Among the class I hyperuniform systems, \emph{stealthy hyperuniform systems} formally correspond to the $\alpha \to \infty$ limit. In SHU systems the structure factor does not vanish only for $|\mathbf{k}| \to 0$, but rather for a finite interval $0 \leq |\mathbf{k}| < K$ at small wave numbers,
\begin{equation}
\label{eq:stalthy_Sk}
    S(\mathbf{k}) = 0 \qquad \mathrm{for} \quad 0< k \leq K.
\end{equation}
Notice that perfect crystals are SHU systems, as their structure factor vanishes for all wave numbers $|\mathbf{k}| < |\mathbf{k}_{\mathrm{Bragg}}|$ smaller than the position of the first Bragg peak. 

\begin{figure}
    \centering
    \includegraphics[width=\linewidth]{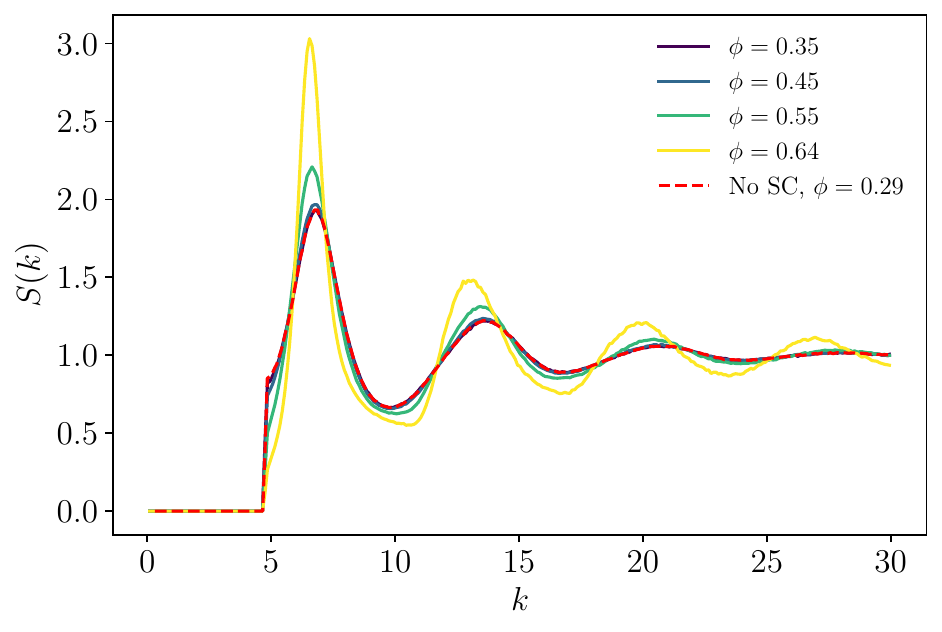}
    \caption{Structure factor for disordered 2D SHU packings with $\chi = 0.45$ and different values of the packing fraction $\phi$. All the curves are obtained using the collective-coordinate optimization procedure: the solid curves are generated with the addition of a soft-core repulsion, allowing tuning of the packing fraction, whereas the dashed red curve has no soft-core repulsion.}
    \label{fig:Sk_chi0.45}
\end{figure}

Such systems display another important feature called the bounded-hole property. In a many-particle system, a “hole'' is a spherical region of the Euclidean space $\mathbb{R}^d$ of a certain radius without particle centers. 
It has been conjectured~\cite{zhang_can_2017} and subsequently rigorously proved~\cite{ghosh_generalized_2017} that disordered SHU configurations across any dimension cannot tolerate arbitrarily large holes in the infinite system-size limit. On the other hand, in a typical disordered system, the probability of finding a hole of arbitrarily large size in the thermodynamic limit is non-vanishing.

The properties of disordered SHU systems depend on the value of $K$, {\it i.e.} the number $M(K)$ of independently constrained wave vectors. It is convenient to introduce the stealthiness parameter $\chi$, defined as
\begin{equation}
\label{eq:chi1}
    \chi = \frac{M(K)}{d(N-1)},
\end{equation}
which gives a measure of the relative fraction of constrained degrees of freedom compared to the total number of degrees of freedom $d(N-1)$, once the translational degrees of freedom are removed. In the thermodynamic limit, Eq.~\eqref{eq:chi1} becomes~\cite{torquato_ensemble_2015}
\begin{equation}
\label{eq:stealthy_per}
    \chi = \frac{v_1(K)}{2 \, \rho \, d \, (2\pi)^d},\qquad v_1(R) = \frac{\pi^{d/2}}{\Gamma(1+d/2)}R^d,  
\end{equation}
being $v_1(R)$ the volume of a $d$-dimensional hypersphere of radius $R$, and $\rho$ is the number density. 
As we will discuss in Sec.~\ref{sec:backgr:collcoor}, SHU configurations can be obtained as ground states of a potential energy.
Such ground states are highly degenerate and markedly disordered for small values of $\chi \gtrsim 0$, as few constraints are applied to the system. For larger values of $\chi$, the dimensionality $d_C$ of the configuration space reduces and becomes zero for $\chi = 1/2$, according to the equation $d_C = d(1-2\chi)$ for $0 \leq \chi \leq 1/2$~\cite{torquato_ensemble_2015}. 
For $\chi = 1/2$, there can be transitions to a crystal phase, depending on the value of $d$. Notice that periodic configurations belong to the ground state manifold, but constitute a zero-measure set, and therefore cannot be reached through the collective-coordinate optimization procedure discussed in the next section.


\subsection{Collective-coordinate optimization with additional soft-core repulsion}
\label{sec:backgr:collcoor}

In the present work, we will obtain disordered SHU configurations by using a modified version of the Collective-Coordinate Optimization (CCO) procedure~\cite{torquato_ensemble_2015}. 
Following Ref.~\cite{zhang_can_2017}, we consider the following potential energy
\begin{equation}
\label{eq:soft_core}
    \Phi(\mathbf{r}^N) = \frac{N}{2V_{\mathcal{F}}} \sum_{\mathbf{k}} \tilde{v}(\mathbf{k}) S(\mathbf{k}) + \sum_{i<j} u(r_{ij}),
\end{equation}
with $\tilde{v}(\mathbf{k}) = V(k) \Theta(K-k)$.
In the above equation, $S(\mathbf{k})$ is the single-configuration structure factor, which by definition is always positive, and $\Theta(x)$ is the Heaviside step function. The pair potential $u(r)$ penalizes any configuration possessing particles closer than $\sigma$. We will choose the pair potential $u(r) = u_0 \, (1 - r/\sigma)^2 \ \Theta(\sigma - r)$. By choosing $V(k) \geq 0$ ({\it e.g.} $V(k) = 1$), we have that $\Phi(\mathbf{r}^N) \geq 0$ and, in particular, $\Phi(\mathbf{r}^N) = 0$ only if $S(0\leq \mathbf{k} <K) = 0$ and the minimum pair distance is $\sigma$. Therefore, the ground states of the potential $\Phi(\mathbf{r}^N)$ are SHU configurations with the desired value of $K$ and minimum pair distance $\sigma$. 

In this work, we will employ the same potential used in Ref.~\cite{Kim_2025_DenseSphere}, although the specific form of the potential is not expected to change the qualitative properties of the ground state configurations. However, different choices of $u(r)$ could change the numerical efficiency of the procedure. With our selection, the potential is continuous and differentiable at $r=\sigma$.

The addition of the repulsive potential adds a constraint to the ground state manifold, and therefore not all values of $\sigma$ are allowed: for a given value of $\chi$, by increasing $\sigma$ the dimensionality of the ground state manifold reduces and can ultimately become zero at $\sigma = \sigma_c$. 
The dependence of $\sigma_c$ on $\chi$ and spacial dimension $d$ has been investigated in Ref.~\cite{Kim_2025_DenseSphere}. 
In particular, $\sigma_c$ decreases with increasing $\chi$, meaning that packing fraction monotonically increases with decreasing $\chi$.
This dependence arises because a decrease in $\chi$ (or, equivalently, the degrees of freedom $M(K)$ constrained by the stealthy hyperuniform condition) increases the maximum number of constraints from particles that are effectively in contact. Such a change consequently leads to a decrease in the void space between particles and a higher maximum packing fraction; see Ref. \cite{Kim_2025_DenseSphere} for details.

As in previous implementations of the collective-coordinate procedure~\cite{torquato_ensemble_2015,Zh15a,Kim_2025_DenseSphere}, we numerically obtain the point configurations by minimizing the potential energy $\Phi$ in Eq.~\eqref{eq:soft_core} using the low-storage Broyden-Fletcher-Goldfarb-Shanno (L-BFGS) algorithm~\cite{Liu89}.
For given values of $d$, $\chi$, $N$, and $\sigma$, we draw a random initial configuration and run the minimization algorithm till when $\Phi < 7\chi \times 10^{-20}$.

The statistical properties of the ground state configurations obtained through the optimization procedure are affected by the choice of $V(k)$ and initial condition~\cite{Zh15a}. For example, ordered periodic configurations belong to the ground state manifold, but cannot be reached from random initial conditions (often denoted as high-temperature initial configurations), as they are entropically suppressed. In this work, we will consider random initial conditions with unit number density $\rho = 1$. Given a point configuration, we then obtain a sphere packing by decorating all points with spheres of radius $a=\sigma/2$.

Kim and Torquato~\cite{Kim_2025_DenseSphere} showed that, without soft-core repulsion, the average maximum packing fraction $\phi_{\mathrm{max}}$ vanishes as $N\to \infty$, while the inclusion of the soft-core repulsion allows one to achieve unprecedented large values of $\phi_{\mathrm{max}}$ independently of $N$.
In this work, we make use of this result and, for a given value of $d$ and $\chi$, we consider the maximal packing fraction $\phi_{\mathrm{max}} = \rho \, v_1(\sigma_c/2)$ for fixed number density $\rho$. Figure~\ref{fig:Sk_chi0.45} demonstrates the structure factors of such 2D SHU packings with a fixed value of $\chi=0.45$ for a range of packing fractions up to the maximal value. 
When $\phi$ is sufficiently small, the structure factor of such a packing becomes nearly identical to those of SHU point patterns without soft-core repulsion.


\subsection{Two-phase media}
\label{sec:backgr:twophase}

In the following, we briefly introduce the two-phase media formalism, which will be instrumental in the study of particulate composites.
Let us consider a medium constituted of two phases $i=1, \ 2$. In our case, phase $i=1$ will be the matrix filling the space between spheres and phase $i=2$ the spheres themselves, but the discussion presented in this Section is valid in general.
We define the indicator function
\begin{equation}
    \mathcal{I}^{(i)}(\mathbf{x}) = 
    \begin{cases}
        1 \quad \mathrm{if \,} \mathbf{x}\in \mathcal{V}_i\\
        0 \quad \mathrm{otherwise},
    \end{cases}
\end{equation}
being $\mathcal{V}_i$ the region of space occupied by phase $i$, with volume fraction $\phi_i$~\cite{To02a}.
We can then define the $n$-point correlation function for phase $i$ as $S_n^{(i)} =\langle \prod_{j=1}^n \mathcal{I}^{(i)}(\mathbf{x}_j) \rangle $, where the average is performed over the realization ensemble. For statistically homogeneous media, it follows immediately that $S_1^{(i)} = \phi_i$. The autocovariance function $\chi_{_{V}}(\mathbf{r})$ for statistically homogeneous systems is defined through the two-point function as
\begin{equation}
\label{eq:autocov}
    \chi_{_{V}}(\mathbf{r}) \equiv S_2^{(1)}(\mathbf{r}) - \phi_1^2 = S_2^{(2)}(\mathbf{r}) - \phi_2^2,
\end{equation}
which is the same for both phases. Notice that $\chi_{_{V}}(\mathbf{0}) = \phi_1 \phi_2$. The spectral density is defined as the Fourier transform of the autocorrelation function
\begin{equation}
    \tilde{\chi}_{_{V}}(\mathbf{k}) = \int_{\mathbb{R}^d} d \mathbf{r} \, \chi_{_{V}}(\mathbf{r}) \, e^{-i \mathbf{k}\cdot \mathbf{r}}
\end{equation}
and $\tilde{\chi}_{_{V}}(\mathbf{k}) \geq 0$ for all $\mathbf{k}$. 

Let us now specialize to the case of isotropic packings of identical spheres of radius $a$; the spectral density is directly related to the structure factor $S(\mathbf{k})$~\cite{To02a,To16b} and depends only on the wave number $k = |\mathbf{k}|$
\begin{equation}
\label{eq:spectr_dens_1}
    \tilde{\chi}_{_{V}}(k) = \phi_2 \, \tilde{\alpha}_2(k;a) \, S(k),
\end{equation}
where
\begin{equation}
\label{eq:alpha_tilde}
    \tilde{\alpha}_2(k;a) = \frac{1}{v_1(a)} \left(\frac{2\pi a}{k} \right)^d J^2_{d/2}(ka).
\end{equation}
$v_1(a) = \pi^{d/2} a^d / \Gamma(1+d/2)$ is the $d$-dimensional volume of a sphere of radius $a$ and $\phi_2 = \rho \, v_1(a)$. 
Consequently, in SHU packings, $\tilde{\chi}_{_{V}}(k) = 0$ for $0 < k \leq K$.
The large-$k$ behavior is given by $\tilde{\chi}_{_{V}}(k) \sim \gamma(d)s/k^{d+1}$~\cite{To21d}, where $\gamma(d)$ is a dimension-dependent constant and $s$ is the specific surface.

\subsection{Diffusion Spreadability}
\label{sec:back:spread}

Consider a two-phase medium and suppose a solute is present initially only in phase 2 -- the spheres in our case -- and that, as time passes, the solute diffuses from phase 2 to a liquid phase (phase 1). Also, for simplicity, we consider the case in which the diffusion coefficient $D$ is the same in both phases at any time $t$~\footnote{This condition can be relaxed to the general case of different phase diffusion constants without affecting the long-time behavior, as discussed in Ref.~\cite{To21d}.}.
The time-dependent spreadability $\mathcal{S}(t)$ is defined as the fraction of the total amount of solute present that has diffused into phase 1 at time $t$. Given two different microstructures, the one with a larger value of $\mathcal{S}(t)$ at time $t$ spreads diffusion information more rapidly. From the definition, it follows immediately that $\mathcal{S}(\infty) = \phi_1$.

It was demonstrated by Torquato~\cite{To21d} that the excess spreadability $\mathcal{S}(\infty) - \mathcal{S}(t)$ of a two-phase medium in $d$-dimensional Euclidean space $\mathbb{R}^d$ can be expressed exactly as
\begin{equation}
\label{eq:exc_spread}
    \mathcal{S}(\infty) - \mathcal{S}(t) = \frac{1}{(2\pi)^d \phi_2} \int_{\mathbb{R}^d} \tilde{\chi}_{_{V}}(\mathbf{k}) e^{-k^2 D t}d \mathbf{k}.
\end{equation}
Note that only the one-point (i.e., $\phi_i$) and two-point (i.e., $\tilde{\chi}_{_{V}}(\mathbf{k})$) correlations are involved in Eq.~\eqref{eq:exc_spread}.

As shown in Fig. 2 (see also Ref. [52]), if $\tilde{\chi}_{_{V}}(\mathbf{k}) \sim B |\mathbf{k}|^{\alpha}$ in the limit $|\mathbf{k}| \to 0$, the long time excess spreadability is given by
\begin{align}
    \mathcal{S}(\infty) - \mathcal{S}(t) =& \frac{C}{(Dt/a^2)^{(d+\alpha)/2}}\\
    &+ o\left( (Dt/a^2)^{-(d+\alpha)/2} \right),\quad Dt/a^2 \gg 1,
\end{align}
where
\begin{equation}
    C= B \ \Gamma \left(\frac{d+\alpha}{2} \right) \frac{\phi_2}{2^d \pi^{d/2} \Gamma(d/2)}.
\end{equation}
Consequently, the larger the value of $\alpha$ for fixed space dimensionality $d$, the faster the power-law approach of the spreadability to its asymptotic value $\phi_1$. In stealthy systems, as we discussed, $\alpha = +\infty$, and the asymptotic value is approached exponentially fast~\cite{To21d}:
\begin{equation}
\label{eq:spread_stealthy}
    \mathcal{S}(\infty) - \mathcal{S}(t) \sim \left[\frac{d v_1(1)}{2(2\pi)^d} \tilde{\alpha}_2(K;a) S(K)\right] \frac{e^{-K^2 Dt}}{K^2 Dt},
\end{equation}
for $Dt/a^2 \gg 1$.

Remarkably, the spreadability is closely related to nuclear magnetic resonance (NMR) and diffusion magnetic resonance imaging (dMRI) measurements. Indeed, it has been shown by Torquato~\cite{To21d} that the NMR pulsed field gradient spin-echo (PFGSE) amplitude $\mathcal{M}(\mathbf{k},t)$ of a fluid-saturated porous medium~\cite{Mit92b,Mi93,Or02} can be related to the excess spreadability in Eq.~\eqref{eq:exc_spread} via the mapping $\mathcal{S}(\infty) - \mathcal{S}(t) \to \mathcal{M}(\mathbf{0},t) - \phi_2$ and $D \to \mathcal{D}(t)$, being $\mathcal{D}(t)$ the effective time-dependent diffusion coefficient of the porous medium. In addition, the transverse relaxation rate $\dv*{R_2}{t}$ shows long-time asymptotic behaviors analogous to $\mathcal{S}(\infty)-\mathcal{S}(t)$ \cite{Ruh_2023_Observation}.
In the context of dMRI, it has been argued~\cite{No14} and then experimentally verified~\cite{Le20} that the time-dependent diffusion coefficient $\mathcal{D}(t)$ scales as $\mathcal{D}(t) - \mathcal{D}_e \sim C/t^{(d+\alpha)/2}$. 
Such a long-time behavior coincides with that of the excess spreadability, suggesting the mapping $\mathcal{S}(t) \to \mathcal{D}(t)$ and $\mathcal{S}(\infty) \to \mathcal{D}_e = \mathcal{D}(\infty)$. 
These remarkable connections show the experimental relevance of the diffusion spreadability in NMR and dMRI measurements in physical and biological porous media.

\subsection{Effective Dynamic Dielectric Constant}
\label{sec:back:EDDC}

\begin{figure}
    \centering
    \includegraphics[width=0.45\linewidth]{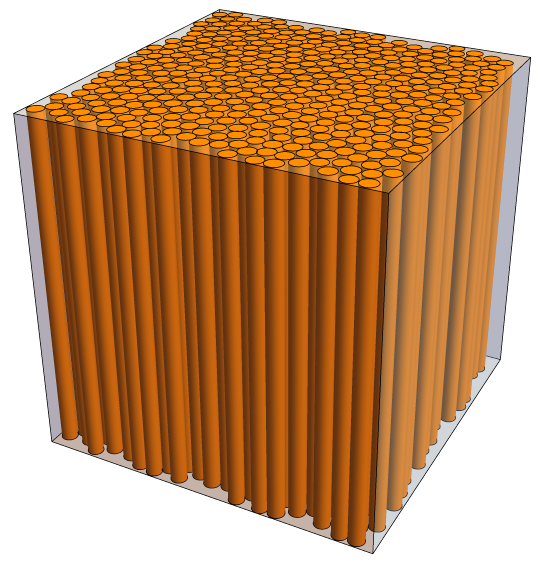}
    \includegraphics[width=0.45\linewidth]{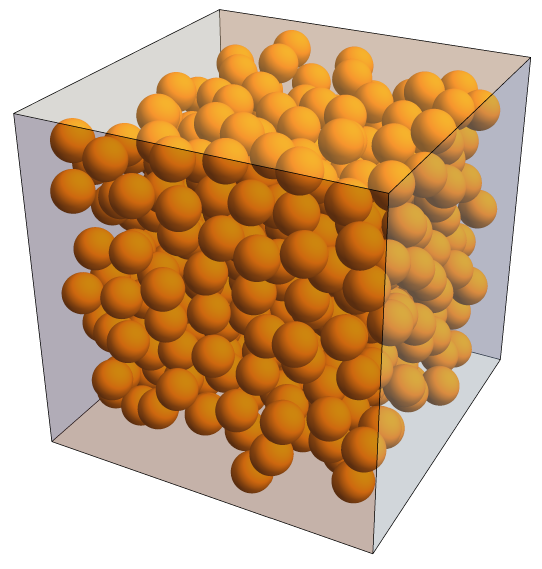}
    \caption{(Left) 3D transversely isotropic medium, obtained from a 2D SHU packing. (Right) 3D fully isotropic medium, consisting of a 3D SHU packing.}
    \label{fig:2_3Dmedia}
\end{figure}

In this Section, we will briefly review the nonlocal strong-contrast expansion of the effective dynamic dielectric constant tensor $\mathbf{\epsilon}_e(\mathbf{k}_q,\omega)$ in 3D anisotropic two-phase media. We denote with $\mathbf{E}_0(\mathbf{x})$ the incident electric field, which is assumed to be a plane electric wave with frequency $\omega$ and wavevector $\mathbf{k}_q$ in phase $q$. We will also assume that both phases are nonmagnetic, have real-valued and frequency-independent dielectric constants, and are dielectrically isotropic. Consequently, we have the dispersion relation $k(\omega) \equiv |\mathbf{k}_q(\omega)| = \sqrt{\epsilon_q} \ \omega/c$. We can thus drop the dependence on $\omega$ of the effective dynamic dielectric constant tensor, $\mathbf{\epsilon}_e(\mathbf{k}_q,\omega) = \mathbf{\epsilon}_e(\mathbf{k}_q)$.

The nonlocal strong-contrast expansion is obtained starting from the dyadic Green's function $\mathbf{G}^{(q)}(\mathbf{r})$ for the electric field in the reference phase $q$ of the time-harmonic vector wave equation. It can be written as~\cite{To02a}
\begin{equation}
    \mathbf{G}^{(q)}(\mathbf{r}) = -\mathbf{D}^{(q)} \delta(\mathbf{r}) + \mathbf{H}^{(q)}(\mathbf{r}),
\end{equation}
where the singular part at $\mathbf{r}=0$ has been isolated from the regular part $\mathbf{H}^{(q)}(\mathbf{r})$, and the source dyadic tensor $\mathbf{D}^{(q)}$ accounts for the contribution around the singularity. The regular part of the Green's function, for isotropic media, takes the form
\begin{align}
\mathbf{H}^{(q)}(\mathbf{r}) =& \frac{e^{ikr}}{\epsilon_q 4\pi r^3} 
\left[
\left(-1 + i k r + (kr)^2\right)\mathbf{I}\right.\\ \nonumber
&+ 
\left.\left(3 - 3ikr - (kr)^2\right)\hat{\mathbf{r}}\hat{\mathbf{r}}
\right],
\end{align}
with $\hat{\mathbf{r}} \equiv \mathbf{r}/|\mathbf{r}|$.

The nonlocal strong-contrast expansion is \emph{not} a series expansion for the effective dielectric constant $\mathbf{\epsilon}_e(\mathbf{k}_q)$, but rather for the effective tensor $\left[ L_e^{(q)}(\mathbf{k_q}) \right]^{-1}$, where
\begin{equation}
    L_e^{(q)}(\mathbf{k_q}) \equiv \left[\epsilon_e(\mathbf{k_q}) - \epsilon_q\mathbf{I}\right] 
\cdot \left\{ \mathbf{I} + \mathbf{D}^{(q)} 
\cdot \left[\epsilon_e(\mathbf{k_q}) - \epsilon_q\mathbf{I}\right] \right\}^{-1}.
\end{equation}
The expansion reads
\begin{equation}
\label{eq:NLSCE}
    \phi_p L_p^{(q)} \cdot \left[L_e^{(q)}(\mathbf{k_q})\right]^{-1} \cdot \phi_p L_p^{(q)}
= \phi_p L_p^{(q)} - \sum_{n=2}^\infty \mathcal{A}_n^{(p)}(\mathbf{k_q}),
\end{equation}
where we have defined
\begin{equation}
    L_p^{(q)} \equiv (\epsilon_p - \epsilon_q) 
\left[ \mathbf{I} + \mathbf{D}^{(q)} (\epsilon_p - \epsilon_q) \right]^{-1}.
\end{equation}
The advantage of the expansion Eq.~\eqref{eq:NLSCE} for a linear fractional transformation of $\mathbf{\epsilon}_e(\mathbf{k}_q)$ compared to a standard series expansion for $\mathbf{\epsilon}_e(\mathbf{k}_q)$ is due to the rapid convergence obtained by truncating the left-hand side of Eq.~\eqref{eq:NLSCE} at low order. It provides good approximations for the exact series to all orders of higher-order functionals in terms of lower-order diagrams~\cite{To21a}.

In the following paragraphs, we provide a brief review of the second-order approximations that we used in this work. We refer the reader to Refs.~\cite{To21a,Kim_2024_extraordinary} for the analytical derivations and further details.

\subsubsection{Transversely isotropic media}
\label{sec:transviso}

Let us consider first the case of 3D transversely isotropic media (see Fig.~\ref{fig:2_3Dmedia} (left)), whose transverse sections are a realization of a 2D packing of disks. We report the results for the effective dynamic dielectric constant of a normally incident wave with $\mathbf{k}_q = k_q \hat{\mathbf{y}}$, being $\hat{\mathbf{y}}$ the unit vector in the $y$-direction. The label $q = 1,2 $ denotes the phase in the medium. As a consequence of the symmetries of the problem, the effective dielectric constant tensor can be decomposed as $\mathbf{\epsilon}_e(k_q) = \epsilon^{TM}_e(k_q)\ \hat{\mathbf{z}}\hat{\mathbf{z}} + \epsilon^{TE}_e(k_q)\ (\mathbf{I}-\hat{\mathbf{z}}\hat{\mathbf{z}})$ into two orthogonal components $\epsilon^{TM}_e(k_q)$ and $\epsilon^{TE}_e(k_q)$, denoting the transverse magnetic (TM) and transverse electric (TE) polarizations, respectively.

By considering contributions only up to the second-order terms, the components of $\mathbf{\epsilon}_e(k_q)$ take the form
\begin{widetext}
\begin{equation}
\label{eq:TM_2D}
    \frac{\epsilon_e^{TM}(k_q)}{\epsilon_q} = 
    1 + \frac{\phi_p^2 \left[(\epsilon_p + \epsilon_q)\beta_{pq}\right]}
    {\phi_p - A_2^{TM}(k_\ast^{TM}, \langle \epsilon \rangle) \left[(\epsilon_p + \epsilon_q)\beta_{pq}\right] 
    + 2\phi_p^2\beta_{pq}}
\end{equation}

\begin{equation}
\label{eq:TE_2D}
    \frac{\epsilon_e^{TE}(k_q)}{\epsilon_q} = 
    1 + \frac{\phi_p(1 - \phi_p\beta_{pq}) 
    - A_2^{TE}\left(k_\ast^{TE}; \epsilon_{BG}^{(2D)}\right)\left[2\epsilon_q\beta_{pq}\right]}
    {2\epsilon_q\beta_{pq}}.
\end{equation}
where $\beta_{pq} = (\epsilon_p - \epsilon_q)/(\epsilon_p + (d-1)\epsilon_q)$ (we set $d=2$), $k_*^{TE} \equiv k_q \sqrt{\epsilon_{BG}^{(2D)}/\epsilon_q}$, $k_*^{TM} \equiv k_q \sqrt{\langle \epsilon \rangle/\epsilon_q}$, and $\epsilon_{BG}^{(2D)}$ is the Bruggeman approximation for two-phase media~\cite{To02a,bruggeman_berechnung_1935}.
The second-order coefficients $A_2^{TM}(k_q; \epsilon_q)$ and $A_2^{TE}(k_q; \epsilon_q)$ are given by
\begin{align*}
\label{eq:A_2D}
    A_2^{TM}(k_q; \epsilon_q) &= 2 A_2^{TE}(k_q; \epsilon_q) = -\frac{\pi}{2\epsilon_q}F^{(2D)}(k_q) \\
    &= \frac{1}{\epsilon_q} \left\{ \frac{k_q^2}{\pi^2} \int_0^{\pi/2} d\phi 
    \left[ \text{p.v.} \int_0^{\infty} dq \frac{2q \tilde{\chi}_{_{V}}(q)}{q^2 - (2k_q \cos\phi)^2} \right] 
    + i \frac{k_q^2}{\pi} \int_0^{\pi/2} \tilde{\chi}_{_{V}}(2k_q \cos\phi) d\phi \right\}.
\end{align*}
\end{widetext}
In the above equation, the nonlocal attenuation function of a 2D statistically isotropic two-phase medium $F^{(2D)}(k)$ has been introduced~\cite{To21a}. Notice that, similarly to what happens for diffusion spreadability, only the one- and two-body correlations enter these approximations through the packing fraction and spectral density.

\subsubsection{Fully isotropic media}
\label{sec:fullyiso}

Let us now consider the case of macroscopically isotropic media~\cite{To21a}, as the one depicted in Fig.~\ref{fig:2_3Dmedia} (right). In this case, the effective dielectric constant tensor reduces to a scalar, namely $\epsilon_e(\mathbf{k_q}) = \mathrm{Tr}[\mathbf{\epsilon}_e(\mathbf{k_q})]/d$. It has been shown in Ref.~\cite{To21a} that the two-point level approximation for the effective dielectric constant for macroscopically isotropic media is given by
\begin{equation}
\label{eq:3D_str_contr}
    \frac{\epsilon_e(k_q)}{\epsilon_q} = 
    1 + \frac{3 \ \beta_{pq} \phi_p^2}{\phi_p (1 - \beta_{pq} \phi_p) + 
    \sqrt{2\pi} \beta_{pq} 
    F^{(3D)}\left( \sqrt{\frac{\epsilon^{(3D)}_{\mathrm{BG}}}{\epsilon_q}} k_q\right)}.
\end{equation}
In the above expression, $F(\mathbf{Q})$ is given by
\begin{equation}
    F^{(\mathrm{3D)}}({\bf Q}) = -\frac{Q^2}{4\sqrt{2 \pi^7}} \int \frac{\tilde{\chi}_{V} (\mathbf{Q})}{|\mathbf{q} + \mathbf{Q}|^2 - Q^2} d\mathbf{q}
\end{equation}
and is referred to as the nonlocal attenuation function, while $\beta_{pq} = (\epsilon_p - \epsilon_q)/(\epsilon_p + (d-1) \ \epsilon_q)$ as before, having now $d=3$. $\epsilon_{BG}^{(3D)}$ is the Bruggeman approximation for two-phase media~\cite{To02a,bruggeman_berechnung_1935} in 3D. Once again, we notice that only correlations up to the two-point level are entered in Eq.~\eqref{eq:3D_str_contr}.


\section{Results}
\label{sec:results}

Here we provide results for the diffusive transport and optical physical properties and discuss cross-property relations between them.
In Sec.~\ref{sec:res:spectr_dens}, we will begin by showing the dependence of the spectral density of SHU packings on the packing fraction, as this will directly impact the behavior of the diffusion spreadability, discussed in Sec.~\ref{sec:diff_spread}, and the effective dynamic dielectric constant, discussed in Sec.~\ref{sec:dieleconst}. In Sec.~\ref{sec:crossprop}, we obtain cross-property relations between them.

It is also important to emphasize that, as shown in Ref.~\cite{Kim_2025_DenseSphere}, the highest packing fractions can be reached for small values of $\chi$, and therefore both for 2D and 3D systems we show the results obtained at small-$\chi$ values. In addition, for 2D media, we also address the high-$\chi$ limit, as in this regime nearly optimal effective properties can be achieved, as we will argue in Sec.~\ref{sec:conclusions}.
For the results at small $\chi$ ($\chi = 0.0025$) we will show packing fractions up to $\phi = 0.8$, even though it is possible to reach even higher packing fractions, up to $\phi_c \simeq 0.86$. This choice is motivated by the fact that, as discussed in Ref.~\cite{Kim_2025_DenseSphere}, at small $\chi$ and largest achievable values of $\phi$, the Bragg-peak structure of the triangular lattice appears, the system becomes polycrystalline and thus ceases to be disordered.
We refer the reader to the Supplemental Material for a discussion of the effect of different microstructures (generated with different values of $\sigma$ in the soft-core repulsion) at fixed packing fractions.

\subsection{Spectral Densities}
\label{sec:res:spectr_dens}

We start by considering the spectral density, introduced in Sec:~\ref{sec:backgr:twophase}. Its behavior depends on the structure factor, encoding microstructural information, and the packing fraction.
\begin{figure}
    \centering
    \includegraphics[width=\linewidth]{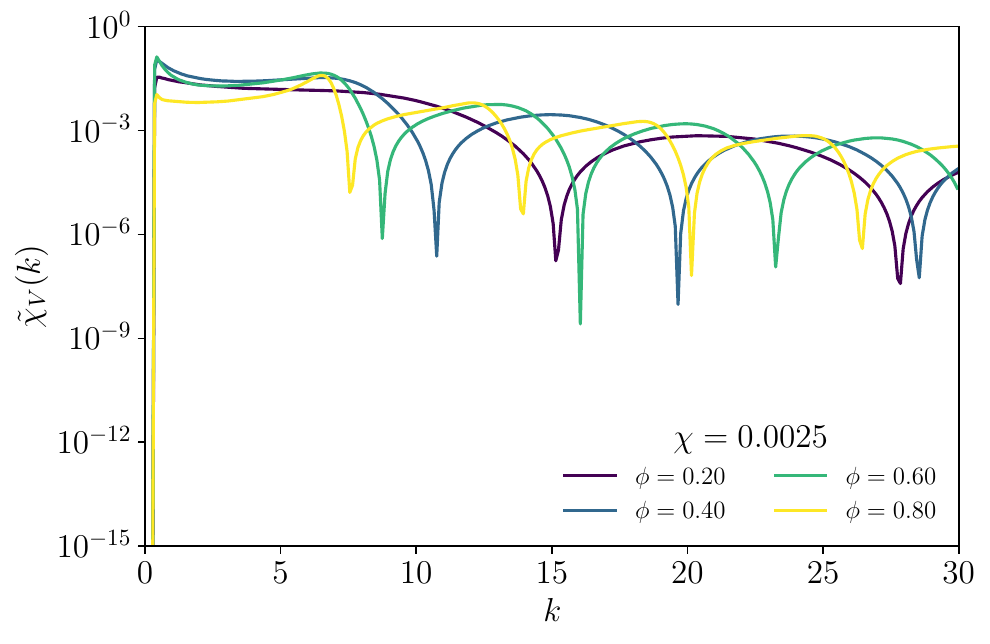}
    \includegraphics[width=\linewidth]{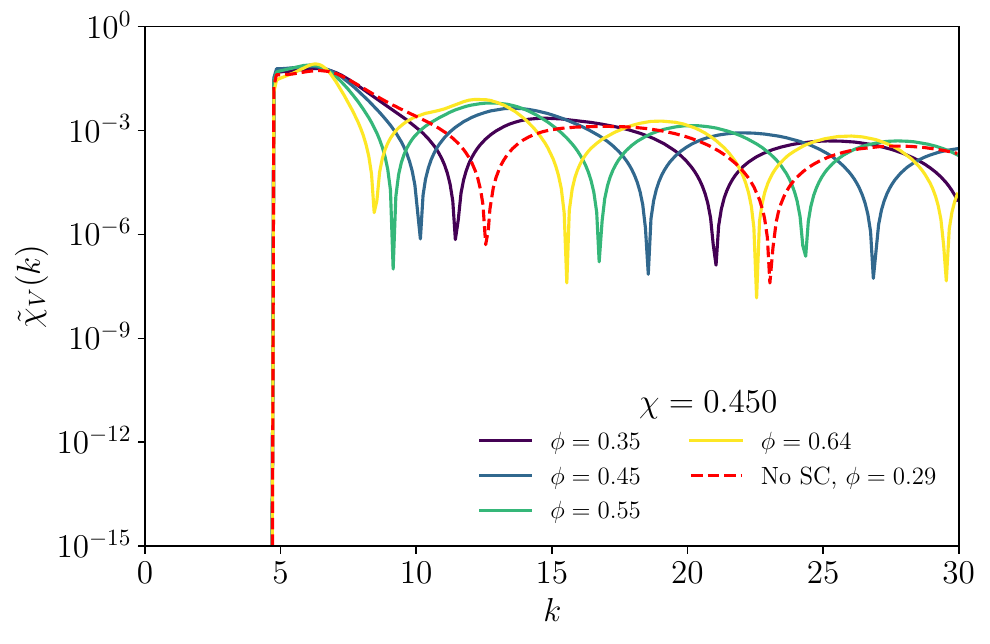}
    \caption{Spectral density of a $2D$ SHU packing of $N=4000$ disks (solid curves), with $\chi = 0.0025$ (top panel) and $\chi = 0.45$ (bottom panel) and various values of $\phi$, as reported in the legends. In all cases, the disk radius $a=\sigma/2$ is the largest allowed value.
    The red dashed curve corresponds to the results obtained without soft-core repulsion. By increasing the packing fraction, the position of the local minima moves towards smaller wave numbers, as dictated by Eq.~\eqref{eq:alpha_tilde}, with consequences for the physical properties, as we discuss below.}
    \label{fig:spectr_dens_0.45}
\end{figure}
\begin{figure}
    \centering
    \includegraphics[width=\linewidth]{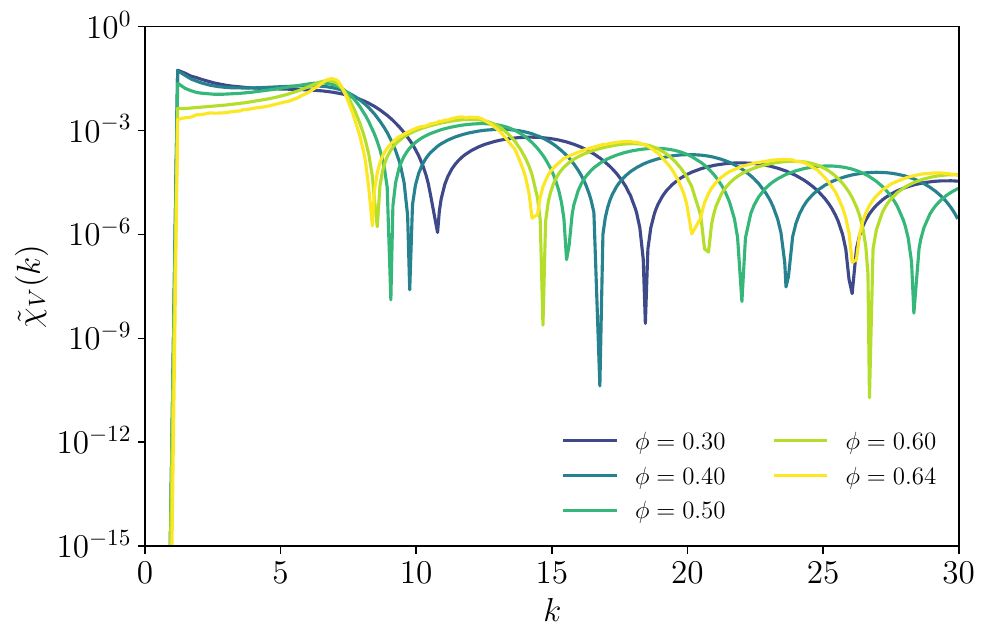}
    \caption{Spectral density of a 3D SHU packing of $N=400$ spheres, with $\chi = 0.0025$ and various values of $\phi$, ranging from $\phi=0.2$ to $\phi=0.64$, as reported in the color bar. In all cases, the disk radius $a=\sigma/2$ is the largest allowed value. Also in $3D$, the position of the local minima changes with $\phi$, according to Eq.~\eqref{eq:alpha_tilde}.}
    \label{fig:spectr_dens_0.002}
\end{figure}
We show in Fig.~\ref{fig:spectr_dens_0.45} the spectral density for a 2D SHU disk packing, for $\chi = 0.0025$ and $\chi = 0.45$ and different packing fractions $\phi$.
In all the curves obtained using the soft-core repulsion, we have set $a=\sigma/2$, the maximum allowed radius for the particles. 

As expected, for a fixed value of $\chi$, the spectral density vanishes for the same range of wave number, as imposed by the structure factor in Eq.~\eqref{eq:spectr_dens_1}.
Let us also notice that, as the packing fraction $\phi$ increases, the position of the first local minimum in $\tilde{\chi}_{_{V}}(k)$ shifts to smaller values of $k$. 
This can be easily understood from Eq.~\eqref{eq:alpha_tilde} for $\tilde{\alpha}_2(k;a)$: the dependence on $\phi$ enters through $a$ in the argument of the Bessel function $J^2_{d/2}(ka)$, acting as a scaling factor for the wave number $k$. 

In Fig.~\ref{fig:spectr_dens_0.45} we also report the result obtained without soft-core repulsion for $\chi = 0.45$. At finite system size, we observe that the resulting spectral density coincides with the results obtained at finite $\sigma$, with an effective packing fraction $\phi \sim 0.29$ for $\chi = 0.45$ and $N=4000$. This is in agreement with the results of Ref.~\cite{Kim_2025_DenseSphere}, as for large-$\chi$ it is expected that, at finite size, it is possible to generate packings with non-vanishing $\phi$. However, in the thermodynamic limit, Ref.~\cite{Kim_2025_DenseSphere} shows that the maximum packing fraction vanishes. At small $\chi$, instead, even at finite size it is not possible to generate a sphere packing with a non-vanishing $\phi$, and therefore we do not report the comparison for $\chi = 0.0025$.

We report in Fig.~\ref{fig:spectr_dens_0.002} the spectral density for a 3D SHU sphere packing at a small value of $\chi = 0.0025$, for different values of the packing fraction. Also in this case the same argument presented for the 2D case applies, and higher packing fractions lead to local minima moving to smaller values of $k$.

In the Supplemental Material, we consider microstructures having the same values of $\chi$ and $\phi$, but different values of $\sigma$ in the generating soft-core repulsion. We show that the sole effect of the microstructure, encoded in the structure factor, is considerably milder than the effect generated by different packing fractions.
The results we presented for the spectral density will determine the behavior of the physical properties, which we will discuss in the next sections.

\subsection{Diffusion Spreadability}
\label{sec:diff_spread}

Following the theoretical foundations presented in Sec.~\ref{sec:back:spread}, we now discuss the properties of the excess diffusion spreadability. We numerically obtain it by performing the integration in Eq.~\eqref{eq:exc_spread} and using the spectral density result discussed in Sec.~\ref{sec:res:spectr_dens}. 
\begin{figure}
    \centering
    \includegraphics[width=\linewidth]{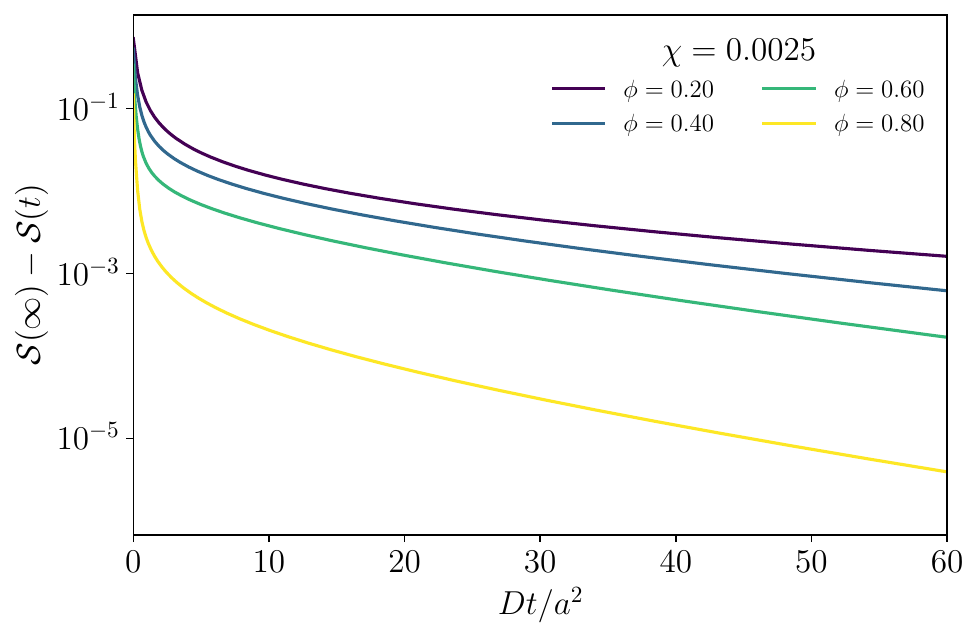}
    \includegraphics[width=\linewidth]{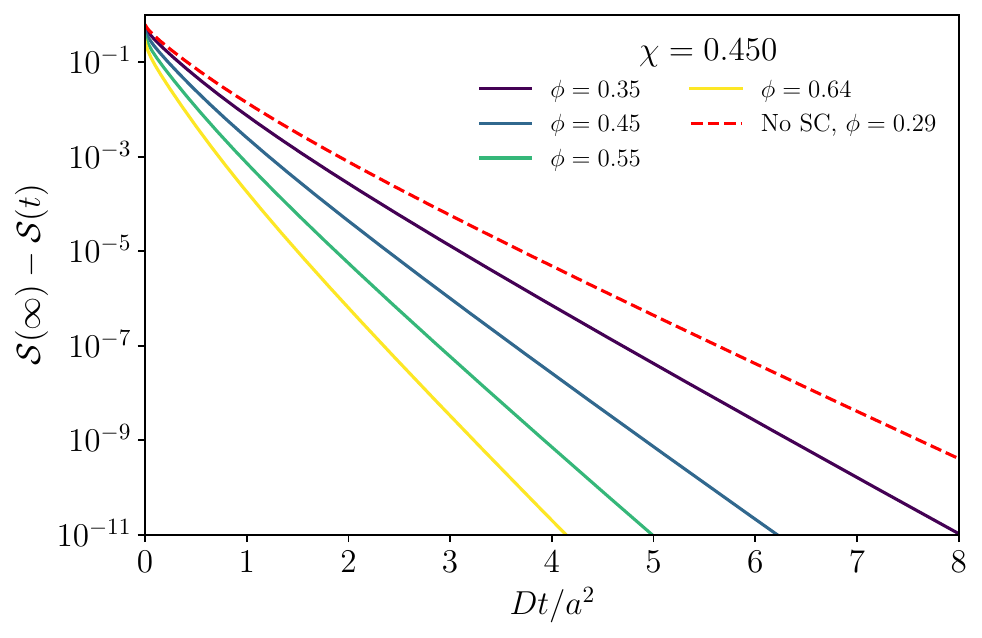}
    \caption{Excess diffusion spreadability for 2D SHU packings having $\chi=0.0025$ (top panel) and $\chi=0.45$ (bottom panel) and packing fraction. The solid lines are obtained using the soft-core repulsion, while the red dashed line is obtained without soft-core repulsion. Higher packing fractions correspond to more efficient spreading. Higher values of $\chi$ correspond to larger decay rates of the excess spreadability, as dictated by Eq.~\eqref{eq:spread_stealthy}.}
    \label{fig:2Ddiffspread}
\end{figure}

\begin{figure}
    \centering
    \includegraphics[width=\linewidth]{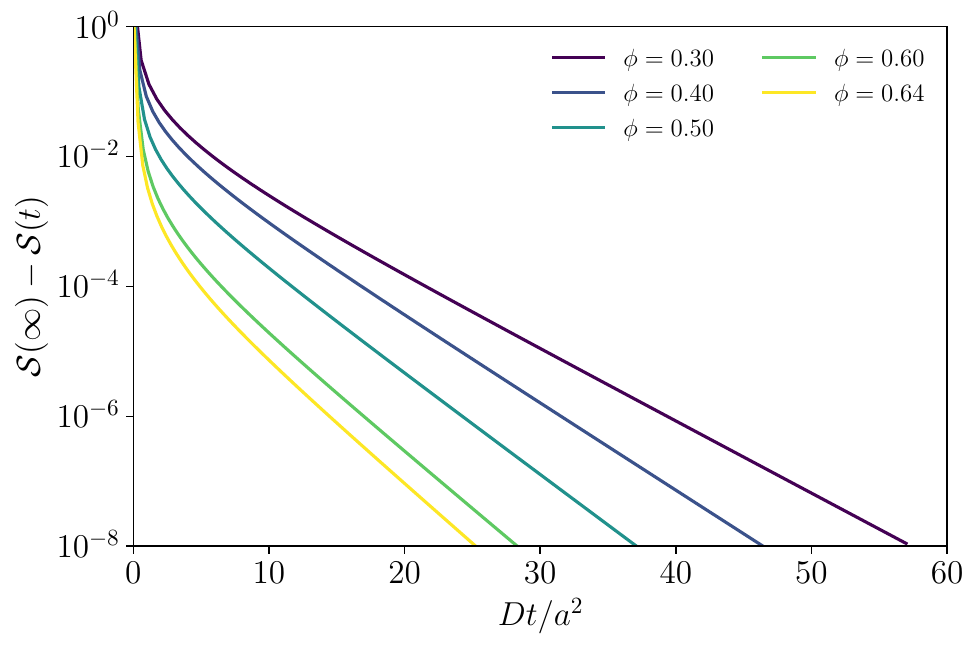}
    \caption{Excess diffusion spreadability for 3D SHU packings having $\chi=0.0025$ and packing fraction. Higher packing fractions correspond to faster decay in dimensionless time $Dt/a^2$, as visible from the different slopes for varying $\phi$.}
    \label{fig:3Ddiffspread}
\end{figure}

In Fig.~\ref{fig:2Ddiffspread} we show the excess diffusion spreadability $\mathcal{S}(\infty) - \mathcal{S}(t)$ as a function of the dimensionless time $Dt/a^2$, for a two-phase medium consisting of a 2D SHU disk packing embedded in a matrix. We report the result for $\chi=0.0025$ and $\chi = 0.45$.
Comparing the behavior at small and large $\chi$, we can note that at small $\chi$ the decay is considerably slower than at large $\chi$. This is a direct consequence of Eq.~\eqref{eq:spread_stealthy}, as small $\chi$ corresponds to small $K$ and, therefore, a slower decay rate.
As we can readily observe, higher packing fractions allow for a faster decay of the excess spreadability, meaning a faster spreading of solute from the disks (phase 2) to the matrix (phase 1). 
Notice that in Fig.~\ref{fig:2Ddiffspread} and Fig.~\ref{fig:3Ddiffspread} we show the excess diffusion spreadability as a function of the dimensionless time $Dt/a^2$; with this choice, the rate of the exponential decay of $\mathcal{S}(\infty) - \mathcal{S}(t)$ is $K^2 a^2$ (see also Eq.~\eqref{eq:xi_spead}), and thus is larger for larger $\phi$.

This result can be understood as follows: for large packing fractions, the volume fraction of phase 1 is reduced, as well as the typical size of the spaces between spheres. This leads to a more efficient spreading of a solution from phase 2 to phase 1. Moreover, as demonstrated in the Supplemental Material, for fixed packing fraction $\phi$, higher values of $\sigma$ provide a more efficient solute spreading, although the difference is considerably smaller than for different packing fractions.

In Fig.~\ref{fig:3Ddiffspread}, we report the results for the excess spreadability in a 3D SHU packing of identical spheres embedded in a matrix phase. As for the 2D case, the reported curves are obtained for different packing fractions with $\chi$ fixed. Also, in this case, we can observe the great advantage in the resulting spreadability gained in using the soft-core repulsion to reach higher packing fractions.

\subsection{Effective Dynamic Dielectric Constant}
\label{sec:dieleconst}

\begin{figure*}
    \centering
    \includegraphics[width=0.49\linewidth]{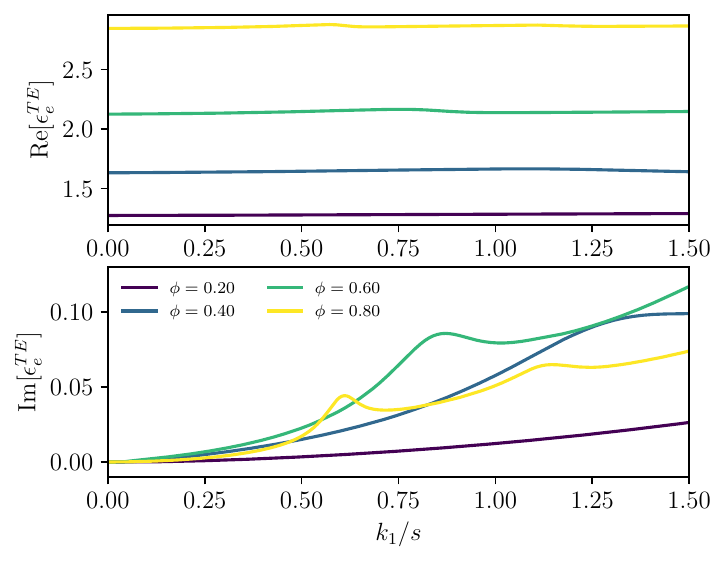}
    \includegraphics[width=0.49\linewidth]{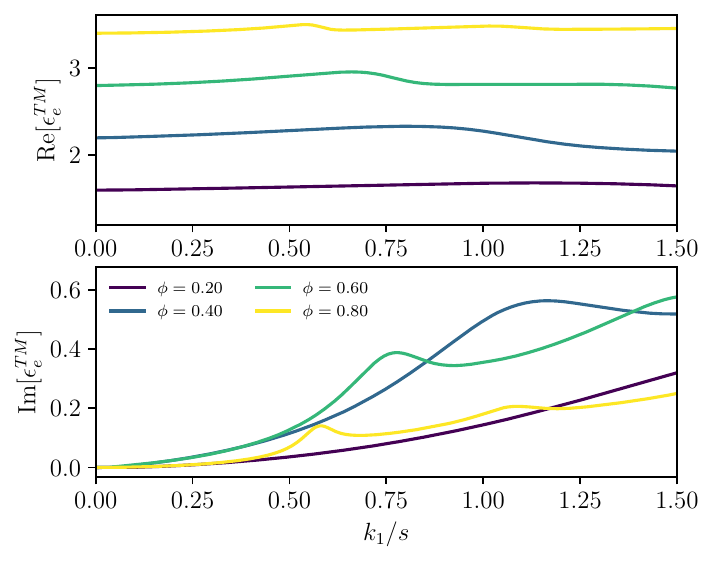}
    \caption{Effective dynamic dielectric constant for a transversely isotropic medium with contrast ratio $\epsilon_2/\epsilon_1 = 4$ and $\chi = 0.0025$. The upper panel shows the real and imaginary parts of the transverse electric (TE) polarization and the lower panel shows the real and imaginary parts of the transverse magnetic (TM) polarization. The wave number on the horizontal axis has been rescaled with the specific surface $s$, to make it dimensionless. The imaginary part of the TE and TM polarizations vanishes at small wave numbers, signaling the perfect transparency for the corresponding wavelengths.}
    \label{fig:EDDC_0.002}
\end{figure*}
\begin{figure*}
    \centering
    \includegraphics[width=0.49\linewidth]{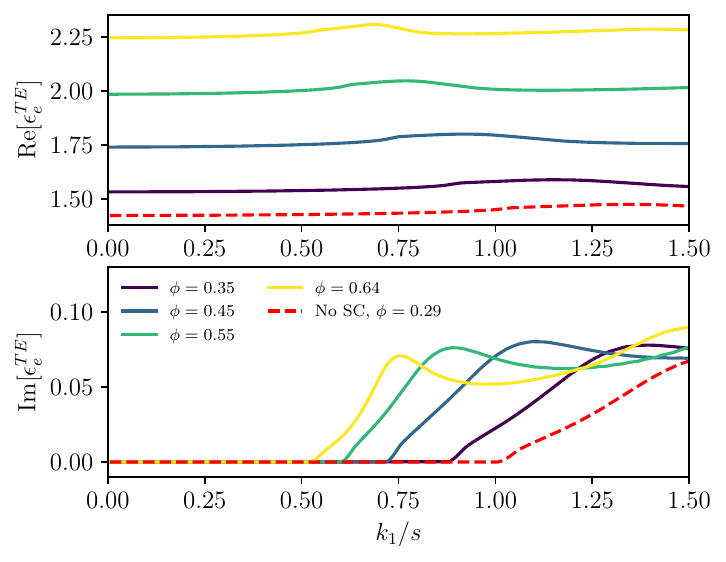}
    \includegraphics[width=0.49\linewidth]{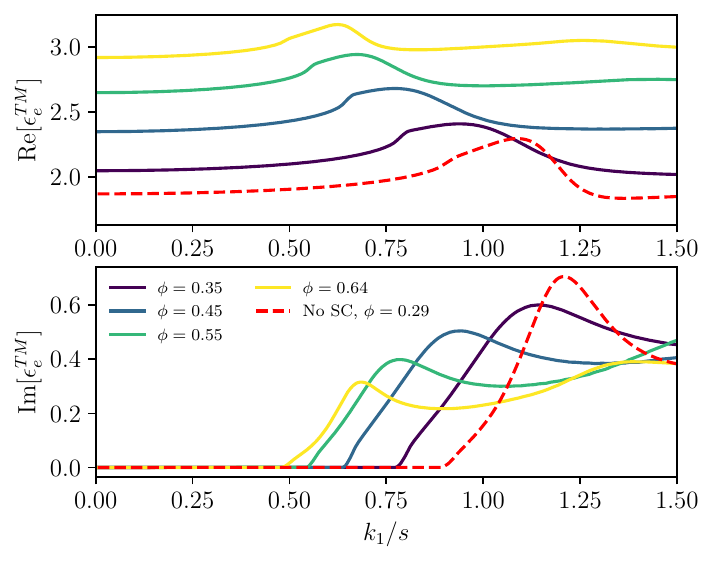}
    \caption{Effective dynamic dielectric constant for a transversely isotropic medium with contrast ratio $\epsilon_2/\epsilon_1 = 4$ and $\chi = 0.45$. The upper panel shows the real and imaginary parts of the transverse electric (TE) polarization and the lower panel shows the real and imaginary parts of the transverse magnetic (TM) polarization. The wave number on the horizontal axis has been rescaled with the specific surface $s$, to make it dimensionless. The solid curves are obtained using the soft-core repulsion, and the corresponding packing fraction $\phi$ is indicated in the color bar, whereas the dashed red curve is without soft-core repulsion, and corresponds to a packing fraction $\phi \simeq 0.29$. The imaginary part of the TE and TM polarizations vanishes at small wave numbers, signaling the perfect transparency for the corresponding wavelengths.}
    \label{fig:EDDC_0.45}
\end{figure*}

In Fig.~\ref{fig:EDDC_0.002} and Fig.~\ref{fig:EDDC_0.45}, we show the results for the TE and TM polarizations of the effective dynamic dielectric constant for a 3D transversely isotropic medium. The results shown are obtained by numerically evaluating the second-order approximation of the nonlocal strong-contrast expansion, as reported in Eqs.~\eqref{eq:TM_2D}, \eqref{eq:TE_2D}. As discussed in Ref.~\cite{To21a}, the second-order approximations provide an excellent estimate for the numerical results obtained using finite-difference time-domain (FDTD) simulations, up to $k_1 \rho^{1/d} \lesssim 1.5$. In Fig.~\ref{fig:EDDC_0.002} and Fig.~\ref{fig:EDDC_0.45} we report the results up to $k_1 / s \sim 1.5$, to show the qualitative behavior of the first attenuation peak, being $s$ the specific surface. 

Let us consider the 2D results for the imaginary parts of both $\epsilon_e^{TM}(k)$ and $\epsilon_e^{TE}(k)$; we can observe that they vanish exactly for a range of values of the wave number $0\leq k_1 < K_T$. This is a property of stealthy hyperuniform media, and it physically corresponds to perfect transparency for the corresponding range of wave numbers~\cite{To21a,Kim_2024_extraordinary}. 
\begin{equation}
    \frac{K_T}{K} = \frac{1}{2 \sqrt{\epsilon_*/\epsilon_q}}
\end{equation}
with $\epsilon_* = \langle \epsilon \rangle$ for TM polarization and $\epsilon_* = \epsilon_{BG}^{(2D)}$ for TE polarization.
We can observe that, in agreement with Refs.~\cite{To21a,Kim_2024_extraordinary}, the size of the transparency region depends on the packing fraction. In particular, $K_T/K \leq 1/2$ and $K_T = K/2$ for $\phi \to 0^+$.
As a consequence, small $\chi$ leads to a small transparency interval and the soft-core repulsion allows us to tune the transparency interval of the material for a given $\chi$, by changing the packing fraction $\phi$. In particular, the larger the value of $\phi$, the smaller the transparency interval. Intuitively, as the volume fraction of the phase with a higher dielectric constant slightly increases, the scattering events occur more frequently, and thus, wave propagation is impeded further, resulting in a reduced transparency interval.
We also note that for given values of $\chi$ and $\phi$, the transparency interval $K_T$ decreases as the contrast ratio $\epsilon_2/\epsilon_1$ increases because the strength of individual scattering events increases, leading to further wave propagation impedance.

\begin{figure}
    \centering
    \includegraphics[width=\linewidth]{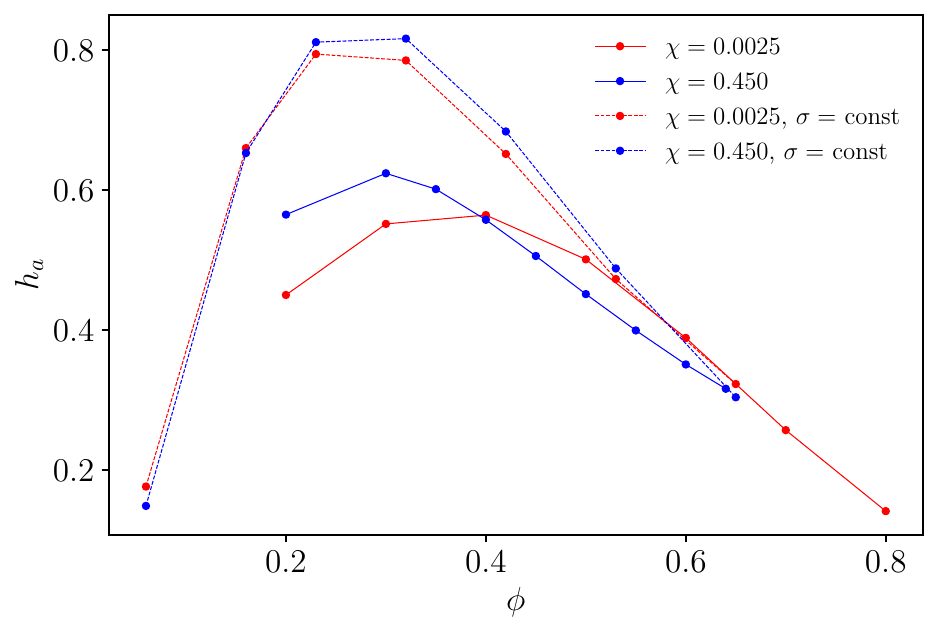}
    \caption{Height of the first attenuation peak, denoted with $h_a$, as a function of packing fraction for the TM polarization in a 3D transversely isotropic SHU medium. Solid lines are obtained for different values of $\sigma$ and by setting $a = \sigma/2$, whereas dashed lines are obtained for fixed $\sigma$ and different values of $a<\sigma/2$.}
    \label{fig:abs_peak}
\end{figure}

We also note that, by changing the packing fraction $\phi$, the height of the first attenuation peak, denoted as $h_a$, changes. In principle, this behavior can be a consequence of the change in the packing fraction or the microstructure, as both $\phi$ and $\sigma$ are varied in Fig.~\ref{fig:EDDC_0.45}. We analyze the two effects separately in the Supplemental Material. For concreteness, let us consider the TM polarization and let us compare the values of $h_a$ for different values of $\phi$ and $\chi$. We report the results in Fig.~\ref{fig:abs_peak}. 
We observe that $h_a$ decreases for $\phi \gtrsim 0.3$, which is the volume fraction range of Fig.~\ref{fig:EDDC_0.002} and Fig.~\ref{fig:EDDC_0.45}, also when $\sigma$ is kept fixed (i.e. for fixed microstructure) and the packing fraction is varied, as shown in Fig.~18 in the SM. The dependence of $h_a$ on $\phi$ reported in Fig.~18 in the SM can be qualitatively understood as follows: if one of the two phases is predominant compared to the other, the dissipation due to scattering occurs less. Consequently, by increasing $\phi$ towards the maximal allowed value (for a given $\chi$), the height of the attenuation peak is reduced. The same phenomenon occurs for $\phi \lesssim 0.3$, as shown in Fig.~18 in the SM, as the other phase becomes predominant.
We also notice that different microstructures lead to different attenuation properties (see Fig.~17 in the SM), even if the effect is not as pronounced as for different packing fractions.

\begin{figure*}
    \centering
    \includegraphics[width=0.49\linewidth]{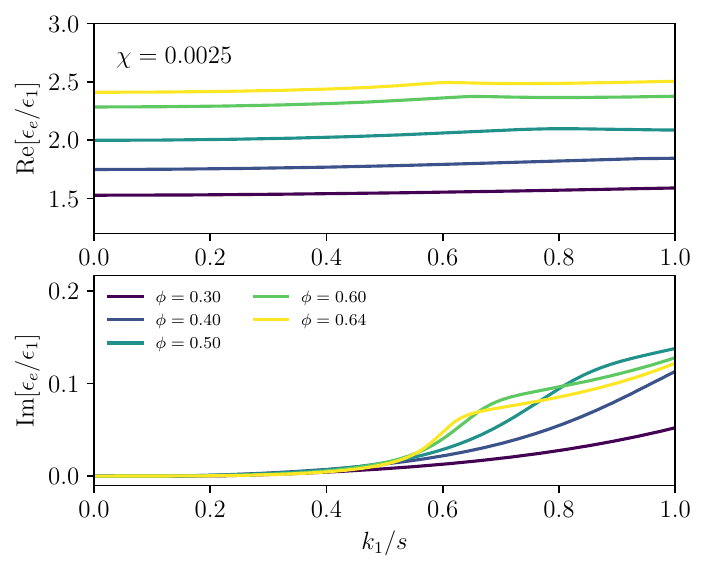}
    \includegraphics[width=0.49\linewidth]{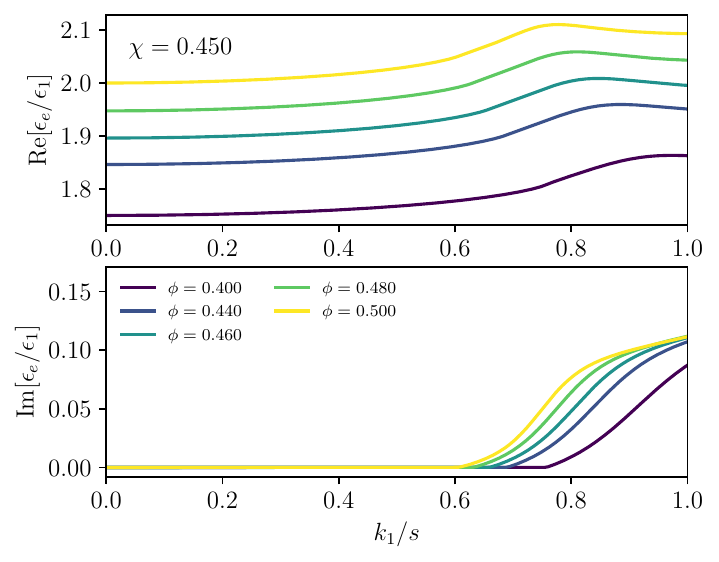}
    \caption{Effective dynamic dielectric constant for a fully isotropic 3D media with contrast ratio $\epsilon_2/\epsilon_1 = 4$ and $\chi = 0.0025$. Different curves represent different values of the packing fraction, as indicated by the color bar. The wave number on the horizontal axis has been rescaled with the specific surface so that it is dimensional. As noticed for the two-dimensional case, also in $3D$ the size of the transparency interval, as determined by the range of wave number with vanishing imaginary part of the effective dielectric constant, changes with packing fraction $\phi$.}
    \label{fig:EDDC_3D}
\end{figure*}

We show in Fig.~\ref{fig:EDDC_3D} the numerical results for the effective dynamic dielectric constant in 3D fully isotropic media, computed using the second-order strong-contrast approximation, Eq.~\eqref{eq:3D_str_contr}. In particular, we show results for $\chi = 0.0025$.
We can observe that, also for fully isotropic stealthy media, there is a perfect transparency interval, whose size changes with $\chi$ and $\phi$. In particular, 
\begin{equation}
    \frac{K_T}{K} = \frac{1}{2 \sqrt{\epsilon_{BG}^{(3D)}/\epsilon_q}}
\end{equation}
and thus for low values of $\chi$, corresponding to small $K$, the size of the transparency interval is also small.
In addition, the size of the transparency interval $K_T$ decreases with $\phi$.
In analogy to what has been observed for 2D transversely isotropic media, we note that, as $\phi$ increases from intermediate values to the maximally allowed values, the height of the attenuation peak changes, becoming smaller at high values of $\phi$.

\subsection{Cross-property relations}
\label{sec:crossprop}

\begin{figure}
    \centering
    \includegraphics[width=\linewidth]{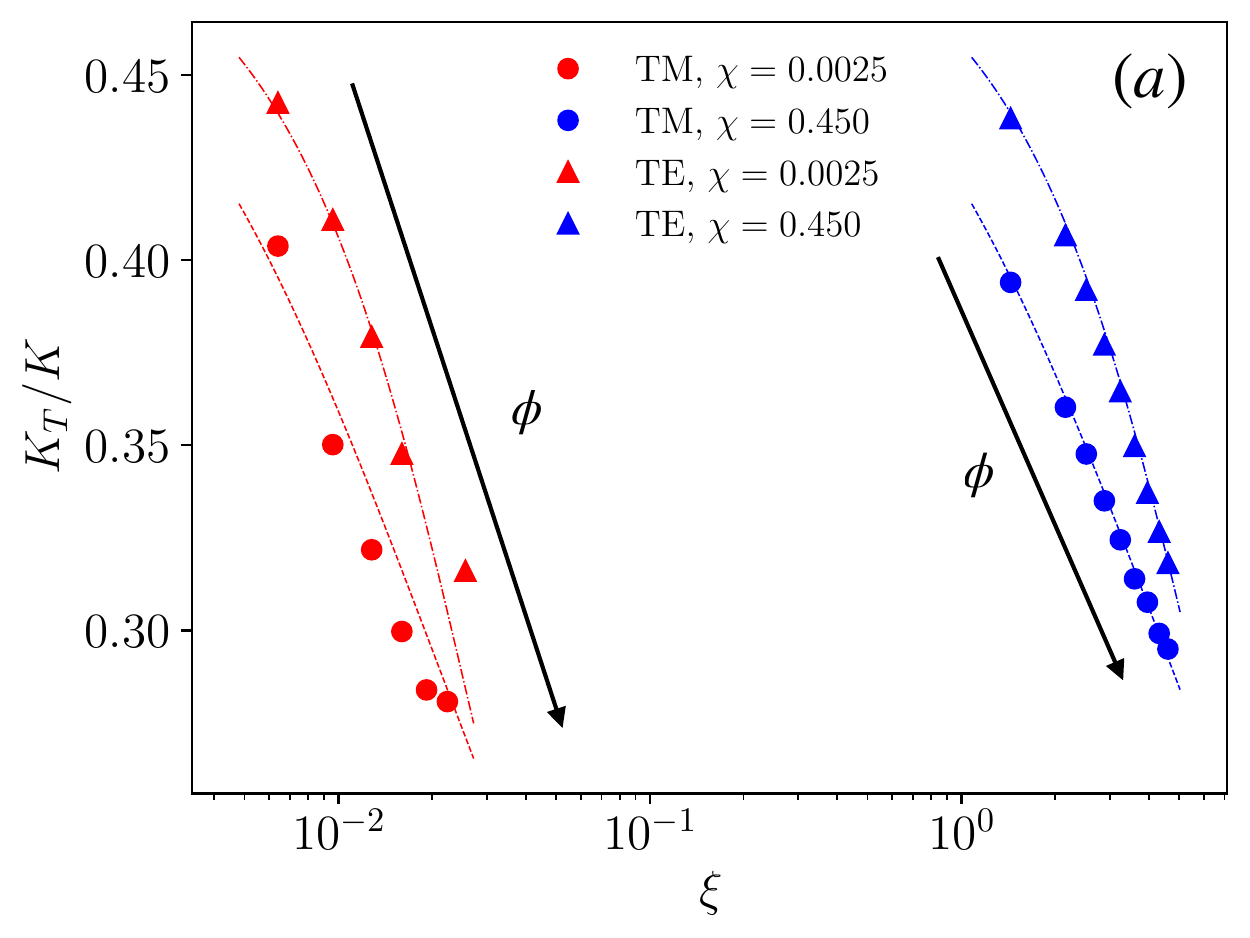}
    \includegraphics[width=\linewidth]{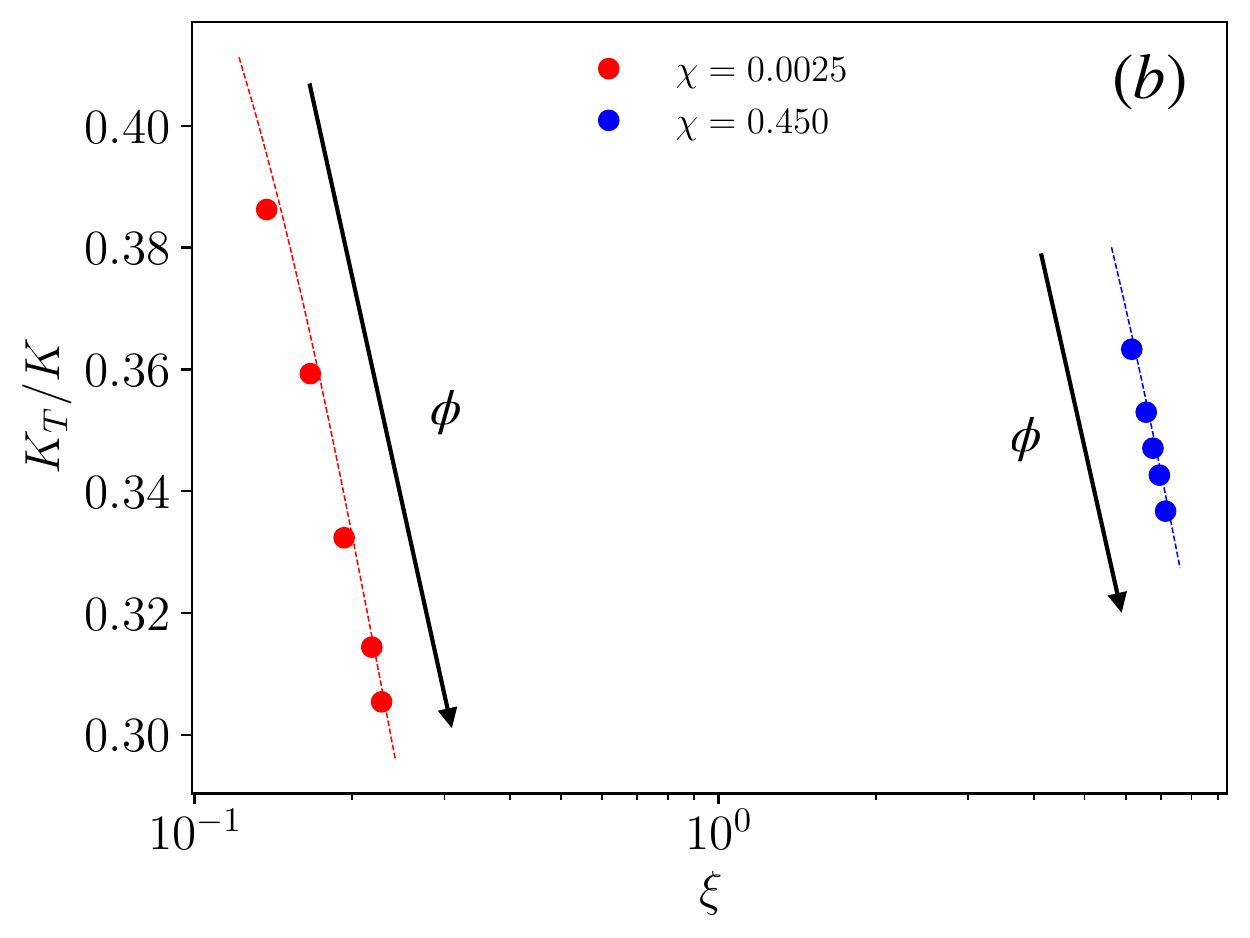}
    \caption{Dependence on $\xi$, defined in Eq.~\eqref{eq:xi_spead}, versus $K_T/K$, being $K_T$ the size of the transparency interval of the TM (dots) and TE (triangles) polarization in a 3D transversely isotropic media $(a)$ and for fully isotropic media $(b)$. Red curves correspond to $\chi = 0.0025$, while blue curves correspond to $\chi = 0.45$. The contrast ratio is $\epsilon_2/\epsilon_1 = 4$. Different points are obtained at different packing fractions, with $\phi$ increasing for decreasing $K_T/K$, as indicated by the arrows. The dashed curves represent the analytical result Eq~\eqref{eq:KTK}. The presence of correlations between these two quantities proves the existence of cross-property relations.}
    \label{fig:cross_props}
\end{figure}

In heterogeneous materials, effective properties manifest common morphological features, meaning that knowledge of one property can provide valuable insights into another. Such cross-property relations enable the rational design of multifunctional heterogeneous materials with desirable properties~\cite{To02a} using inverse techniques~\cite{torquato_inverse_2009}.

In the previous sections, we presented the results for the physical properties of ultradense SHU two-phase media. Specifically, we discussed diffusion spreadability in Sec.~\ref{sec:diff_spread} and the effective dynamic dielectric constant in Sec.~\ref{sec:dieleconst}. 
We highlighted that since both are ultimately determined by the microstructure via the spectral density and the packing fraction.
Indeed, here we provide direct evidence for the existence of cross-property relations between optical and diffusive transport properties in ultradense SHU packings.

To demonstrate the existence of cross-property relations between the diffusion spreadability and effective dynamic dielectric constant, we first need to select, for each of them, a distinctive feature characterizing their behavior. 
In the case of diffusion spreadability, we introduce the long-time spreadability parameter $\xi$, defined as follows:
\begin{equation}
\label{eq:xi_spead}
    \xi = \lim_{Dt/a^2 \to \infty} - \frac{a^2}{D}\frac{d \, \ln[\mathcal{S}(\infty) - \mathcal{S}(t)]}{dt}.
\end{equation}
Notice that this quantity is defined for any two-phase medium and provides a way for experimentally probing whether the medium is stealthy hyperuniform or not. In fact, in the large-time limit, $\xi$ approaches a constant value only if the excess spreadability decays exponentially, while it diverges in all other cases, in which the decay is polynomial (see Fig.~\ref{fig:phase_diag_HU}). In the case of stealthy systems, the long-time spreadability parameter $\xi$ depends
on both $K$ and $\phi$, $\xi = \xi(K, \phi)$ and is given explicitly by
\begin{equation}
\label{eq:xi_stealthy}
    \xi(K,\phi) = K^2 a^2 = K^2 \left[ \phi \, \Gamma(1+d/2) \, / \left(\rho \pi^{d/2}\right) \right]^{2/d}.
\end{equation}
Therefore, the long-time spreadability parameter $\xi$ is experimentally accessible, contains information about the stealthiness of the system, and corresponds to the slope of the curves in Figs.~\ref{fig:2Ddiffspread} and~\ref{fig:3Ddiffspread}, thus containing information about the size of the exclusion region $K$ and packing fraction $\phi$. 
To link $\xi$ to characteristics of the effective dynamic dielectric constant, we choose to consider the size $K_T/K$ of the transparency interval discussed in Sec~\ref{sec:dieleconst}.

We show in Fig.~\ref{fig:cross_props} the correlation between the decay rate of the excess spreadability and the size of the transparency interval. We show two sets of curves for two different values of $\chi$ (red curves for $\chi = 0.0025$ and blue curves for $\chi = 0.45$), and we consider different values of $\phi$. We also report the size of the transparency interval $K_T/K$ both for the TM (dots) and TE (triangles) polarizations in both cases. As the figure clearly shows, there is a negative correlation between the size of the transparency interval $K_T/K$ and the decay rate of the diffusion spreadability. In particular, by increasing the packing fraction, the size of the transparency interval decreases while the rate increases. 

In Fig.~\ref{fig:cross_props} we also report the cross-property relations derived from Eq.~\eqref{eq:xi_stealthy} for $\xi$ and the analytical expressions for the size of transparency interval, that are given by Ref.~\cite{Kim_2024_extraordinary}
\begin{widetext}
\begin{align}
\label{eq:KTK}
\frac{K_{T}}{K}   = 
\begin{cases}
\left[4 + 4\left(\frac{\epsilon_2}{\epsilon_1}-1 \right)\frac{\xi \pi}{K^2} \right]^{-\frac{1}{2}}, & \text{3D TM trans. iso.} \\
\left[ 2 \left( \left(2\frac{\xi \pi}{K^2}-1 \right) \left(\frac{\epsilon_2}{\epsilon_1}-1\right) + \sqrt{ 4\frac{\epsilon_2}{\epsilon_1}+\left(\frac{\epsilon_2}{\epsilon_1}-1\right)^2 \left(2\frac{\xi \pi}{K^2}-1\right)^2 } \right) \right]^{-\frac{1}{2}}, & \text{3D TE trans. iso.} \\
\left[ \left( \left(2-\frac{4\pi}{K^2}\xi^{3/2} -\frac{\epsilon_2}{\epsilon_1} \left(1-\frac{4\pi}{K^2}\xi^{3/2}\right)\right) + \sqrt{8\frac{\epsilon_2}{\epsilon_1}+ \left(2-\frac{4\pi}{K^2}\xi^{3/2} -\frac{\epsilon_2}{\epsilon_1}\left(1-\frac{4\pi}{K^2}\xi^{3/2}\right) \right)^2 } \right) \right]^{-\frac{1}{2}}, & \text{3D fully iso.}
\end{cases}
\end{align}
\end{widetext}
In Fig.~14 of the SM, we show the plots for the analytically obtained $K_T/K$ versus $\xi$ in TM and TE transversely isotropic and in fully isotropic media.

Such cross-property relations have important experimental implications. Suppose, for instance, one desires to measure experimentally the long-time behavior of the diffusion spreadability of a two-phase medium composed of identical spheres with a given packing fraction. 
(Practically, such experiments can be performed via NMR measurements, as noted in Sec.~\ref{sec:back:spread}.)
By computing $\xi$, it is possible to determine whether the system is stealthy hyperuniform or not; if the derivative in Eq.~\eqref{eq:xi_spead} asymptotes to a constant value the system is stealthy, and from Eq.~\eqref{eq:xi_stealthy} it is possible to obtain the size of the exclusion region $K$. Therefore, the value of $\chi$ can be obtained and, as a consequence, by using the cross-property relations, one can immediately derive the range of wavelengths for which such a medium is optically transparent. Furthermore, by obtaining the value $K$ and using Eq~\eqref{eq:spread_stealthy}, one can obtain the value of the structure factor $S(K)$ at the wave number size $K$ of the exclusion region.

We note that a similar result can be obtained by considering other properties. For example, one could consider the dimensionless time for which the excess spreadability reaches a reference value and the height of the first attenuation peak, although in this case, the correlation between the two would be positive or negative depending on the packing fraction, as one can observe from Fig.~\ref{fig:abs_peak}.


\section{Discussion and Conclusions}
\label{sec:conclusions}


In this work, we have initiated the study of the dynamical physical properties of a special class of two-phase media, in which one of the two phases consists of an ultradense disordered SHU packing of identical spheres.
The possibility of reaching such high packing fractions is guaranteed by the employment of the collective-coordinate optimization procedure to generate point configurations with the addition of a soft-core repulsion between particles, enforcing a minimum inter-particle distance.
We examined diffusive transport properties through time-dependent diffusion spreadability and optical properties through the effective dynamic dielectric constant.
In addition, we explicitly demonstrated the existence of cross-property relationships between these two properties for SHU two-phase media.

We have shown that the possibility of increasing the packing fraction to previously unattainable values through the soft-core repulsion has crucial consequences on the physical properties of disordered SHU media. 
In particular, by increasing the packing fraction $\phi$ for a fixed value of the stealthiness parameter $\chi$, we find that the spreading rate of a solute initially contained in one phase increases, becoming more efficient. Large values of $\chi$ provide an excess spreadability decaying faster than for small-$\chi$ values for fixed $\phi$, as theoretically expected. 
We also emphasized the connection between diffusion spreadability and NMR and MRI, making our results particularly relevant also in experimental contexts.
Regarding the effective optical properties of such ultradense disordered SHU two-phase media, we have shown that different packing fractions correspond to different transparency intervals, as encoded by the vanishing imaginary part of the effective dynamic dielectric constant. In addition, different packing fractions lead to different attenuation properties, which we have quantified by considering the height of the first peak in the imaginary part of $\epsilon_e(k)$. We have also addressed different values of $\chi$ in the same range of packing fractions and showed, as theoretically predicted, that the transparency interval increases with $\chi$. 
We then discussed cross-property relations between the time-dependent diffusion spreadability and the effective dynamic dielectric constant. Such cross-property relations are valuable as they allow one to estimate one property based on the measurement of the other~\cite{To02a}.

The results presented in this work have important consequences, both for theoretical reasons and for material design. 
We have discussed two fundamental properties: on the one hand, the addition of the soft-core repulsion allows us to tune the packing fraction of SHU configurations and to reach ultradense sphere packings; on the other hand, we can accurately predict dynamical physical properties from the knowledge of the spectral density alone. 
By combining these two elements with the numerical technique to generate two-phase media with a prescribed spectral density \cite{yu_characterization_2017,chen_designing_2018,iyer_designing_2020,shi_threedimensional_2025}, one can implement an inverse-design approach \cite{torquato_inverse_2009} to engineer and fabricate two-phase media with multiple types of desired properties.
This result, therefore, opens the door to new directions in material design driven by the desired physical property.

Knowledge of the structures of disordered ultradense SHU packings can also be used to design static material properties.
As $\chi$ increases from zero to 0.45 in disordered ultradense SHU packings, the mean contact number per particle decreases linearly with $\chi$, and thus the networks of contacting particles stop percolating for $\chi\gtrsim0.35$~\cite{Kim_2025_DenseSphere}.
Thus, for disordered SHU packings at $\chi=0.45$, all particles are well-separated in a fully connected matrix phase, which is a necessary requirement to attain optimal two-phase structures \cite{To18c,zhang_transport_2016}.
Indeed, Skolnick and Torquato recently showed~\cite{Skolnick_2025_Effective} that SHU packings at $\chi=0.45$ achieve effective conductivities and elastic moduli that nearly coincide with the corresponding Hashin-Shtrikman lower bounds ~\cite{Ha62c,Ha70} across all packing fractions when the particle phase has the higher phase contrast.

\acknowledgments

This work was supported by the Army Research Office
under Cooperative Agreement No. W911NF-22-2-0103.

\section*{Data availability}

Data underlying the results presented in this paper are not publicly available at this time but may be obtained from the authors upon request.


\bibliography{references}

\begin{thebibliography}{74}%
\makeatletter
\providecommand \@ifxundefined [1]{%
 \@ifx{#1\undefined}
}%
\providecommand \@ifnum [1]{%
 \ifnum #1\expandafter \@firstoftwo
 \else \expandafter \@secondoftwo
 \fi
}%
\providecommand \@ifx [1]{%
 \ifx #1\expandafter \@firstoftwo
 \else \expandafter \@secondoftwo
 \fi
}%
\providecommand \natexlab [1]{#1}%
\providecommand \enquote  [1]{``#1''}%
\providecommand \bibnamefont  [1]{#1}%
\providecommand \bibfnamefont [1]{#1}%
\providecommand \citenamefont [1]{#1}%
\providecommand \href@noop [0]{\@secondoftwo}%
\providecommand \href [0]{\begingroup \@sanitize@url \@href}%
\providecommand \@href[1]{\@@startlink{#1}\@@href}%
\providecommand \@@href[1]{\endgroup#1\@@endlink}%
\providecommand \@sanitize@url [0]{\catcode `\\12\catcode `\$12\catcode `\&12\catcode `\#12\catcode `\^12\catcode `\_12\catcode `\%12\relax}%
\providecommand \@@startlink[1]{}%
\providecommand \@@endlink[0]{}%
\providecommand \url  [0]{\begingroup\@sanitize@url \@url }%
\providecommand \@url [1]{\endgroup\@href {#1}{\urlprefix }}%
\providecommand \urlprefix  [0]{URL }%
\providecommand \Eprint [0]{\href }%
\providecommand \doibase [0]{https://doi.org/}%
\providecommand \selectlanguage [0]{\@gobble}%
\providecommand \bibinfo  [0]{\@secondoftwo}%
\providecommand \bibfield  [0]{\@secondoftwo}%
\providecommand \translation [1]{[#1]}%
\providecommand \BibitemOpen [0]{}%
\providecommand \bibitemStop [0]{}%
\providecommand \bibitemNoStop [0]{.\EOS\space}%
\providecommand \EOS [0]{\spacefactor3000\relax}%
\providecommand \BibitemShut  [1]{\csname bibitem#1\endcsname}%
\let\auto@bib@innerbib\@empty
\bibitem [{\citenamefont {Torquato}\ and\ \citenamefont {Stillinger}(2003)}]{To03a}%
  \BibitemOpen
  \bibfield  {author} {\bibinfo {author} {\bibfnamefont {S.}~\bibnamefont {Torquato}}\ and\ \bibinfo {author} {\bibfnamefont {F.~H.}\ \bibnamefont {Stillinger}},\ }\bibfield  {title} {\bibinfo {title} {Local density fluctuations, hyperuniform systems, and order metrics},\ }\href {https://doi.org/10.1103/PhysRevE.68.041113} {\bibfield  {journal} {\bibinfo  {journal} {Phys. Rev. E}\ }\textbf {\bibinfo {volume} {68}},\ \bibinfo {pages} {041113} (\bibinfo {year} {2003})}\BibitemShut {NoStop}%
\bibitem [{\citenamefont {Gabrielli}\ \emph {et~al.}(2002)\citenamefont {Gabrielli}, \citenamefont {Joyce},\ and\ \citenamefont {Labini}}]{Ga02}%
  \BibitemOpen
  \bibfield  {author} {\bibinfo {author} {\bibfnamefont {A.}~\bibnamefont {Gabrielli}}, \bibinfo {author} {\bibfnamefont {M.}~\bibnamefont {Joyce}},\ and\ \bibinfo {author} {\bibfnamefont {F.~S.}\ \bibnamefont {Labini}},\ }\bibfield  {title} {\bibinfo {title} {Glass-like universe: {R}eal-space correlation properties of standard cosmological models},\ }\href {https://doi.org/10.1103/PhysRevD.65.083523} {\bibfield  {journal} {\bibinfo  {journal} {Phys. Rev. D}\ }\textbf {\bibinfo {volume} {65}},\ \bibinfo {pages} {083523} (\bibinfo {year} {2002})}\BibitemShut {NoStop}%
\bibitem [{\citenamefont {Torquato}(2018)}]{To18a}%
  \BibitemOpen
  \bibfield  {author} {\bibinfo {author} {\bibfnamefont {S.}~\bibnamefont {Torquato}},\ }\bibfield  {title} {\bibinfo {title} {Hyperuniform states of matter},\ }\href {https://doi.org/10.1016/j.physrep.2018.03.001} {\bibfield  {journal} {\bibinfo  {journal} {Phys. Rep.}\ }\textbf {\bibinfo {volume} {745}},\ \bibinfo {pages} {1} (\bibinfo {year} {2018})}\BibitemShut {NoStop}%
\bibitem [{\citenamefont {{O{\u g}uz}}\ \emph {et~al.}(2017)\citenamefont {{O{\u g}uz}}, \citenamefont {{Socolar}}, \citenamefont {{Steinhardt}},\ and\ \citenamefont {{Torquato}}}]{Og17}%
  \BibitemOpen
  \bibfield  {author} {\bibinfo {author} {\bibfnamefont {E.~C.}\ \bibnamefont {{O{\u g}uz}}}, \bibinfo {author} {\bibfnamefont {J.~E.~S.}\ \bibnamefont {{Socolar}}}, \bibinfo {author} {\bibfnamefont {P.~J.}\ \bibnamefont {{Steinhardt}}},\ and\ \bibinfo {author} {\bibfnamefont {S.}~\bibnamefont {{Torquato}}},\ }\bibfield  {title} {\bibinfo {title} {Hyperuniformity of quasicrystals},\ }\href@noop {} {\bibfield  {journal} {\bibinfo  {journal} {Phys. Rev. B}\ }\textbf {\bibinfo {volume} {95}},\ \bibinfo {pages} {054119} (\bibinfo {year} {2017})}\BibitemShut {NoStop}%
\bibitem [{\citenamefont {Levine}\ and\ \citenamefont {Steinhardt}(1984)}]{Le84}%
  \BibitemOpen
  \bibfield  {author} {\bibinfo {author} {\bibfnamefont {D.}~\bibnamefont {Levine}}\ and\ \bibinfo {author} {\bibfnamefont {P.~J.}\ \bibnamefont {Steinhardt}},\ }\bibfield  {title} {\bibinfo {title} {Quasicrystals: {A} new class of ordered structures},\ }\href@noop {} {\bibfield  {journal} {\bibinfo  {journal} {Phys. Rev. Lett.}\ }\textbf {\bibinfo {volume} {53}},\ \bibinfo {pages} {2477} (\bibinfo {year} {1984})}\BibitemShut {NoStop}%
\bibitem [{\citenamefont {Zhang}\ \emph {et~al.}(2016{\natexlab{a}})\citenamefont {Zhang}, \citenamefont {Stillinger},\ and\ \citenamefont {Torquato}}]{Zh16a}%
  \BibitemOpen
  \bibfield  {author} {\bibinfo {author} {\bibfnamefont {G.}~\bibnamefont {Zhang}}, \bibinfo {author} {\bibfnamefont {F.~H.}\ \bibnamefont {Stillinger}},\ and\ \bibinfo {author} {\bibfnamefont {S.}~\bibnamefont {Torquato}},\ }\bibfield  {title} {\bibinfo {title} {The perfect glass paradigm: Disordered hyperuniform glasses down to absolute zero},\ }\href {https://doi.org/10.1038/srep36963} {\bibfield  {journal} {\bibinfo  {journal} {Sci. Rep.}\ }\textbf {\bibinfo {volume} {6}},\ \bibinfo {pages} {36963} (\bibinfo {year} {2016}{\natexlab{a}})}\BibitemShut {NoStop}%
\bibitem [{\citenamefont {Levesque}\ \emph {et~al.}(2000)\citenamefont {Levesque}, \citenamefont {Weis},\ and\ \citenamefont {Lebowitz}}]{Le00}%
  \BibitemOpen
  \bibfield  {author} {\bibinfo {author} {\bibfnamefont {D.}~\bibnamefont {Levesque}}, \bibinfo {author} {\bibfnamefont {J.-J.}\ \bibnamefont {Weis}},\ and\ \bibinfo {author} {\bibfnamefont {J.}~\bibnamefont {Lebowitz}},\ }\bibfield  {title} {\bibinfo {title} {Charge fluctuations in the two-dimensional one-component plasma},\ }\href@noop {} {\bibfield  {journal} {\bibinfo  {journal} {J. Stat. Phys.}\ }\textbf {\bibinfo {volume} {100}},\ \bibinfo {pages} {209} (\bibinfo {year} {2000})}\BibitemShut {NoStop}%
\bibitem [{\citenamefont {{Hexner}}\ and\ \citenamefont {{Levine}}(2015)}]{He15}%
  \BibitemOpen
  \bibfield  {author} {\bibinfo {author} {\bibfnamefont {D.}~\bibnamefont {{Hexner}}}\ and\ \bibinfo {author} {\bibfnamefont {D.}~\bibnamefont {{Levine}}},\ }\bibfield  {title} {\bibinfo {title} {{Hyperuniformity of critical absorbing states}},\ }\href {https://doi.org/10.1103/PhysRevLett.114.110602} {\bibfield  {journal} {\bibinfo  {journal} {Phys. Rev. Lett.}\ }\textbf {\bibinfo {volume} {114}},\ \bibinfo {pages} {110602} (\bibinfo {year} {2015})}\BibitemShut {NoStop}%
\bibitem [{\citenamefont {Wiese}(2024)}]{Wiese2024Hyperuniformity}%
  \BibitemOpen
  \bibfield  {author} {\bibinfo {author} {\bibfnamefont {K.~J.}\ \bibnamefont {Wiese}},\ }\bibfield  {title} {\bibinfo {title} {Hyperuniformity in the manna model, conserved directed percolation and depinning},\ }\href {https://doi.org/10.1103/PhysRevLett.133.067103} {\bibfield  {journal} {\bibinfo  {journal} {Phys. Rev. Lett.}\ }\textbf {\bibinfo {volume} {133}},\ \bibinfo {pages} {067103} (\bibinfo {year} {2024})}\BibitemShut {NoStop}%
\bibitem [{\citenamefont {Ma}\ and\ \citenamefont {Torquato}(2019)}]{Ma19}%
  \BibitemOpen
  \bibfield  {author} {\bibinfo {author} {\bibfnamefont {Z.}~\bibnamefont {Ma}}\ and\ \bibinfo {author} {\bibfnamefont {S.}~\bibnamefont {Torquato}},\ }\bibfield  {title} {\bibinfo {title} {Hyperuniformity of generalized random organization models},\ }\href {https://doi.org/10.1103/PhysRevE.99.022115} {\bibfield  {journal} {\bibinfo  {journal} {Phys. Rev. E}\ }\textbf {\bibinfo {volume} {99}},\ \bibinfo {pages} {022115} (\bibinfo {year} {2019})}\BibitemShut {NoStop}%
\bibitem [{\citenamefont {Torquato}\ \emph {et~al.}(2000)\citenamefont {Torquato}, \citenamefont {Truskett},\ and\ \citenamefont {Debenedetti}}]{To00b}%
  \BibitemOpen
  \bibfield  {author} {\bibinfo {author} {\bibfnamefont {S.}~\bibnamefont {Torquato}}, \bibinfo {author} {\bibfnamefont {T.~M.}\ \bibnamefont {Truskett}},\ and\ \bibinfo {author} {\bibfnamefont {P.~G.}\ \bibnamefont {Debenedetti}},\ }\bibfield  {title} {\bibinfo {title} {Is random close packing of spheres well defined?},\ }\href {https://doi.org/10.1103/PhysRevLett.84.2064} {\bibfield  {journal} {\bibinfo  {journal} {Phys. Rev. Lett.}\ }\textbf {\bibinfo {volume} {84}},\ \bibinfo {pages} {2064} (\bibinfo {year} {2000})}\BibitemShut {NoStop}%
\bibitem [{\citenamefont {Maher}\ \emph {et~al.}(2023)\citenamefont {Maher}, \citenamefont {Jiao},\ and\ \citenamefont {Torquato}}]{Ma23}%
  \BibitemOpen
  \bibfield  {author} {\bibinfo {author} {\bibfnamefont {C.~E.}\ \bibnamefont {Maher}}, \bibinfo {author} {\bibfnamefont {Y.}~\bibnamefont {Jiao}},\ and\ \bibinfo {author} {\bibfnamefont {S.}~\bibnamefont {Torquato}},\ }\bibfield  {title} {\bibinfo {title} {Hyperuniformity of maximally random jammed packings of hyperspheres across spatial dimensions},\ }\href {https://doi.org/10.1103/PhysRevE.108.064602} {\bibfield  {journal} {\bibinfo  {journal} {Phys. Rev. E}\ }\textbf {\bibinfo {volume} {108}},\ \bibinfo {pages} {064602} (\bibinfo {year} {2023})}\BibitemShut {NoStop}%
\bibitem [{\citenamefont {Torquato}\ \emph {et~al.}(2008)\citenamefont {Torquato}, \citenamefont {Scardicchio},\ and\ \citenamefont {Zachary}}]{To08b}%
  \BibitemOpen
  \bibfield  {author} {\bibinfo {author} {\bibfnamefont {S.}~\bibnamefont {Torquato}}, \bibinfo {author} {\bibfnamefont {A.}~\bibnamefont {Scardicchio}},\ and\ \bibinfo {author} {\bibfnamefont {C.~E.}\ \bibnamefont {Zachary}},\ }\bibfield  {title} {\bibinfo {title} {Point processes in arbitrary dimension from {F}ermionic gases, random matrix theory, and number theory},\ }\href {https://doi.org/10.1088/1742-5468/2008/11/p11019} {\bibfield  {journal} {\bibinfo  {journal} {J. Stat. Mech.: Theory Exp.}\ }\textbf {\bibinfo {volume} {2008}},\ \bibinfo {pages} {P11019}}\BibitemShut {NoStop}%
\bibitem [{\citenamefont {Montgomery}(1973)}]{Mon73}%
  \BibitemOpen
  \bibfield  {author} {\bibinfo {author} {\bibfnamefont {H.~L.}\ \bibnamefont {Montgomery}},\ }\bibfield  {title} {\bibinfo {title} {The pair correlation of zeros of the zeta function},\ }in\ \href@noop {} {\emph {\bibinfo {booktitle} {Proc. Symp. Pure math}}},\ Vol.~\bibinfo {volume} {24}\ (\bibinfo {year} {1973})\ pp.\ \bibinfo {pages} {181--193}\BibitemShut {NoStop}%
\bibitem [{\citenamefont {{Lesanovsky}}\ and\ \citenamefont {{Garrahan}}(2014)}]{Le14}%
  \BibitemOpen
  \bibfield  {author} {\bibinfo {author} {\bibfnamefont {I.}~\bibnamefont {{Lesanovsky}}}\ and\ \bibinfo {author} {\bibfnamefont {J.~P.}\ \bibnamefont {{Garrahan}}},\ }\bibfield  {title} {\bibinfo {title} {{Out-of-equilibrium structures in strongly interacting Rydberg gases with dissipation}},\ }\href {https://doi.org/10.1103/PhysRevA.90.011603} {\bibfield  {journal} {\bibinfo  {journal} {Phys. Rev. A}\ }\textbf {\bibinfo {volume} {90}},\ \bibinfo {pages} {011603} (\bibinfo {year} {2014})}\BibitemShut {NoStop}%
\bibitem [{\citenamefont {Lei}\ \emph {et~al.}(2019)\citenamefont {Lei}, \citenamefont {Ciamarra},\ and\ \citenamefont {Ni}}]{Le19b}%
  \BibitemOpen
  \bibfield  {author} {\bibinfo {author} {\bibfnamefont {Q.-L.}\ \bibnamefont {Lei}}, \bibinfo {author} {\bibfnamefont {M.~P.}\ \bibnamefont {Ciamarra}},\ and\ \bibinfo {author} {\bibfnamefont {R.}~\bibnamefont {Ni}},\ }\bibfield  {title} {\bibinfo {title} {Nonequilibrium strongly hyperuniform fluids of circle active particles with large local density fluctuations},\ }\href@noop {} {\bibfield  {journal} {\bibinfo  {journal} {Sci. Adv.}\ }\textbf {\bibinfo {volume} {5}} (\bibinfo {year} {2019})}\BibitemShut {NoStop}%
\bibitem [{\citenamefont {Backofen}\ \emph {et~al.}(2024)\citenamefont {Backofen}, \citenamefont {Altawil}, \citenamefont {Salvalaglio},\ and\ \citenamefont {Voigt}}]{backofen_nonequilibrium_2024}%
  \BibitemOpen
  \bibfield  {author} {\bibinfo {author} {\bibfnamefont {R.}~\bibnamefont {Backofen}}, \bibinfo {author} {\bibfnamefont {A.~Y.~A.}\ \bibnamefont {Altawil}}, \bibinfo {author} {\bibfnamefont {M.}~\bibnamefont {Salvalaglio}},\ and\ \bibinfo {author} {\bibfnamefont {A.}~\bibnamefont {Voigt}},\ }\bibfield  {title} {\bibinfo {title} {Nonequilibrium hyperuniform states in active turbulence},\ }\href {https://doi.org/10.1073/pnas.2320719121} {\bibfield  {journal} {\bibinfo  {journal} {Proc. Natl. Acad. Sci. U.S.A.}\ }\textbf {\bibinfo {volume} {121}},\ \bibinfo {pages} {e2320719121} (\bibinfo {year} {2024})}\BibitemShut {NoStop}%
\bibitem [{\citenamefont {Jiao}\ \emph {et~al.}(2014)\citenamefont {Jiao}, \citenamefont {Lau}, \citenamefont {Hatzikirou}, \citenamefont {Meyer-Hermann}, \citenamefont {Corbo},\ and\ \citenamefont {Torquato}}]{Ji14}%
  \BibitemOpen
  \bibfield  {author} {\bibinfo {author} {\bibfnamefont {Y.}~\bibnamefont {Jiao}}, \bibinfo {author} {\bibfnamefont {T.}~\bibnamefont {Lau}}, \bibinfo {author} {\bibfnamefont {H.}~\bibnamefont {Hatzikirou}}, \bibinfo {author} {\bibfnamefont {M.}~\bibnamefont {Meyer-Hermann}}, \bibinfo {author} {\bibfnamefont {J.~C.}\ \bibnamefont {Corbo}},\ and\ \bibinfo {author} {\bibfnamefont {S.}~\bibnamefont {Torquato}},\ }\bibfield  {title} {\bibinfo {title} {Avian photoreceptor patterns represent a disordered hyperuniform solution to a multiscale packing problem},\ }\href {https://doi.org/10.1103/PhysRevE.89.022721} {\bibfield  {journal} {\bibinfo  {journal} {Phys. Rev. E}\ }\textbf {\bibinfo {volume} {89}},\ \bibinfo {pages} {022721} (\bibinfo {year} {2014})}\BibitemShut {NoStop}%
\bibitem [{\citenamefont {Mayer}\ \emph {et~al.}(2015)\citenamefont {Mayer}, \citenamefont {Balasubramanian}, \citenamefont {Mora},\ and\ \citenamefont {Walczak}}]{Ma15}%
  \BibitemOpen
  \bibfield  {author} {\bibinfo {author} {\bibfnamefont {A.}~\bibnamefont {Mayer}}, \bibinfo {author} {\bibfnamefont {V.}~\bibnamefont {Balasubramanian}}, \bibinfo {author} {\bibfnamefont {T.}~\bibnamefont {Mora}},\ and\ \bibinfo {author} {\bibfnamefont {A.~M.}\ \bibnamefont {Walczak}},\ }\bibfield  {title} {\bibinfo {title} {How a well-adapted immune system is organized},\ }\href {https://doi.org/10.1073/pnas.1421827112} {\bibfield  {journal} {\bibinfo  {journal} {Proc. Nat. Acad. Sci. U. S. A.}\ }\textbf {\bibinfo {volume} {112}},\ \bibinfo {pages} {5950} (\bibinfo {year} {2015})}\BibitemShut {NoStop}%
\bibitem [{\citenamefont {Huang}\ \emph {et~al.}(2021)\citenamefont {Huang}, \citenamefont {Hu}, \citenamefont {Yang}, \citenamefont {Liu},\ and\ \citenamefont {Zhang}}]{Hu21}%
  \BibitemOpen
  \bibfield  {author} {\bibinfo {author} {\bibfnamefont {M.}~\bibnamefont {Huang}}, \bibinfo {author} {\bibfnamefont {W.}~\bibnamefont {Hu}}, \bibinfo {author} {\bibfnamefont {S.}~\bibnamefont {Yang}}, \bibinfo {author} {\bibfnamefont {Q.~X.}\ \bibnamefont {Liu}},\ and\ \bibinfo {author} {\bibfnamefont {H.~P.}\ \bibnamefont {Zhang}},\ }\bibfield  {title} {\bibinfo {title} {{Circular swimming motility and disordered hyperuniform state in an algae system}},\ }\href {https://doi.org/10.1073/pnas.2100493118} {\bibfield  {journal} {\bibinfo  {journal} {Proc. Natl. Acad. Sci. U.S.A.}\ }\textbf {\bibinfo {volume} {118}},\ \bibinfo {pages} {e2100493118} (\bibinfo {year} {2021})}\BibitemShut {NoStop}%
\bibitem [{\citenamefont {Ge}(2023)}]{ge_hidden_2023}%
  \BibitemOpen
  \bibfield  {author} {\bibinfo {author} {\bibfnamefont {Z.}~\bibnamefont {Ge}},\ }\bibfield  {title} {\bibinfo {title} {The hidden order of turing patterns in arid and semi-arid vegetation ecosystems},\ }\href {https://doi.org/10.1073/pnas.2306514120} {\bibfield  {journal} {\bibinfo  {journal} {Proc. Natl. Acad. Sci. U.S.A.}\ }\textbf {\bibinfo {volume} {120}},\ \bibinfo {pages} {e2306514120} (\bibinfo {year} {2023})}\BibitemShut {NoStop}%
\bibitem [{\citenamefont {{Lomba}}\ \emph {et~al.}(2018)\citenamefont {{Lomba}}, \citenamefont {{Weis}},\ and\ \citenamefont {{Torquato}}}]{Lo18a}%
  \BibitemOpen
  \bibfield  {author} {\bibinfo {author} {\bibfnamefont {E.}~\bibnamefont {{Lomba}}}, \bibinfo {author} {\bibfnamefont {J.-J.}\ \bibnamefont {{Weis}}},\ and\ \bibinfo {author} {\bibfnamefont {S.}~\bibnamefont {{Torquato}}},\ }\bibfield  {title} {\bibinfo {title} {Disordered multihyperuniformity derived from binary plasmas},\ }\href {https://doi.org/10.1103/PhysRevE.97.010102} {\bibfield  {journal} {\bibinfo  {journal} {Phys. Rev. E}\ }\textbf {\bibinfo {volume} {97}},\ \bibinfo {pages} {010102(R)} (\bibinfo {year} {2018})}\BibitemShut {NoStop}%
\bibitem [{\citenamefont {Christogeorgos}\ \emph {et~al.}(2024)\citenamefont {Christogeorgos}, \citenamefont {Okon},\ and\ \citenamefont {Hao}}]{christogeorgos2024computational}%
  \BibitemOpen
  \bibfield  {author} {\bibinfo {author} {\bibfnamefont {O.}~\bibnamefont {Christogeorgos}}, \bibinfo {author} {\bibfnamefont {E.}~\bibnamefont {Okon}},\ and\ \bibinfo {author} {\bibfnamefont {Y.}~\bibnamefont {Hao}},\ }\bibfield  {title} {\bibinfo {title} {A computational model for generating multihyperuniform distributions for realistic antenna array and metasurface designs},\ }\href@noop {} {\bibfield  {journal} {\bibinfo  {journal} {EPJ Applied Metamaterials}\ }\textbf {\bibinfo {volume} {11}},\ \bibinfo {pages} {5} (\bibinfo {year} {2024})}\BibitemShut {NoStop}%
\bibitem [{\citenamefont {Uche}\ \emph {et~al.}(2004)\citenamefont {Uche}, \citenamefont {Stillinger},\ and\ \citenamefont {Torquato}}]{Uc04b}%
  \BibitemOpen
  \bibfield  {author} {\bibinfo {author} {\bibfnamefont {O.~U.}\ \bibnamefont {Uche}}, \bibinfo {author} {\bibfnamefont {F.~H.}\ \bibnamefont {Stillinger}},\ and\ \bibinfo {author} {\bibfnamefont {S.}~\bibnamefont {Torquato}},\ }\bibfield  {title} {\bibinfo {title} {Constraints on collective density variables: Two dimensions},\ }\href {https://doi.org/10.1103/PhysRevE.70.046122} {\bibfield  {journal} {\bibinfo  {journal} {Phys. Rev. E}\ }\textbf {\bibinfo {volume} {70}},\ \bibinfo {pages} {046122} (\bibinfo {year} {2004})}\BibitemShut {NoStop}%
\bibitem [{\citenamefont {Batten}\ \emph {et~al.}(2008)\citenamefont {Batten}, \citenamefont {Stillinger},\ and\ \citenamefont {Torquato}}]{batten_classical_2008}%
  \BibitemOpen
  \bibfield  {author} {\bibinfo {author} {\bibfnamefont {R.}~\bibnamefont {Batten}}, \bibinfo {author} {\bibfnamefont {F.}~\bibnamefont {Stillinger}},\ and\ \bibinfo {author} {\bibfnamefont {S.}~\bibnamefont {Torquato}},\ }\bibfield  {title} {\bibinfo {title} {Classical disordered ground states: Super-ideal gases and stealth and equi-luminous materials},\ }\href {https://doi.org/10.1063/1.2961314} {\bibfield  {journal} {\bibinfo  {journal} {J. Appl. Phys.}\ }\textbf {\bibinfo {volume} {104}},\ \bibinfo {pages} {033504} (\bibinfo {year} {2008})}\BibitemShut {NoStop}%
\bibitem [{\citenamefont {Torquato}\ \emph {et~al.}(2015)\citenamefont {Torquato}, \citenamefont {Zhang},\ and\ \citenamefont {Stillinger}}]{torquato_ensemble_2015}%
  \BibitemOpen
  \bibfield  {author} {\bibinfo {author} {\bibfnamefont {S.}~\bibnamefont {Torquato}}, \bibinfo {author} {\bibfnamefont {G.}~\bibnamefont {Zhang}},\ and\ \bibinfo {author} {\bibfnamefont {F.~H.}\ \bibnamefont {Stillinger}},\ }\bibfield  {title} {\bibinfo {title} {Ensemble theory for stealthy hyperuniform disordered ground states},\ }\href {https://doi.org/10.1103/PhysRevX.5.021020} {\bibfield  {journal} {\bibinfo  {journal} {Phys. Rev. X}\ }\textbf {\bibinfo {volume} {5}},\ \bibinfo {pages} {021020} (\bibinfo {year} {2015})}\BibitemShut {NoStop}%
\bibitem [{\citenamefont {Zhang}\ \emph {et~al.}(2015)\citenamefont {Zhang}, \citenamefont {Stillinger},\ and\ \citenamefont {Torquato}}]{Zh15a}%
  \BibitemOpen
  \bibfield  {author} {\bibinfo {author} {\bibfnamefont {G.}~\bibnamefont {Zhang}}, \bibinfo {author} {\bibfnamefont {F.}~\bibnamefont {Stillinger}},\ and\ \bibinfo {author} {\bibfnamefont {S.}~\bibnamefont {Torquato}},\ }\bibfield  {title} {\bibinfo {title} {Ground states of stealthy hyperuniform potentials: {I}. {E}ntropically favored configurations},\ }\href {https://doi.org/10.1103/PhysRevE.92.022119} {\bibfield  {journal} {\bibinfo  {journal} {Phys. Rev. E}\ }\textbf {\bibinfo {volume} {92}},\ \bibinfo {pages} {022119} (\bibinfo {year} {2015})}\BibitemShut {NoStop}%
\bibitem [{\citenamefont {Zhang}\ \emph {et~al.}(2017)\citenamefont {Zhang}, \citenamefont {Stillinger},\ and\ \citenamefont {Torquato}}]{zhang_can_2017}%
  \BibitemOpen
  \bibfield  {author} {\bibinfo {author} {\bibfnamefont {G.}~\bibnamefont {Zhang}}, \bibinfo {author} {\bibfnamefont {F.~H.}\ \bibnamefont {Stillinger}},\ and\ \bibinfo {author} {\bibfnamefont {S.}~\bibnamefont {Torquato}},\ }\bibfield  {title} {\bibinfo {title} {Can exotic disordered ``stealthy" particle configurations tolerate arbitrarily large holes?},\ }\href {https://doi.org/10.1039/c7sm01028a} {\bibfield  {journal} {\bibinfo  {journal} {Soft Matter}\ }\textbf {\bibinfo {volume} {13}},\ \bibinfo {pages} {6197} (\bibinfo {year} {2017})}\BibitemShut {NoStop}%
\bibitem [{\citenamefont {Ghosh}\ and\ \citenamefont {Lebowitz}(2018)}]{ghosh_generalized_2017}%
  \BibitemOpen
  \bibfield  {author} {\bibinfo {author} {\bibfnamefont {S.}~\bibnamefont {Ghosh}}\ and\ \bibinfo {author} {\bibfnamefont {J.~L.}\ \bibnamefont {Lebowitz}},\ }\bibfield  {title} {\bibinfo {title} {Generalized stealthy hyperuniform processes: maximal rigidity and the bounded holes conjecture},\ }\href {https://doi.org/10.1007/s00220-018-3226-5} {\bibfield  {journal} {\bibinfo  {journal} {Commun. Math. Phys.}\ }\textbf {\bibinfo {volume} {363}},\ \bibinfo {pages} {97} (\bibinfo {year} {2018})}\BibitemShut {NoStop}%
\bibitem [{\citenamefont {{Leseur}}\ \emph {et~al.}(2016)\citenamefont {{Leseur}}, \citenamefont {{Pierrat}},\ and\ \citenamefont {{Carminati}}}]{Le16}%
  \BibitemOpen
  \bibfield  {author} {\bibinfo {author} {\bibfnamefont {O.}~\bibnamefont {{Leseur}}}, \bibinfo {author} {\bibfnamefont {R.}~\bibnamefont {{Pierrat}}},\ and\ \bibinfo {author} {\bibfnamefont {R.}~\bibnamefont {{Carminati}}},\ }\bibfield  {title} {\bibinfo {title} {{High-density hyperuniform materials can be transparent}},\ }\href {https://doi.org/10.1364/optica.3.000763} {\bibfield  {journal} {\bibinfo  {journal} {Optica}\ }\textbf {\bibinfo {volume} {3}},\ \bibinfo {pages} {763} (\bibinfo {year} {2016})}\BibitemShut {NoStop}%
\bibitem [{\citenamefont {{Froufe-P{\'e}rez}}\ \emph {et~al.}(2017)\citenamefont {{Froufe-P{\'e}rez}}, \citenamefont {{Engel}}, \citenamefont {{Jos{\'e} S{\'a}enz}},\ and\ \citenamefont {{Scheffold}}}]{Fr17}%
  \BibitemOpen
  \bibfield  {author} {\bibinfo {author} {\bibfnamefont {L.~S.}\ \bibnamefont {{Froufe-P{\'e}rez}}}, \bibinfo {author} {\bibfnamefont {M.}~\bibnamefont {{Engel}}}, \bibinfo {author} {\bibfnamefont {J.}~\bibnamefont {{Jos{\'e} S{\'a}enz}}},\ and\ \bibinfo {author} {\bibfnamefont {F.}~\bibnamefont {{Scheffold}}},\ }\bibfield  {title} {\bibinfo {title} {{Band gap formation and Anderson localization in disordered photonic materials with structural correlationss}},\ }\href {https://doi.org/10.1073/pnas.1705130114} {\bibfield  {journal} {\bibinfo  {journal} {Proc. Nat. Acad. Sci.}\ }\textbf {\bibinfo {volume} {114}},\ \bibinfo {pages} {9570–} (\bibinfo {year} {2017})}\BibitemShut {NoStop}%
\bibitem [{\citenamefont {Torquato}\ and\ \citenamefont {Kim}(2021)}]{To21a}%
  \BibitemOpen
  \bibfield  {author} {\bibinfo {author} {\bibfnamefont {S.}~\bibnamefont {Torquato}}\ and\ \bibinfo {author} {\bibfnamefont {J.}~\bibnamefont {Kim}},\ }\bibfield  {title} {\bibinfo {title} {Nonlocal effective electromagnetic wave characteristics of composite media: Beyond the quasistatic regime},\ }\href {https://doi.org/10.1103/PhysRevX.11.021002} {\bibfield  {journal} {\bibinfo  {journal} {Phys. Rev. X}\ }\textbf {\bibinfo {volume} {11}},\ \bibinfo {pages} {021002} (\bibinfo {year} {2021})}\BibitemShut {NoStop}%
\bibitem [{\citenamefont {{Froufe-P{\'e}rez}}\ \emph {et~al.}(2023)\citenamefont {{Froufe-P{\'e}rez}}, \citenamefont {Aubry}, \citenamefont {Scheffold},\ and\ \citenamefont {Magkiriadou}}]{Fr23}%
  \BibitemOpen
  \bibfield  {author} {\bibinfo {author} {\bibfnamefont {L.~S.}\ \bibnamefont {{Froufe-P{\'e}rez}}}, \bibinfo {author} {\bibfnamefont {G.~J.}\ \bibnamefont {Aubry}}, \bibinfo {author} {\bibfnamefont {F.}~\bibnamefont {Scheffold}},\ and\ \bibinfo {author} {\bibfnamefont {S.}~\bibnamefont {Magkiriadou}},\ }\bibfield  {title} {\bibinfo {title} {Bandgap fluctuations and robustness in two-dimensional hyperuniform dielectric materials},\ }\href {https://doi.org/10.1364/OE.484232} {\bibfield  {journal} {\bibinfo  {journal} {Opt. Express}\ }\textbf {\bibinfo {volume} {31}},\ \bibinfo {pages} {18509} (\bibinfo {year} {2023})}\BibitemShut {NoStop}%
\bibitem [{\citenamefont {Klatt}\ \emph {et~al.}(2022)\citenamefont {Klatt}, \citenamefont {Steinhardt},\ and\ \citenamefont {Torquato}}]{Kl22}%
  \BibitemOpen
  \bibfield  {author} {\bibinfo {author} {\bibfnamefont {M.~A.}\ \bibnamefont {Klatt}}, \bibinfo {author} {\bibfnamefont {P.~J.}\ \bibnamefont {Steinhardt}},\ and\ \bibinfo {author} {\bibfnamefont {S.}~\bibnamefont {Torquato}},\ }\bibfield  {title} {\bibinfo {title} {Wave propagation and band tails of two-dimensional disordered systems in the thermodynamic limit},\ }\href {https://doi.org/10.1073/pnas.2213633119} {\bibfield  {journal} {\bibinfo  {journal} {Proc. Nat. Acad. Sci.}\ }\textbf {\bibinfo {volume} {119}},\ \bibinfo {pages} {e2213633119} (\bibinfo {year} {2022})}\BibitemShut {NoStop}%
\bibitem [{\citenamefont {Kim}\ and\ \citenamefont {Torquato}(2023)}]{kim_effective_2023}%
  \BibitemOpen
  \bibfield  {author} {\bibinfo {author} {\bibfnamefont {J.}~\bibnamefont {Kim}}\ and\ \bibinfo {author} {\bibfnamefont {S.}~\bibnamefont {Torquato}},\ }\bibfield  {title} {\bibinfo {title} {Effective electromagnetic wave properties of disordered stealthy hyperuniform layered media beyond the quasistatic regime},\ }\href {https://doi.org/10.1364/OPTICA.489797} {\bibfield  {journal} {\bibinfo  {journal} {Optica}\ }\textbf {\bibinfo {volume} {10}},\ \bibinfo {pages} {965} (\bibinfo {year} {2023})}\BibitemShut {NoStop}%
\bibitem [{\citenamefont {Alha{\"i}tz}\ \emph {et~al.}(2023)\citenamefont {Alha{\"i}tz}, \citenamefont {Conoir},\ and\ \citenamefont {{Valier-Brasier}}}]{alhaitz_experimental_2023}%
  \BibitemOpen
  \bibfield  {author} {\bibinfo {author} {\bibfnamefont {L.}~\bibnamefont {Alha{\"i}tz}}, \bibinfo {author} {\bibfnamefont {J.-M.}\ \bibnamefont {Conoir}},\ and\ \bibinfo {author} {\bibfnamefont {T.}~\bibnamefont {{Valier-Brasier}}},\ }\bibfield  {title} {\bibinfo {title} {Experimental evidence of isotropic transparency and complete band gap formation for ultrasound propagation in stealthy hyperuniform media},\ }\href {https://doi.org/10.1103/PhysRevE.108.065001} {\bibfield  {journal} {\bibinfo  {journal} {Phys. Rev. E}\ }\textbf {\bibinfo {volume} {108}},\ \bibinfo {pages} {065001} (\bibinfo {year} {2023})}\BibitemShut {NoStop}%
\bibitem [{\citenamefont {Kim}\ and\ \citenamefont {Torquato}(2024{\natexlab{a}})}]{Kim_2024_extraordinary}%
  \BibitemOpen
  \bibfield  {author} {\bibinfo {author} {\bibfnamefont {J.}~\bibnamefont {Kim}}\ and\ \bibinfo {author} {\bibfnamefont {S.}~\bibnamefont {Torquato}},\ }\bibfield  {title} {\bibinfo {title} {Extraordinary optical and transport properties of disordered stealthy hyperuniform two-phase media},\ }\href {https://doi.org/10.1088/1361-648X/ad2802} {\bibfield  {journal} {\bibinfo  {journal} {Journal of Physics: Condensed Matter}\ }\textbf {\bibinfo {volume} {36}},\ \bibinfo {pages} {225701} (\bibinfo {year} {2024}{\natexlab{a}})}\BibitemShut {NoStop}%
\bibitem [{\citenamefont {Bigourdan}\ \emph {et~al.}(2019)\citenamefont {Bigourdan}, \citenamefont {Pierrat},\ and\ \citenamefont {Carminati}}]{Bi19}%
  \BibitemOpen
  \bibfield  {author} {\bibinfo {author} {\bibfnamefont {F.}~\bibnamefont {Bigourdan}}, \bibinfo {author} {\bibfnamefont {R.}~\bibnamefont {Pierrat}},\ and\ \bibinfo {author} {\bibfnamefont {R.}~\bibnamefont {Carminati}},\ }\bibfield  {title} {\bibinfo {title} {Enhanced absorption of waves in stealth hyperuniform disordered media},\ }\href {https://doi.org/10.1364/oe.27.008666} {\bibfield  {journal} {\bibinfo  {journal} {Optics Express}\ }\textbf {\bibinfo {volume} {27}},\ \bibinfo {pages} {8666} (\bibinfo {year} {2019})}\BibitemShut {NoStop}%
\bibitem [{\citenamefont {Merkel}\ \emph {et~al.}(2024)\citenamefont {Merkel}, \citenamefont {Stappers}, \citenamefont {Ray}, \citenamefont {Denz},\ and\ \citenamefont {Imbrock}}]{merkel_stealthy_2023}%
  \BibitemOpen
  \bibfield  {author} {\bibinfo {author} {\bibfnamefont {M.}~\bibnamefont {Merkel}}, \bibinfo {author} {\bibfnamefont {M.}~\bibnamefont {Stappers}}, \bibinfo {author} {\bibfnamefont {D.}~\bibnamefont {Ray}}, \bibinfo {author} {\bibfnamefont {C.}~\bibnamefont {Denz}},\ and\ \bibinfo {author} {\bibfnamefont {J.}~\bibnamefont {Imbrock}},\ }\bibfield  {title} {\bibinfo {title} {Stealthy hyperuniform surface structures for efficiency enhancement of organic solar cells},\ }\href {https://doi.org/10.1002/adpr.202300256} {\bibfield  {journal} {\bibinfo  {journal} {Adv. Photonics Res.}\ }\textbf {\bibinfo {volume} {n/a}},\ \bibinfo {pages} {2300256} (\bibinfo {year} {2024})}\BibitemShut {NoStop}%
\bibitem [{\citenamefont {Sgrignuoli}\ \emph {et~al.}(2022)\citenamefont {Sgrignuoli}, \citenamefont {Torquato},\ and\ \citenamefont {Dal~Negro}}]{Sg22}%
  \BibitemOpen
  \bibfield  {author} {\bibinfo {author} {\bibfnamefont {F.}~\bibnamefont {Sgrignuoli}}, \bibinfo {author} {\bibfnamefont {S.}~\bibnamefont {Torquato}},\ and\ \bibinfo {author} {\bibfnamefont {L.}~\bibnamefont {Dal~Negro}},\ }\bibfield  {title} {\bibinfo {title} {Subdiffusive wave transport and weak localization transition in three-dimensional stealthy hyperuniform disordered systems},\ }\href {https://doi.org/doi.org/10.1103/PhysRevB.105.064204} {\bibfield  {journal} {\bibinfo  {journal} {Phys. Rev. B}\ }\textbf {\bibinfo {volume} {105}},\ \bibinfo {pages} {064204} (\bibinfo {year} {2022})}\BibitemShut {NoStop}%
\bibitem [{\citenamefont {Scheffold}\ \emph {et~al.}(2022)\citenamefont {Scheffold}, \citenamefont {Haberko}, \citenamefont {Magkiriadou},\ and\ \citenamefont {{Froufe-P{\'e}rez}}}]{Sc22}%
  \BibitemOpen
  \bibfield  {author} {\bibinfo {author} {\bibfnamefont {F.}~\bibnamefont {Scheffold}}, \bibinfo {author} {\bibfnamefont {J.}~\bibnamefont {Haberko}}, \bibinfo {author} {\bibfnamefont {S.}~\bibnamefont {Magkiriadou}},\ and\ \bibinfo {author} {\bibfnamefont {L.~S.}\ \bibnamefont {{Froufe-P{\'e}rez}}},\ }\bibfield  {title} {\bibinfo {title} {Transport through amorphous photonic materials with localization and bandgap regimes},\ }\href {https://doi.org/10.1103/PhysRevLett.129.157402} {\bibfield  {journal} {\bibinfo  {journal} {Phys. Rev. Lett.}\ }\textbf {\bibinfo {volume} {129}},\ \bibinfo {pages} {157402} (\bibinfo {year} {2022})}\BibitemShut {NoStop}%
\bibitem [{\citenamefont {Gkantzounis}\ \emph {et~al.}(2017)\citenamefont {Gkantzounis}, \citenamefont {Amoah},\ and\ \citenamefont {Florescu}}]{Gk17}%
  \BibitemOpen
  \bibfield  {author} {\bibinfo {author} {\bibfnamefont {G.}~\bibnamefont {Gkantzounis}}, \bibinfo {author} {\bibfnamefont {T.}~\bibnamefont {Amoah}},\ and\ \bibinfo {author} {\bibfnamefont {M.}~\bibnamefont {Florescu}},\ }\bibfield  {title} {\bibinfo {title} {Hyperuniform disordered phononic structures},\ }\href {https://doi.org/10.1103/PhysRevB.95.094120} {\bibfield  {journal} {\bibinfo  {journal} {Phys. Rev. B}\ }\textbf {\bibinfo {volume} {95}},\ \bibinfo {pages} {094120} (\bibinfo {year} {2017})}\BibitemShut {NoStop}%
\bibitem [{\citenamefont {Romero-Garc{\'\i}a}\ \emph {et~al.}(2019)\citenamefont {Romero-Garc{\'\i}a}, \citenamefont {Lamothe}, \citenamefont {Theocharis}, \citenamefont {Richoux},\ and\ \citenamefont {Garc{\'\i}a-Raffi}}]{Ro19}%
  \BibitemOpen
  \bibfield  {author} {\bibinfo {author} {\bibfnamefont {V.}~\bibnamefont {Romero-Garc{\'\i}a}}, \bibinfo {author} {\bibfnamefont {N.}~\bibnamefont {Lamothe}}, \bibinfo {author} {\bibfnamefont {G.}~\bibnamefont {Theocharis}}, \bibinfo {author} {\bibfnamefont {O.}~\bibnamefont {Richoux}},\ and\ \bibinfo {author} {\bibfnamefont {L.~M.}\ \bibnamefont {Garc{\'\i}a-Raffi}},\ }\bibfield  {title} {\bibinfo {title} {Stealth acoustic materials},\ }\href {https://doi.org/10.1103/PhysRevApplied.11.054076} {\bibfield  {journal} {\bibinfo  {journal} {Phys. Rev. Appl.}\ }\textbf {\bibinfo {volume} {11}},\ \bibinfo {pages} {054076} (\bibinfo {year} {2019})}\BibitemShut {NoStop}%
\bibitem [{\citenamefont {Rohfritsch}\ \emph {et~al.}(2020)\citenamefont {Rohfritsch}, \citenamefont {Conoir}, \citenamefont {Valier-Brasier},\ and\ \citenamefont {Marchiano}}]{Roh20}%
  \BibitemOpen
  \bibfield  {author} {\bibinfo {author} {\bibfnamefont {A.}~\bibnamefont {Rohfritsch}}, \bibinfo {author} {\bibfnamefont {J.-M.}\ \bibnamefont {Conoir}}, \bibinfo {author} {\bibfnamefont {T.}~\bibnamefont {Valier-Brasier}},\ and\ \bibinfo {author} {\bibfnamefont {R.}~\bibnamefont {Marchiano}},\ }\bibfield  {title} {\bibinfo {title} {Impact of particle size and multiple scattering on the propagation of waves in stealthy-hyperuniform media},\ }\href {https://doi.org/10.1103/PhysRevE.102.053001} {\bibfield  {journal} {\bibinfo  {journal} {Phys. Rev. E}\ }\textbf {\bibinfo {volume} {102}},\ \bibinfo {pages} {053001} (\bibinfo {year} {2020})}\BibitemShut {NoStop}%
\bibitem [{\citenamefont {Zhang}\ \emph {et~al.}(2019)\citenamefont {Zhang}, \citenamefont {Chu}, \citenamefont {Giddens}, \citenamefont {Wu},\ and\ \citenamefont {Hao}}]{Zh19}%
  \BibitemOpen
  \bibfield  {author} {\bibinfo {author} {\bibfnamefont {H.}~\bibnamefont {Zhang}}, \bibinfo {author} {\bibfnamefont {H.}~\bibnamefont {Chu}}, \bibinfo {author} {\bibfnamefont {H.}~\bibnamefont {Giddens}}, \bibinfo {author} {\bibfnamefont {W.}~\bibnamefont {Wu}},\ and\ \bibinfo {author} {\bibfnamefont {Y.}~\bibnamefont {Hao}},\ }\bibfield  {title} {\bibinfo {title} {Experimental demonstration of luneburg lens based on hyperuniform disordered media},\ }\href {https://doi.org/10.1063/1.5055295} {\bibfield  {journal} {\bibinfo  {journal} {Appl. Phys. Lett.}\ }\textbf {\bibinfo {volume} {114}},\ \bibinfo {pages} {053507} (\bibinfo {year} {2019})}\BibitemShut {NoStop}%
\bibitem [{\citenamefont {Christogeorgos}\ \emph {et~al.}(2021)\citenamefont {Christogeorgos}, \citenamefont {Zhang}, \citenamefont {Cheng},\ and\ \citenamefont {Hao}}]{Ch21}%
  \BibitemOpen
  \bibfield  {author} {\bibinfo {author} {\bibfnamefont {O.}~\bibnamefont {Christogeorgos}}, \bibinfo {author} {\bibfnamefont {H.}~\bibnamefont {Zhang}}, \bibinfo {author} {\bibfnamefont {Q.}~\bibnamefont {Cheng}},\ and\ \bibinfo {author} {\bibfnamefont {Y.}~\bibnamefont {Hao}},\ }\bibfield  {title} {\bibinfo {title} {Extraordinary {{Directive Emission}} and {{Scanning}} from an {{Array}} of {{Radiation Sources}} with {{Hyperuniform Disorder}}},\ }\href {https://doi.org/10.1103/PhysRevApplied.15.014062} {\bibfield  {journal} {\bibinfo  {journal} {Phys. Rev. Appl.}\ }\textbf {\bibinfo {volume} {15}},\ \bibinfo {pages} {014062} (\bibinfo {year} {2021})}\BibitemShut {NoStop}%
\bibitem [{\citenamefont {Tang}\ \emph {et~al.}(2023)\citenamefont {Tang}, \citenamefont {Wang}, \citenamefont {Wang}, \citenamefont {Gao}, \citenamefont {Li}, \citenamefont {Liang}, \citenamefont {Sebbah}, \citenamefont {Li}, \citenamefont {Zhang},\ and\ \citenamefont {Shi}}]{tang2023hyperuniform}%
  \BibitemOpen
  \bibfield  {author} {\bibinfo {author} {\bibfnamefont {K.}~\bibnamefont {Tang}}, \bibinfo {author} {\bibfnamefont {Y.}~\bibnamefont {Wang}}, \bibinfo {author} {\bibfnamefont {S.}~\bibnamefont {Wang}}, \bibinfo {author} {\bibfnamefont {D.}~\bibnamefont {Gao}}, \bibinfo {author} {\bibfnamefont {H.}~\bibnamefont {Li}}, \bibinfo {author} {\bibfnamefont {X.}~\bibnamefont {Liang}}, \bibinfo {author} {\bibfnamefont {P.}~\bibnamefont {Sebbah}}, \bibinfo {author} {\bibfnamefont {Y.}~\bibnamefont {Li}}, \bibinfo {author} {\bibfnamefont {J.}~\bibnamefont {Zhang}},\ and\ \bibinfo {author} {\bibfnamefont {J.}~\bibnamefont {Shi}},\ }\bibfield  {title} {\bibinfo {title} {Hyperuniform disordered parametric loudspeaker array},\ }\href {https://doi.org/10.1103/PhysRevApplied.19.054035} {\bibfield  {journal} {\bibinfo  {journal} {Phys. Rev. Appl.}\ }\textbf {\bibinfo {volume} {19}},\ \bibinfo {pages} {054035} (\bibinfo {year} {2023})}\BibitemShut {NoStop}%
\bibitem [{\citenamefont {Tamraoui}\ \emph {et~al.}(2023)\citenamefont {Tamraoui}, \citenamefont {Roux},\ and\ \citenamefont {Liebgott}}]{tamraoui_hyperuniform_2023}%
  \BibitemOpen
  \bibfield  {author} {\bibinfo {author} {\bibfnamefont {M.}~\bibnamefont {Tamraoui}}, \bibinfo {author} {\bibfnamefont {E.}~\bibnamefont {Roux}},\ and\ \bibinfo {author} {\bibfnamefont {H.}~\bibnamefont {Liebgott}},\ }\bibfield  {title} {\bibinfo {title} {Hyperuniform disordered sparse array for 3d ultrasound imaging},\ }in\ \href {https://doi.org/10.1109/IUS51837.2023.10308368} {\emph {\bibinfo {booktitle} {2023 IEEE International Ultrasonics Symposium (IUS)}}}\ (\bibinfo {year} {2023})\ pp.\ \bibinfo {pages} {1--4}\BibitemShut {NoStop}%
\bibitem [{\citenamefont {Granchi}\ \emph {et~al.}(2023)\citenamefont {Granchi}, \citenamefont {Lodde}, \citenamefont {Stokkereit}, \citenamefont {Spalding}, \citenamefont {{van Veldhoven}}, \citenamefont {Sapienza}, \citenamefont {Fiore}, \citenamefont {Gurioli}, \citenamefont {Florescu},\ and\ \citenamefont {Intonti}}]{granchi_nearfield_2023}%
  \BibitemOpen
  \bibfield  {author} {\bibinfo {author} {\bibfnamefont {N.}~\bibnamefont {Granchi}}, \bibinfo {author} {\bibfnamefont {M.}~\bibnamefont {Lodde}}, \bibinfo {author} {\bibfnamefont {K.}~\bibnamefont {Stokkereit}}, \bibinfo {author} {\bibfnamefont {R.}~\bibnamefont {Spalding}}, \bibinfo {author} {\bibfnamefont {P.~J.}\ \bibnamefont {{van Veldhoven}}}, \bibinfo {author} {\bibfnamefont {R.}~\bibnamefont {Sapienza}}, \bibinfo {author} {\bibfnamefont {A.}~\bibnamefont {Fiore}}, \bibinfo {author} {\bibfnamefont {M.}~\bibnamefont {Gurioli}}, \bibinfo {author} {\bibfnamefont {M.}~\bibnamefont {Florescu}},\ and\ \bibinfo {author} {\bibfnamefont {F.}~\bibnamefont {Intonti}},\ }\bibfield  {title} {\bibinfo {title} {Near-field imaging of optical nanocavities in hyperuniform disordered materials},\ }\href {https://doi.org/10.1103/PhysRevB.107.064204} {\bibfield  {journal} {\bibinfo  {journal} {Phys. Rev. B}\ }\textbf {\bibinfo {volume} {107}},\ \bibinfo {pages} {064204} (\bibinfo {year} {2023})}\BibitemShut {NoStop}%
\bibitem [{\citenamefont {Torquato}\ and\ \citenamefont {Chen}(2018)}]{To18c}%
  \BibitemOpen
  \bibfield  {author} {\bibinfo {author} {\bibfnamefont {S.}~\bibnamefont {Torquato}}\ and\ \bibinfo {author} {\bibfnamefont {D.}~\bibnamefont {Chen}},\ }\bibfield  {title} {\bibinfo {title} {Multifunctional hyperuniform cellular networks: optimality, anisotropy and disorder},\ }\href {https://doi.org/10.1088/2399-7532/aaca91} {\bibfield  {journal} {\bibinfo  {journal} {Multifunct. Mater.}\ }\textbf {\bibinfo {volume} {1}},\ \bibinfo {pages} {015001} (\bibinfo {year} {2018})}\BibitemShut {NoStop}%
\bibitem [{\citenamefont {Zhang}\ \emph {et~al.}(2016{\natexlab{b}})\citenamefont {Zhang}, \citenamefont {Stillinger},\ and\ \citenamefont {Torquato}}]{zhang_transport_2016}%
  \BibitemOpen
  \bibfield  {author} {\bibinfo {author} {\bibfnamefont {G.}~\bibnamefont {Zhang}}, \bibinfo {author} {\bibfnamefont {F.}~\bibnamefont {Stillinger}},\ and\ \bibinfo {author} {\bibfnamefont {S.}~\bibnamefont {Torquato}},\ }\bibfield  {title} {\bibinfo {title} {Transport, geometrical, and topological properties of stealthy disordered hyperuniform two-phase systems},\ }\href {https://doi.org/10.1063/1.4972862} {\bibfield  {journal} {\bibinfo  {journal} {J. Chem. Phys.}\ }\textbf {\bibinfo {volume} {145}},\ \bibinfo {pages} {244109} (\bibinfo {year} {2016}{\natexlab{b}})}\BibitemShut {NoStop}%
\bibitem [{\citenamefont {Kim}\ and\ \citenamefont {Torquato}(2025)}]{Kim_2025_DenseSphere}%
  \BibitemOpen
  \bibfield  {author} {\bibinfo {author} {\bibfnamefont {J.}~\bibnamefont {Kim}}\ and\ \bibinfo {author} {\bibfnamefont {S.}~\bibnamefont {Torquato}},\ }\bibfield  {title} {\bibinfo {title} {Ultradense sphere packings derived from disordered stealthy hyperuniform ground states},\ }\href {https://arxiv.org/abs/2504.16924} {\bibfield  {journal} {\bibinfo  {journal} {J. Chem. Phys., in press; arXiv:2504.16924}\ } (\bibinfo {year} {2025})}\BibitemShut {NoStop}%
\bibitem [{\citenamefont {Torquato}(2021)}]{To21d}%
  \BibitemOpen
  \bibfield  {author} {\bibinfo {author} {\bibfnamefont {S.}~\bibnamefont {Torquato}},\ }\bibfield  {title} {\bibinfo {title} {Diffusion spreadability as a probe of the microstructure of complex media across length scales},\ }\href {https://doi.org/10.1103/PhysRevE.104.054102} {\bibfield  {journal} {\bibinfo  {journal} {Phys. Rev. E}\ }\textbf {\bibinfo {volume} {104}},\ \bibinfo {pages} {054102} (\bibinfo {year} {2021})}\BibitemShut {NoStop}%
\bibitem [{\citenamefont {Kim}\ and\ \citenamefont {Torquato}(2024{\natexlab{b}})}]{kim_accurate_2024}%
  \BibitemOpen
  \bibfield  {author} {\bibinfo {author} {\bibfnamefont {J.}~\bibnamefont {Kim}}\ and\ \bibinfo {author} {\bibfnamefont {S.}~\bibnamefont {Torquato}},\ }\bibfield  {title} {\bibinfo {title} {Theoretical prediction of the effective dynamic dielectric constant of disordered hyperuniform anisotropic composites beyond the long-wavelength regime},\ }\href {https://doi.org/10.1364/OME.507918} {\bibfield  {journal} {\bibinfo  {journal} {Opt. Mater. Express}\ }\textbf {\bibinfo {volume} {14}},\ \bibinfo {pages} {194} (\bibinfo {year} {2024}{\natexlab{b}})}\BibitemShut {NoStop}%
\bibitem [{\citenamefont {Zachary}\ and\ \citenamefont {Torquato}(2009)}]{Za09}%
  \BibitemOpen
  \bibfield  {author} {\bibinfo {author} {\bibfnamefont {C.~E.}\ \bibnamefont {Zachary}}\ and\ \bibinfo {author} {\bibfnamefont {S.}~\bibnamefont {Torquato}},\ }\bibfield  {title} {\bibinfo {title} {Hyperuniformity in point patterns and two-phase heterogeneous media},\ }\href {https://doi.org/10.1088/1742-5468/2009/12/P12015} {\bibfield  {journal} {\bibinfo  {journal} {J. Stat. Mech.: Theory \& Exp.}\ }\textbf {\bibinfo {volume} {2009}},\ \bibinfo {pages} {P12015} (\bibinfo {year} {2009})}\BibitemShut {NoStop}%
\bibitem [{\citenamefont {Liu}\ and\ \citenamefont {Nocedal}(1989)}]{Liu89}%
  \BibitemOpen
  \bibfield  {author} {\bibinfo {author} {\bibfnamefont {D.~C.}\ \bibnamefont {Liu}}\ and\ \bibinfo {author} {\bibfnamefont {J.}~\bibnamefont {Nocedal}},\ }\bibfield  {title} {\bibinfo {title} {On the limited memory {BFGS} method for large scale optimization},\ }\href {https://doi.org/10.1007/BF01589116} {\bibfield  {journal} {\bibinfo  {journal} {Math. Program.}\ }\textbf {\bibinfo {volume} {45}},\ \bibinfo {pages} {503} (\bibinfo {year} {1989})}\BibitemShut {NoStop}%
\bibitem [{\citenamefont {Torquato}(2002)}]{To02a}%
  \BibitemOpen
  \bibfield  {author} {\bibinfo {author} {\bibfnamefont {S.}~\bibnamefont {Torquato}},\ }\href@noop {} {\emph {\bibinfo {title} {Random Heterogeneous Materials: Microstructure and Macroscopic Properties}}}\ (\bibinfo  {publisher} {Springer-Verlag},\ \bibinfo {address} {New York},\ \bibinfo {year} {2002})\BibitemShut {NoStop}%
\bibitem [{\citenamefont {Torquato}(2016)}]{To16b}%
  \BibitemOpen
  \bibfield  {author} {\bibinfo {author} {\bibfnamefont {S.}~\bibnamefont {Torquato}},\ }\bibfield  {title} {\bibinfo {title} {Disordered hyperuniform heterogeneous materials},\ }\href {https://doi.org/10.1088/0953-8984/28/41/414012} {\bibfield  {journal} {\bibinfo  {journal} {J. Phys.: Cond. Mat}\ }\textbf {\bibinfo {volume} {28}},\ \bibinfo {pages} {414012} (\bibinfo {year} {2016})}\BibitemShut {NoStop}%
\bibitem [{Note1()}]{Note1}%
  \BibitemOpen
  \bibinfo {note} {This condition can be relaxed to the general case of different phase diffusion constants without affecting the long-time behavior, as discussed in Ref.~\cite {To21d}.}\BibitemShut {Stop}%
\bibitem [{\citenamefont {Mitra}\ \emph {et~al.}(1992)\citenamefont {Mitra}, \citenamefont {Sen}, \citenamefont {Schwartz},\ and\ \citenamefont {Le~Doussal}}]{Mit92b}%
  \BibitemOpen
  \bibfield  {author} {\bibinfo {author} {\bibfnamefont {P.~P.}\ \bibnamefont {Mitra}}, \bibinfo {author} {\bibfnamefont {P.~N.}\ \bibnamefont {Sen}}, \bibinfo {author} {\bibfnamefont {L.~M.}\ \bibnamefont {Schwartz}},\ and\ \bibinfo {author} {\bibfnamefont {P.}~\bibnamefont {Le~Doussal}},\ }\bibfield  {title} {\bibinfo {title} {Diffusion propagator as a probe of the structure of porous media},\ }\href@noop {} {\bibfield  {journal} {\bibinfo  {journal} {Phys. Rev. Lett.}\ }\textbf {\bibinfo {volume} {68}},\ \bibinfo {pages} {3555} (\bibinfo {year} {1992})}\BibitemShut {NoStop}%
\bibitem [{\citenamefont {Mitra}\ \emph {et~al.}(1993)\citenamefont {Mitra}, \citenamefont {Sen},\ and\ \citenamefont {Schwartz}}]{Mi93}%
  \BibitemOpen
  \bibfield  {author} {\bibinfo {author} {\bibfnamefont {P.~P.}\ \bibnamefont {Mitra}}, \bibinfo {author} {\bibfnamefont {P.~N.}\ \bibnamefont {Sen}},\ and\ \bibinfo {author} {\bibfnamefont {L.~M.}\ \bibnamefont {Schwartz}},\ }\bibfield  {title} {\bibinfo {title} {Short-time behavior of the diffusion coefficient as a geometrical probe of porous media},\ }\href@noop {} {\bibfield  {journal} {\bibinfo  {journal} {Phys. Rev. B}\ }\textbf {\bibinfo {volume} {47}},\ \bibinfo {pages} {8565} (\bibinfo {year} {1993})}\BibitemShut {NoStop}%
\bibitem [{\citenamefont {{\O}ren}\ \emph {et~al.}(2002)\citenamefont {{\O}ren}, \citenamefont {Antonsen}, \citenamefont {Ruesl{\aa}tten},\ and\ \citenamefont {Bakke}}]{Or02}%
  \BibitemOpen
  \bibfield  {author} {\bibinfo {author} {\bibfnamefont {P.~E.}\ \bibnamefont {{\O}ren}}, \bibinfo {author} {\bibfnamefont {F.}~\bibnamefont {Antonsen}}, \bibinfo {author} {\bibfnamefont {H.~G.}\ \bibnamefont {Ruesl{\aa}tten}},\ and\ \bibinfo {author} {\bibfnamefont {S.}~\bibnamefont {Bakke}},\ }\bibfield  {title} {\bibinfo {title} {Numerical simulations of {NMR} responses for improved interpretations of nmr measurements in reservoir rocks},\ }in\ \href@noop {} {\emph {\bibinfo {booktitle} {SPE Annual Technical Conference and Exhibition}}}\ (\bibinfo {organization} {Society of Petroleum Engineers},\ \bibinfo {address} {San Antonio, Texas},\ \bibinfo {year} {2002})\ pp.\ \bibinfo {pages} {1--10}\BibitemShut {NoStop}%
\bibitem [{\citenamefont {Ruh}\ \emph {et~al.}(2023)\citenamefont {Ruh}, \citenamefont {Emerich}, \citenamefont {Scherer}, \citenamefont {Novikov},\ and\ \citenamefont {Kiselev}}]{Ruh_2023_Observation}%
  \BibitemOpen
  \bibfield  {author} {\bibinfo {author} {\bibfnamefont {A.}~\bibnamefont {Ruh}}, \bibinfo {author} {\bibfnamefont {P.}~\bibnamefont {Emerich}}, \bibinfo {author} {\bibfnamefont {H.}~\bibnamefont {Scherer}}, \bibinfo {author} {\bibfnamefont {D.~S.}\ \bibnamefont {Novikov}},\ and\ \bibinfo {author} {\bibfnamefont {V.~G.}\ \bibnamefont {Kiselev}},\ }\bibfield  {title} {\bibinfo {title} {Observation of magnetic structural universality and jamming transition with nmr},\ }\href {https://doi.org/10.1016/j.jmr.2023.107476} {\bibfield  {journal} {\bibinfo  {journal} {J. Magn. Reson.}\ }\textbf {\bibinfo {volume} {353}},\ \bibinfo {pages} {107476} (\bibinfo {year} {2023})}\BibitemShut {NoStop}%
\bibitem [{\citenamefont {Novikov}\ \emph {et~al.}(2014)\citenamefont {Novikov}, \citenamefont {Jensen}, \citenamefont {Helpern},\ and\ \citenamefont {Fieremans}}]{No14}%
  \BibitemOpen
  \bibfield  {author} {\bibinfo {author} {\bibfnamefont {D.~S.}\ \bibnamefont {Novikov}}, \bibinfo {author} {\bibfnamefont {J.~H.}\ \bibnamefont {Jensen}}, \bibinfo {author} {\bibfnamefont {J.~A.}\ \bibnamefont {Helpern}},\ and\ \bibinfo {author} {\bibfnamefont {E.}~\bibnamefont {Fieremans}},\ }\bibfield  {title} {\bibinfo {title} {Revealing mesoscopic structural universality with diffusion},\ }\href@noop {} {\bibfield  {journal} {\bibinfo  {journal} {Proc. Nat. Acad. Sci.}\ }\textbf {\bibinfo {volume} {111}},\ \bibinfo {pages} {5088} (\bibinfo {year} {2014})}\BibitemShut {NoStop}%
\bibitem [{\citenamefont {Lee}\ \emph {et~al.}(2020)\citenamefont {Lee}, \citenamefont {Papaioannou}, \citenamefont {Novikov},\ and\ \citenamefont {Fieremans}}]{Le20}%
  \BibitemOpen
  \bibfield  {author} {\bibinfo {author} {\bibfnamefont {H.-H.}\ \bibnamefont {Lee}}, \bibinfo {author} {\bibfnamefont {A.}~\bibnamefont {Papaioannou}}, \bibinfo {author} {\bibfnamefont {D.~S.}\ \bibnamefont {Novikov}},\ and\ \bibinfo {author} {\bibfnamefont {E.}~\bibnamefont {Fieremans}},\ }\bibfield  {title} {\bibinfo {title} {In vivo observation and biophysical interpretation of time-dependent diffusion in human cortical gray matter},\ }\href@noop {} {\bibfield  {journal} {\bibinfo  {journal} {NeuroImage}\ }\textbf {\bibinfo {volume} {222}},\ \bibinfo {pages} {117054} (\bibinfo {year} {2020})}\BibitemShut {NoStop}%
\bibitem [{\citenamefont {Bruggeman}(1935)}]{bruggeman_berechnung_1935}%
  \BibitemOpen
  \bibfield  {author} {\bibinfo {author} {\bibfnamefont {D.~A.~G.}\ \bibnamefont {Bruggeman}},\ }\bibfield  {title} {\bibinfo {title} {Berechnung verschiedener physikalischer konstanten von heterogenen substanzen. i. dielektrizit\"atskonstanten und leitf\"ahigkeiten der mischk\"orper aus isotropen substanzen},\ }\href {https://doi.org/10.1002/andp.19354160705} {\bibfield  {journal} {\bibinfo  {journal} {Annalen der Physik}\ }\textbf {\bibinfo {volume} {416}},\ \bibinfo {pages} {636} (\bibinfo {year} {1935})}\BibitemShut {NoStop}%
\bibitem [{\citenamefont {Torquato}(2009)}]{torquato_inverse_2009}%
  \BibitemOpen
  \bibfield  {author} {\bibinfo {author} {\bibfnamefont {S.}~\bibnamefont {Torquato}},\ }\bibfield  {title} {\bibinfo {title} {Inverse optimization techniques for targeted self-assembly},\ }\href {https://doi.org/10.1039/B814211B} {\bibfield  {journal} {\bibinfo  {journal} {Soft Matter}\ }\textbf {\bibinfo {volume} {5}},\ \bibinfo {pages} {1157} (\bibinfo {year} {2009})}\BibitemShut {NoStop}%
\bibitem [{\citenamefont {Yu}\ \emph {et~al.}(2017)\citenamefont {Yu}, \citenamefont {Zhang}, \citenamefont {Wang}, \citenamefont {Lee}, \citenamefont {Dong}, \citenamefont {Odom}, \citenamefont {Sun},\ and\ \citenamefont {Chen}}]{yu_characterization_2017}%
  \BibitemOpen
  \bibfield  {author} {\bibinfo {author} {\bibfnamefont {S.}~\bibnamefont {Yu}}, \bibinfo {author} {\bibfnamefont {Y.}~\bibnamefont {Zhang}}, \bibinfo {author} {\bibfnamefont {C.}~\bibnamefont {Wang}}, \bibinfo {author} {\bibfnamefont {W.-k.}\ \bibnamefont {Lee}}, \bibinfo {author} {\bibfnamefont {B.}~\bibnamefont {Dong}}, \bibinfo {author} {\bibfnamefont {T.~W.}\ \bibnamefont {Odom}}, \bibinfo {author} {\bibfnamefont {C.}~\bibnamefont {Sun}},\ and\ \bibinfo {author} {\bibfnamefont {W.}~\bibnamefont {Chen}},\ }\bibfield  {title} {\bibinfo {title} {Characterization and design of functional quasi-random nanostructured materials using spectral density function},\ }\bibfield  {journal} {\bibinfo  {journal} {Journal of Mechanical Design}\ }\textbf {\bibinfo {volume} {139}},\ \href {https://doi.org/10.1115/1.4036582} {10.1115/1.4036582} (\bibinfo {year} {2017})\BibitemShut {NoStop}%
\bibitem [{\citenamefont {Chen}\ and\ \citenamefont {Torquato}(2018)}]{chen_designing_2018}%
  \BibitemOpen
  \bibfield  {author} {\bibinfo {author} {\bibfnamefont {D.}~\bibnamefont {Chen}}\ and\ \bibinfo {author} {\bibfnamefont {S.}~\bibnamefont {Torquato}},\ }\bibfield  {title} {\bibinfo {title} {Designing disordered hyperuniform two-phase materials with novel physical properties},\ }\href {https://doi.org/10.1016/j.actamat.2017.09.053} {\bibfield  {journal} {\bibinfo  {journal} {Acta Mater.}\ }\textbf {\bibinfo {volume} {142}},\ \bibinfo {pages} {152} (\bibinfo {year} {2018})}\BibitemShut {NoStop}%
\bibitem [{\citenamefont {Iyer}\ \emph {et~al.}(2020)\citenamefont {Iyer}, \citenamefont {Dulal}, \citenamefont {Zhang}, \citenamefont {Ghumman}, \citenamefont {Chien}, \citenamefont {Balasubramanian},\ and\ \citenamefont {Chen}}]{iyer_designing_2020}%
  \BibitemOpen
  \bibfield  {author} {\bibinfo {author} {\bibfnamefont {A.}~\bibnamefont {Iyer}}, \bibinfo {author} {\bibfnamefont {R.}~\bibnamefont {Dulal}}, \bibinfo {author} {\bibfnamefont {Y.}~\bibnamefont {Zhang}}, \bibinfo {author} {\bibfnamefont {U.~F.}\ \bibnamefont {Ghumman}}, \bibinfo {author} {\bibfnamefont {T.}~\bibnamefont {Chien}}, \bibinfo {author} {\bibfnamefont {G.}~\bibnamefont {Balasubramanian}},\ and\ \bibinfo {author} {\bibfnamefont {W.}~\bibnamefont {Chen}},\ }\bibfield  {title} {\bibinfo {title} {Designing anisotropic microstructures with spectral density function},\ }\href {https://doi.org/10.1016/j.commatsci.2020.109559} {\bibfield  {journal} {\bibinfo  {journal} {Computational Materials Science}\ }\textbf {\bibinfo {volume} {179}},\ \bibinfo {pages} {109559} (\bibinfo {year} {2020})}\BibitemShut {NoStop}%
\bibitem [{\citenamefont {Shi}\ \emph {et~al.}(2025)\citenamefont {Shi}, \citenamefont {Jiao},\ and\ \citenamefont {Torquato}}]{shi_threedimensional_2025}%
  \BibitemOpen
  \bibfield  {author} {\bibinfo {author} {\bibfnamefont {W.}~\bibnamefont {Shi}}, \bibinfo {author} {\bibfnamefont {Y.}~\bibnamefont {Jiao}},\ and\ \bibinfo {author} {\bibfnamefont {S.}~\bibnamefont {Torquato}},\ }\bibfield  {title} {\bibinfo {title} {Three-dimensional construction of hyperuniform, nonhyperuniform, and antihyperuniform disordered heterogeneous materials and their transport properties via spectral density functions},\ }\href {https://doi.org/10.1103/PhysRevE.111.035310} {\bibfield  {journal} {\bibinfo  {journal} {Physical Review E}\ }\textbf {\bibinfo {volume} {111}},\ \bibinfo {pages} {035310} (\bibinfo {year} {2025})}\BibitemShut {NoStop}%
\bibitem [{\citenamefont {Skolnick}\ and\ \citenamefont {Torquato}(2025)}]{Skolnick_2025_Effective}%
  \BibitemOpen
  \bibfield  {author} {\bibinfo {author} {\bibfnamefont {M.}~\bibnamefont {Skolnick}}\ and\ \bibinfo {author} {\bibfnamefont {S.}~\bibnamefont {Torquato}},\ }\bibfield  {title} {\bibinfo {title} {Effective transport and mechanical properties of two-phase materials across the order-disorder spectrum},\ }\href {https://doi.org/https://doi.org/10.1016/j.actamat.2025.120921} {\bibfield  {journal} {\bibinfo  {journal} {Acta Materialia}\ }\textbf {\bibinfo {volume} {290}},\ \bibinfo {pages} {120921} (\bibinfo {year} {2025})}\BibitemShut {NoStop}%
\bibitem [{\citenamefont {Hashin}\ and\ \citenamefont {Shtrikman}(1962)}]{Ha62c}%
  \BibitemOpen
  \bibfield  {author} {\bibinfo {author} {\bibfnamefont {Z.}~\bibnamefont {Hashin}}\ and\ \bibinfo {author} {\bibfnamefont {S.}~\bibnamefont {Shtrikman}},\ }\bibfield  {title} {\bibinfo {title} {A variational approach to the theory of the effective magnetic permeability of multiphase materials},\ }\href@noop {} {\bibfield  {journal} {\bibinfo  {journal} {J. Appl. Phys.}\ }\textbf {\bibinfo {volume} {33}},\ \bibinfo {pages} {3125} (\bibinfo {year} {1962})}\BibitemShut {NoStop}%
\bibitem [{\citenamefont {Hashin}(1970)}]{Ha70}%
  \BibitemOpen
  \bibfield  {author} {\bibinfo {author} {\bibfnamefont {Z.}~\bibnamefont {Hashin}},\ }\bibfield  {title} {\bibinfo {title} {Theory of composite materials},\ }in\ \href@noop {} {\emph {\bibinfo {booktitle} {Mechanics of Composite Materials}}}\ (\bibinfo  {publisher} {Pergamon Press},\ \bibinfo {address} {New York},\ \bibinfo {year} {1970})\BibitemShut {NoStop}%
\end{thebibliography}%

\pagebreak
\widetext
\newpage

\section*{Supplemental Material \\ Dynamical properties of particulate composites derived from ultradense stealthy hyperuniform sphere packings}

\subsection*{Analytic cross-property relations}
\label{app:analit_cross_prop}

In this Appendix, we report the results of the analytic expressions of the cross-property relations discussed in the main text. In Fig. 14 we show the size of the transparency interval $K_T/K$ against the long-time spreadability parameter $\xi$ for the (a) TM and (b) TE polarizations in a transversely isotropic medium and (c) in a fully isotropic medium. The results are obtained by plotting Eq.~(30), with $\phi$ growing for decreasing $K_T/K$. The reported curves are used in the main text, for $\chi = 0.0025$ and $\chi = 0.45$, in comparison with the numerical data.

\begin{figure}[h!]
    \centering
    \includegraphics[height=0.3\linewidth]{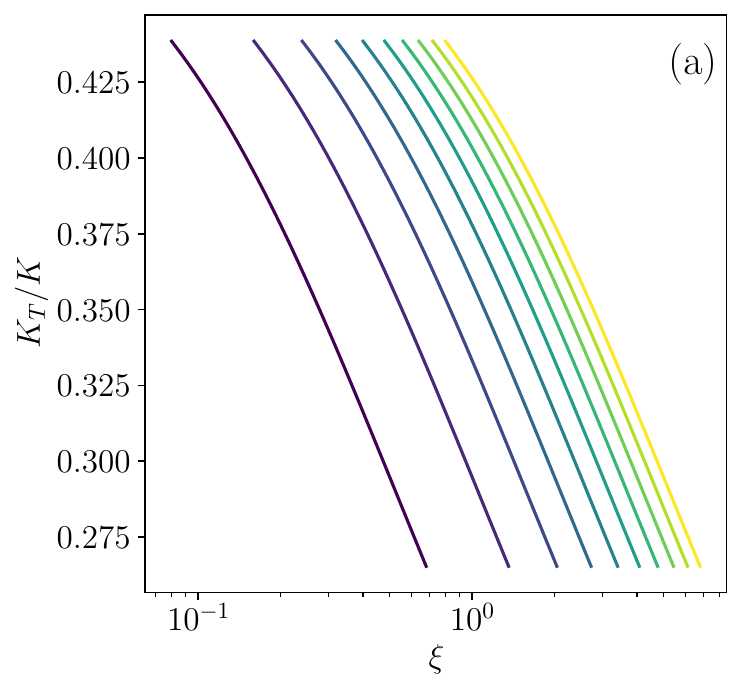}
    \includegraphics[height=0.3\linewidth]{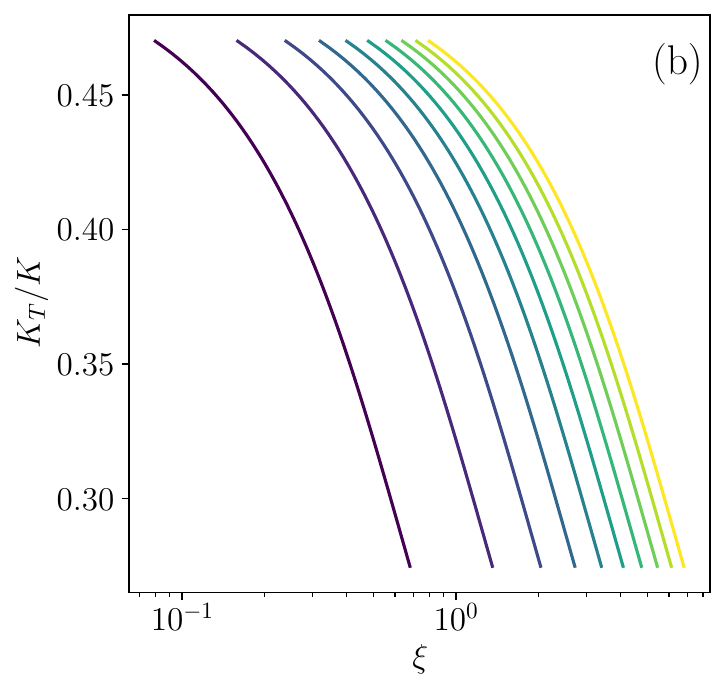}
    \includegraphics[height=0.3\linewidth]{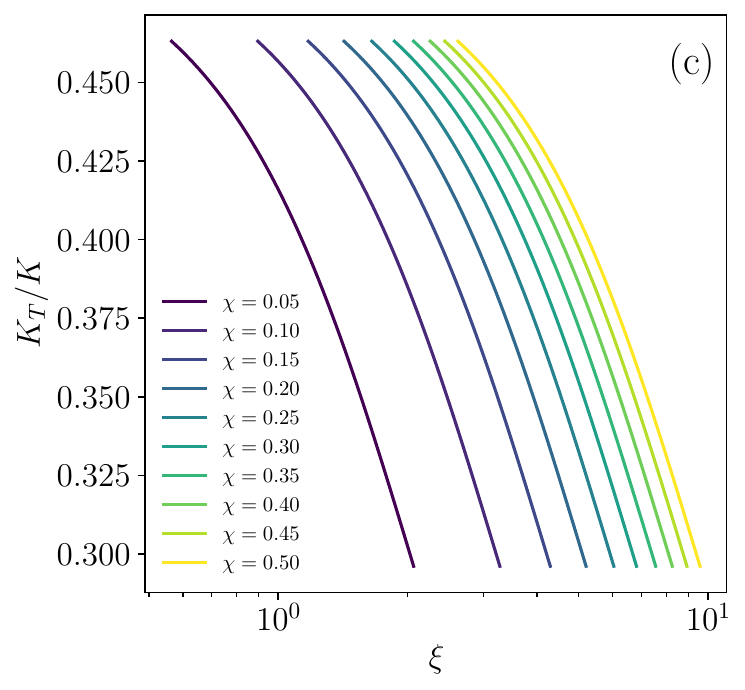}
    \caption{Analytical result for the cross-properties between the long-time spreadability parameter $\xi$ and the size of the optical transparency interval for the (a) TM and (b) TE polarizations in a transversely isotropic medium and (c) in a fully isotropic medium. The contrast ratio is $\epsilon_2/\epsilon_1 = 4$.}
    \label{fig:analit_cross_prop}
\end{figure}

\subsection*{Microstructural dependence of the physical properties}
\label{app:microstr}

In this Appendix, we discuss the effect of different microstructural arrangements, obtained with or without soft-core repulsion, on physical properties. In particular, we will consider the same point configurations analyzed in the main text, but we decorate them with spheres of the same radius. This way, we will be able to compare two-phase media with the same packing fraction and highlight the effect of the microstructure obtained with soft-core repulsion.

\subsubsection*{Autocovariance and Spectral Density}

\begin{figure}[h!]
    \centering
    \includegraphics[height=0.3\linewidth]{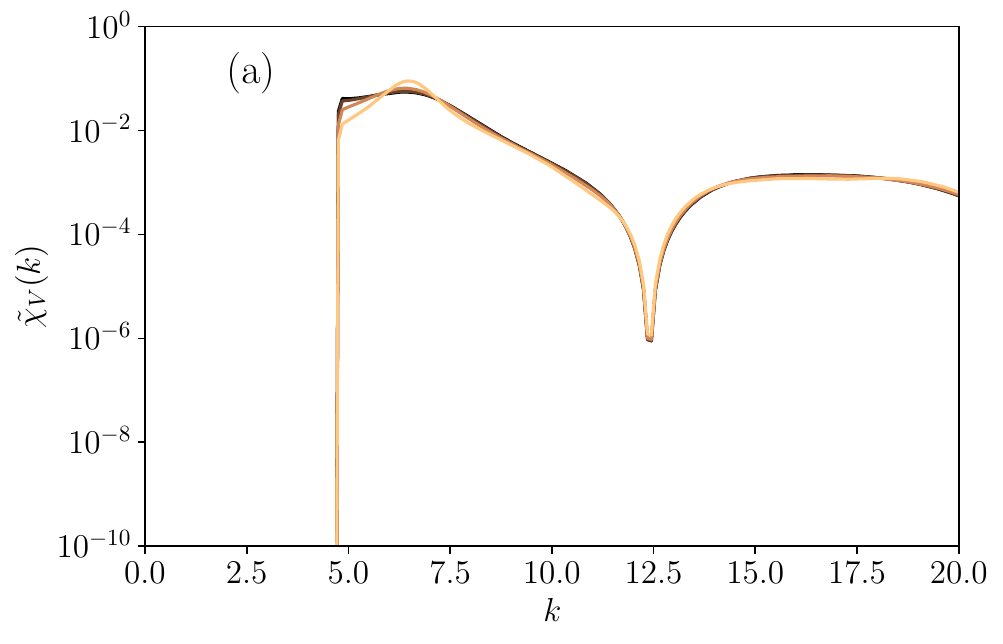}
    \includegraphics[height=0.3\linewidth]{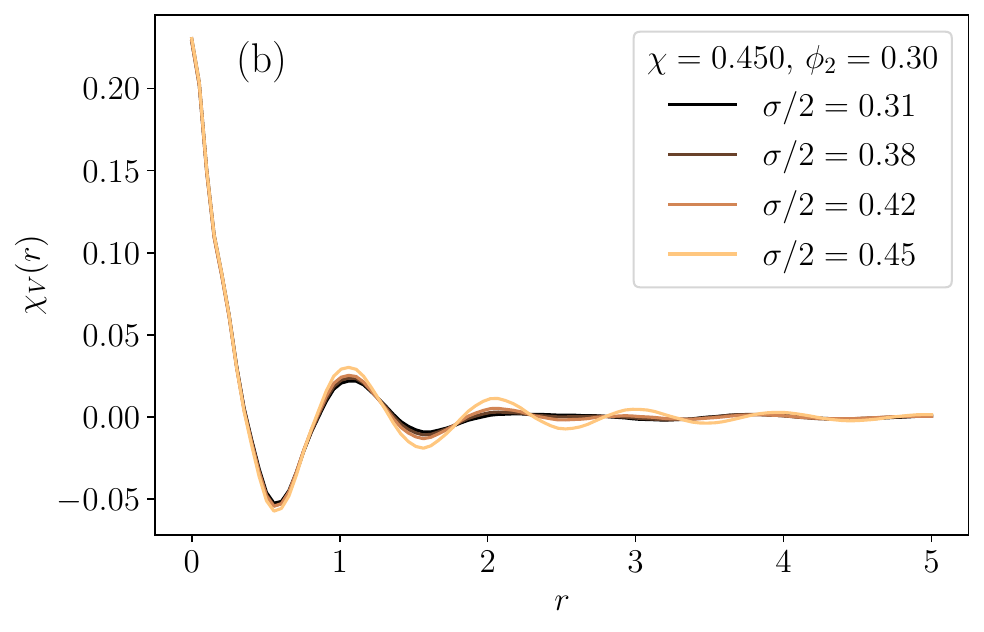}
    \caption{(a) Spectral density of four SHU packings having the same values of $\chi$ and $\phi$, but obtained for different values of $\sigma$. The red curve is obtained setting $a=\sigma/2$, while the blue curve has $a<\sigma/2$. (b) Corresponding autocovariance function for the same SHU packings.}
    \label{fig:spectr_dens_sigma}
\end{figure}

Let us start by considering the autocovariance function and spectral density, defined by Eq.~(8) and Eq.~(10). In Fig.~\ref{fig:spectr_dens_sigma}, we report the results for SHU packings having the same value of $\chi = 0.45$ and the same packing fraction $\phi = 0.3$ but obtained for different values of the soft-core size $\sigma$.

We can notice in Fig.~\ref{fig:spectr_dens_sigma}(a) that the spectral densities $\tilde{\chi}_{_{V}}(k)$ obtained for different values of $\sigma$ almost perfectly coincide: they vanish in the same range of wavenumber (as a consequence of the same value of $\chi$) and the position of the local minima also coincide, differently from what happens at different packing fractions. However, the difference in the microstructure manifests itself at intermediate wavenumbers, in correspondence with the maximum of $\tilde{\chi}_{_{V}}(k)$. In real space, such difference is reflected in the different oscillation periods of the autocovariance function, see Fig.~\ref{fig:spectr_dens_sigma}(b). Such differences are a consequence of the minimum pair distance between spheres, which for large $\sigma$ can be much larger than the sphere diameter.

\subsubsection*{Diffusion Spreadability}

The differences we have observed in the spectral density have direct consequences on the physical properties. In particular, let us observe that, in a small region around $k = K$, $\tilde{\chi}_{_{V}}(k;\sigma_1) > \tilde{\chi}_{_{V}}(k; \sigma_2)$ if $\sigma_1 < \sigma_2$. Therefore, when computing the excess spreadability $\mathcal{S}(\infty)- \mathcal{S}(t)$ according to Eq.~(12), more efficient spreadability is achieved for larger values of $\sigma$. This means that even if the greatest advantage of the soft-core repulsion is that of achieving very large packing fractions, an advantage is obtained even when having a smaller packing fraction than the maximum one, as a consequence of the sole microstructural properties. In particular, we can understand such an advantage as follows; consider two SHU packings with the same $\phi$ and different values of $\sigma = \sigma_1, \, \sigma_2$, with $\sigma_1 = 2a$ and $\sigma_2>\sigma_1$. The packing configuration obtained using $\sigma_1$ will have some spheres touching (see Fig. 1(a)), and some larger holes surrounding other spheres. By keeping the number density fixed, these “holes" will be reduced for larger $\sigma_2$, and this makes the spreading from the spheres to the matrix more efficient.

\begin{figure}[h!]
    \centering
    \includegraphics[width=0.5\linewidth]{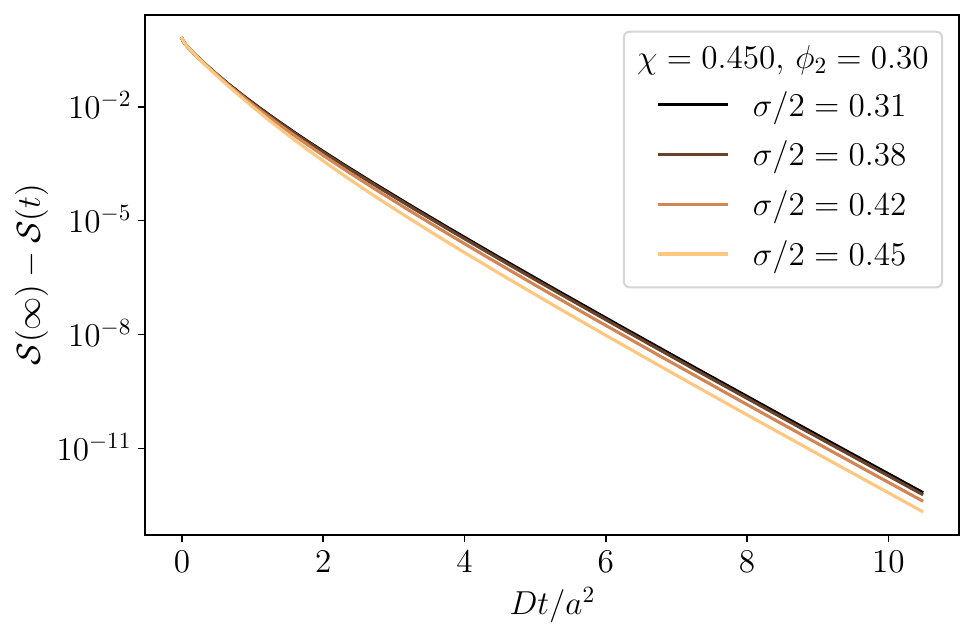}
    \caption{Excess diffusion spreadability for 2D SHU packings with soft-core repulsion having $\chi=0.45$ and $\phi = 0.3$, but different sizes $\sigma$ of the soft-core repulsion. Larger $\sigma$ leads to more efficient spreading.}
    \label{fig:2Ddiffspread_f}
\end{figure}

\subsubsection*{Effective Dynamic Dielectric Constant}

\begin{figure}[h!]
    \centering
    \includegraphics[width=0.49\linewidth]{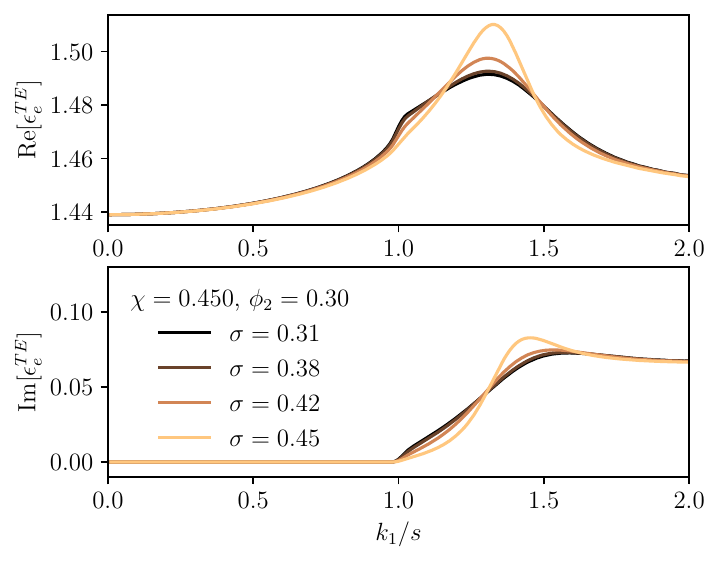}
    \includegraphics[width=0.49\linewidth]{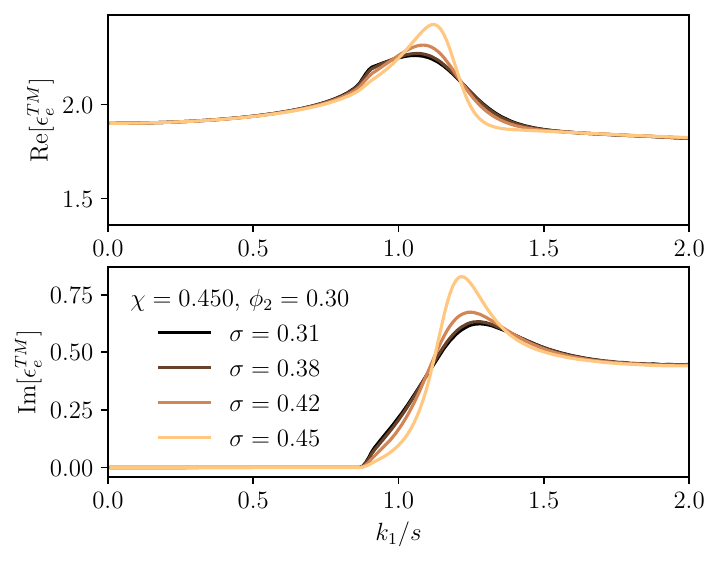}
    \caption{Effective dynamic dielectric constant of 3D transversely isotropic media obtained for the same $\chi = 0.45$ and packing fraction $\phi = 0.3$, but different values of $\sigma$. The left panel shows the real and imaginary part of the TE polarization, while the right panel shows the TM polarization.}
    \label{fig:EDDC_0.45_f}
\end{figure}

We now analyze the effect of the microstructure on the effective dynamic dielectric constant, and we show in Fig.~\ref{fig:EDDC_0.45_f} the results for 3D transversely isotropic media obtained for the same $\chi = 0.45$ and packing fraction $\phi = 0.3$, but different values of $\sigma$. 

As expected, the size of the transparency interval $K_T$ is the same, as all the curves are obtained for the same $\chi$ and $\phi$. On the other hand, the difference in the microstructure is reflected in the height of the attenuation peak, which increases with $\sigma$.

\begin{figure}[h!]
    \centering
    \includegraphics[width=0.49\linewidth]{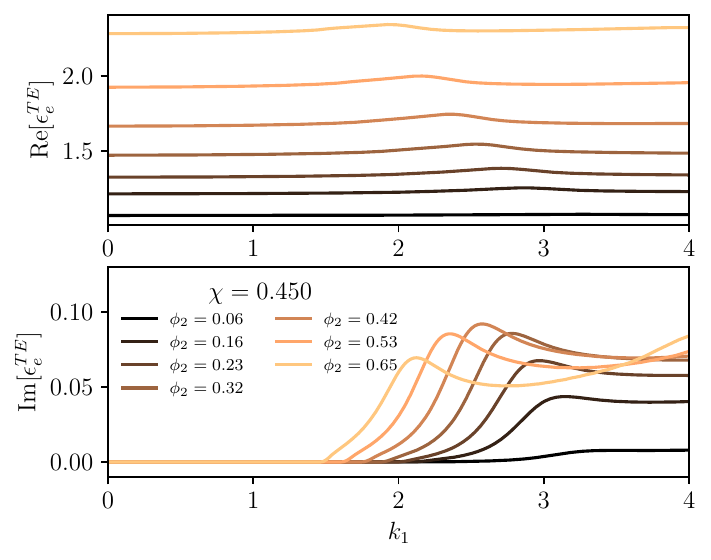}
    \includegraphics[width=0.49\linewidth]{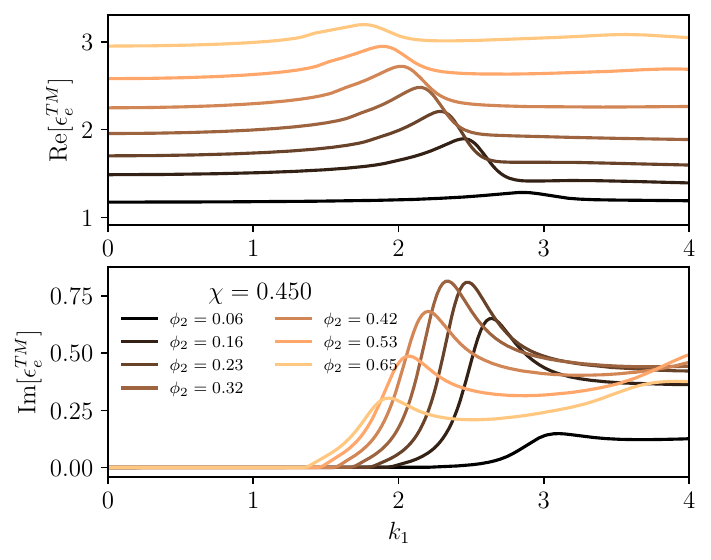}
    \caption{Effective dynamic dielectric constant of 3D transversely isotropic media obtained for the same $\chi = 0.45$ and same value of $\sigma$, but different values of $\phi$. The left panel shows the real and imaginary part of the TE polarization, while the right panel shows the TM polarization.}
    \label{fig:EDDC_phi}
\end{figure}

To complete our analysis, we show in Fig.~\ref{fig:EDDC_phi} the effective dynamic dielectric constant calculated for the same microstructures but with different packing fractions, obtained by decorating the point configurations with disks of various radii. We can notice that the height of the attenuation peak is not monotonic in $\phi$, and the maximum is achieved when the volume fraction of the two phases is comparable. Therefore, the dissipation due to scattering occurs more frequently.

\end{document}